\documentclass[11pt]{article}
\author{
Anand Babu\textsuperscript{1,*},
Rogério Almeida Gouvêa\textsuperscript{1},
Gian-Marco Rignanese\textsuperscript{1,2,*}\\[0.6em]
\small \textsuperscript{1}Institute of Condensed Matter and Nanosciences,\\
\small Université Catholique de Louvain, Louvain-la-Neuve, Belgium.\\
\small \textsuperscript{2}WEL Research Institute, Avenue Pasteur 6, Wavre, Belgium.
}
\usepackage[a4paper,margin=1in]{geometry}
\usepackage{graphicx}
\usepackage{booktabs}
\usepackage{tabularx}
\usepackage{array}
\usepackage{makecell}
\usepackage{amsmath,amssymb}
\usepackage{enumitem}
\usepackage{hyperref}
\usepackage{xurl}
\usepackage{caption}
\usepackage{tikz}
\usepackage{pgfplots}
\usepackage{float}
\pgfplotsset{compat=1.18}
\usetikzlibrary{arrows.meta,positioning,calc,shadows.blur,fit,shadings,decorations.pathmorphing,backgrounds,fadings,shapes.geometric}

\definecolor{virc70}{RGB}{122,209, 81}   %% 0.70 - Latent GD   (yellow-green)
\definecolor{virc60}{RGB}{ 53,183,121}   %% 0.60 - Langevin    (teal-green)
\definecolor{virc35}{RGB}{ 45,108,142}   %% 0.35 - Guided Diff (blue)
\definecolor{virc65}{RGB}{ 87,196,105}   %% 0.65 - Evolutionary(green)
\definecolor{virc25}{RGB}{ 59, 82,139}   %% 0.25 - RL / Policy (blue-purple)
\definecolor{virc72}{RGB}{137,213, 72}   %% 0.72 - Active learning (yellow-green)
\definecolor{qLL}{HTML}{FBF0E2}
\definecolor{qLLtxt}{HTML}{8A6A34}   %% low throughput, local     - warm neutral
\definecolor{qLR}{HTML}{FBEADB}
\definecolor{qLRtxt}{HTML}{B5642A}   %% low throughput, global    - light amber
\definecolor{qUL}{HTML}{E3ECF8}
\definecolor{qULtxt}{HTML}{2C5C93}   %% high throughput, local    - light steel
\definecolor{qUR}{HTML}{E2F0E6}
\definecolor{qURtxt}{HTML}{2E7048}   %% high throughput, global   - light teal

\definecolor{colGD}  {RGB}{ 31,119,180}   %% Latent GD   (tab-blue)
\definecolor{colSGLD}{RGB}{255,127, 14}   %% Langevin    (tab-orange)
\definecolor{colDIFF}{RGB}{ 44,160, 44}   %% Guided Diff (tab-green)
\definecolor{colEVO} {RGB}{214, 39, 40}   %% Evolutionary(tab-red)
\definecolor{colRL}  {RGB}{148,103,189}   %% RL / Policy (tab-purple)
\definecolor{colALL} {RGB}{ 90, 90, 90}   %% All engines (gray)
\definecolor{bandFeas}{RGB}{225,236,246}  %% Feasibility band (light blue)
\definecolor{bandOpt} {RGB}{252,236,224}  %% Optimization band (light orange)
\definecolor{bandEval}{RGB}{229,244,232}  %% Evaluation band   (light green)

\definecolor{cData} {RGB}{ 46,139, 87}   %% encoder / data   (sea-green)
\definecolor{cModel}{RGB}{ 31,119,180}   %% generator        (tab-blue)
\definecolor{cLab}  {RGB}{210,105, 30}   %% lab / char       (chocolate)
\definecolor{cEval} {RGB}{107, 70,193}   %% scheduler / eval (purple)
\definecolor{cMid}  {RGB}{240,243,246}   %% banner fill      (light grey)
\definecolor{fillData} {RGB}{223,243,230}
\definecolor{fillModel}{RGB}{219,232,246}
\definecolor{fillLab}  {RGB}{255,233,214}
\definecolor{fillEval} {RGB}{239,227,251}

\newcommand{\AlgoCloud}[7]{%
  \addplot[
    domain=0:360, samples=120, smooth, forget plot,
    draw=black!18, fill=#5, fill opacity=0.09
  ] ({#1 + #3*cos(x)},{#2 + #4*sin(x)}) \closedcycle;
  \node[font=\bfseries\small\sffamily, text=black!80]
    at (axis cs:#1,{#2 + #4 - #7}) {#6};%
}

\newcolumntype{Y}{>{\raggedright\arraybackslash}X}
\renewcommand{\arraystretch}{1.15}

\usepackage{xcolor}
\usepackage{tcolorbox}
\tcbuselibrary{skins,breakable}
\newif\ifrevisionmode
\revisionmodetrue   % <- set to \revisionmodefalse to produce a clean (black) submission copy
\definecolor{revred}{RGB}{0,0,0}
\definecolor{pipeblue}{RGB}{31,86,168}
\definecolor{pipedk}{RGB}{18,58,120}
\definecolor{pipelt}{RGB}{225,236,250}
\definecolor{failred}{RGB}{196,40,45}
\definecolor{faildk}{RGB}{150,26,32}
\definecolor{faillt}{RGB}{253,236,236}
\definecolor{evgreen}{RGB}{31,120,60}
\definecolor{evdk}{RGB}{20,88,44}
\definecolor{evlt}{RGB}{233,246,236}
\definecolor{metricblue}{RGB}{40,90,160}
\definecolor{dotblue}{RGB}{40,90,175}
\definecolor{dotdk}{RGB}{22,58,130}
\definecolor{cutgray}{RGB}{140,140,140}
\ifrevisionmode
  \newcommand{\revised}[1]{{\color{revred}#1}}
  \newcommand{\cut}[1]{}% removed text is fully deleted (no grey marker) in revision mode
\else
  \newcommand{\revised}[1]{#1}
  \newcommand{\cut}[1]{}
\fi
\newtcolorbox{revbox}[1][]{breakable,colback=revred!3,colframe=revred!55,
  boxrule=0.6pt,left=6pt,right=6pt,top=5pt,bottom=5pt,fonttitle=\bfseries,#1}
\title{\textbf{Towards Automated Discovery:}\\\textbf{A Review of Generative Models, Multimodal Learning and Closed-Loop Workflows in Inverse Materials Design}}
\date{}
\begin{document}
\maketitle

\begin{abstract}
Inverse materials design is shifting materials discovery from forward prediction to targeted proposal of candidates that satisfy objectives under physical constraints. Here, we review recent advances in generative crystal structure modeling, multimodal learning, and closed-loop design pipelines for crystalline solids. We survey how modern generators learn chemical-structural priors from large databases to enable controllable sampling of periodic structures, and compare leading model classes including variational autoencoders, normalizing flows, autoregressive formulations, and diffusion models. \revised{Across these families, we trace where feasibility constraints and physical priors enter the workflow, from representation choices and training objectives to sampling-time guidance and post-generation screening and relaxation.} We also discuss how multimodal learning fuses diverse materials modalities, including crystal structures, thermodynamic, electronic information, microscopy, spectroscopy, processing context, and scientific text, to construct \revised{shared multimodal materials representations of}\cut{a more universal, transferable representation of} chemical space. In addition, diverse inverse-design strategies are examined, above all those that integrate conditional generation with latent optimization, Bayesian optimization, reinforcement learning, and active learning. We set out recurring failure modes, such as surrogate exploitation, diversity collapse, distribution shift, and the stability-synthesizability gap, and outline discovery-grade evaluation practices based on staged reporting of validity, novelty, uniqueness, stability, and cost\revised{. To support credible discovery claims, we define a nine-rung discovery-credibility ladder that fixes what evidence a claim at each level requires, and a multi-level minimum reporting standard that we propose new-method papers and benchmarks adopt: declared matching tolerances and database snapshots; uniqueness, training-set memorization and external rediscovery reported as three separate quantities; novelty as a continuous distance distribution in place of a binary rate; the full energy-above-hull distribution with its functional and hull version; relaxation-survival and dynamical-stability rates; and the validation cost per credible hit. On this standard, a headline validity or S.U.N.\ rate reported without these disclosures should be treated as uninformative}.
\end{abstract}

\section*{Progress and Potential}
AI-enabled materials discovery is opening a path from prediction-centered workflows toward automated discovery platforms that can propose, evaluate, and refine candidate materials. Generative modeling, multimodal learning, active learning, and inverse design of materials each contribute to this transition, but their full potential depends on how they are connected within a validation-aware discovery workflow. This review brings these components together and shows how candidate generation, target specification, validation cost, and closed-loop feedback can be treated as parts of a unified materials discovery framework.
This integration is crucial because faster screening alone is not enough. AI workflows can generate candidates that appear promising but are unstable, unsynthesizable, poorly grounded in physical constraints, or selected through overoptimized proxy objectives. We discuss the main failure modes, stability-synthesizability gap, and show how they can be diagnosed through staged evaluation. We therefore set out a practical framework for building more reliable automated discovery workflows for future materials challenges and outline an integrated path toward self-driving laboratories.

\section{Introduction}\label{sec:intro}

Materials discovery is undergoing a fundamental transition, moving away from trial-and-error methodologies toward autonomous search strategies conditioned on predefined functional requirements. For decades, the dominant workflow has been forward: propose a composition, structure, compute or measure properties, and iterate. Recent advances in machine learning (ML) and deep learning (DL) are fundamentally reshaping the efficiency of this loop by providing computationally efficient predictive models, uncertainty-aware selection strategies, and generative models that can propose plausible crystals directly. This shift reconfigures materials discovery: instead of screening only what can be enumerated, the search is conditioned on predefined objectives and bounded by physical and practical feasibility. Inverse design redefines this process by asking which candidate(s) can realize a target property profile, thereby recasting materials discovery as a directed search problem, guiding exploration toward rare regions of structure-composition space where useful trade-offs may exist.

This transition was enabled by the data-centric materials stack: high-throughput computation and standardized workflows that made materials evaluation at scale routine, and large open databases that made prior knowledge machine-readable and reusable [\ref{cit:Curtarolo2013}-\ref{cit:Ong2013}]. In parallel, active learning and uncertainty-aware decision-making provided a principled way to allocate expensive simulations and experiments to the most informative candidates [\ref{cit:Lookman2019}-\ref{cit:Hase2021}].

In contrast to traditional screening, generative models attempt to learn a prior over physically plausible structures so that the search begins in regions that are statistically and physically meaningful. Recent work spans multiple model families and conditioning strategies, from deep generative models aimed at crystal structure prediction (CSP)-like generation [\ref{cit:Luo2024}-\ref{cit:Qiu2025}] to more general-purpose generators designed for controllable design across broader chemical spaces [\ref{cit:Zeni2025}-\ref{cit:Song2020}].

An equally important aspect is that materials knowledge is inherently multimodal. The evidence used to understand, design, and validate materials is distributed across many forms, including structural descriptions, microscopy, diffraction patterns, spectra, process logs, thermodynamic data, device curves, and scientific text. Multimodal learning aims to align these signals into representations that are more robust than any single modality, and, crucially for inverse design, more usable for conditioning and constraint specification [\ref{cit:Wu2025}-\ref{cit:Venugopal2024}].

These advances point to a closed-loop paradigm for inverse materials design, in which generative proposal models are coupled to verification, synthesis, and measurement systems, and are updated iteratively in response to both unsuccessful and successful outcomes [\ref{cit:Stach2021}-\ref{cit:FloresLeonar2020}]. Realizing this vision, however, requires progress on several fronts. Inverse-design approaches must satisfy multiple constraints simultaneously, including structural validity, stability, and target functionality, while also remaining robust to distribution shift, dataset bias, and poorly specified objectives. A further challenge is that materials predicted to be stable by density functional theory (DFT) are not always experimentally realizable: some require narrow growth windows, some are limited by kinetic barriers, and others are outcompeted by phases that form more readily. This mismatch makes rigorous evaluation essential. Beyond target-property scores, inverse-design studies should assess validity, novelty, uniqueness, and stability using consistent protocols, so that alternative closed-loop strategies can be compared fairly [\ref{cit:DeBreuck2025}-\ref{cit:Dunn2020}]. As foundation-model approaches enter materials science, these issues will become even more important, because many downstream workflows will rely on shared pretrained priors and representations [\ref{cit:PyzerKnapp2025}-\ref{cit:Batra2021}].

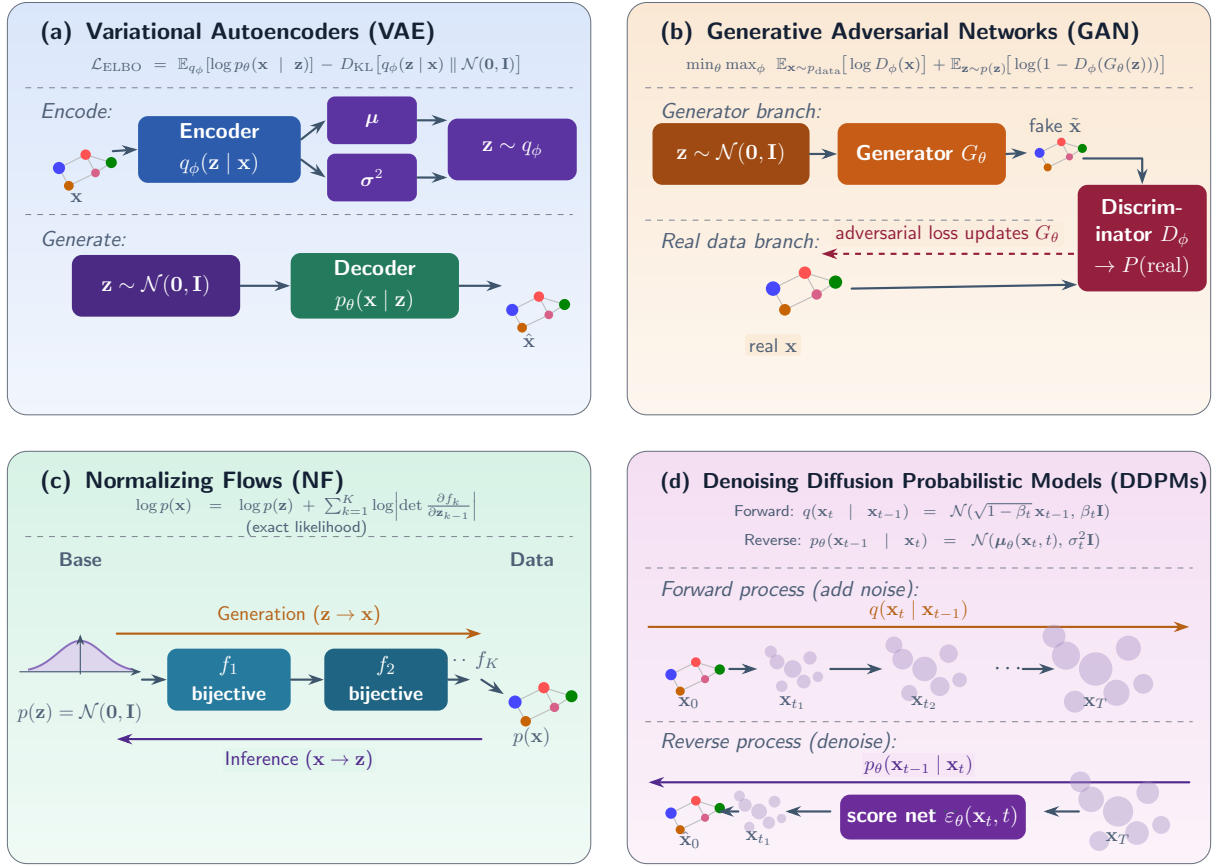
\begin{figure}[t]
\centering
% --- Figure 1 colors ---
\definecolor{inkF}{RGB}{20,32,52}
\definecolor{mutF}{RGB}{78,95,118}
\definecolor{bgVAE}{RGB}{218,230,250}
\definecolor{bgGAN}{RGB}{250,232,212}
\definecolor{bgNF}{RGB}{215,242,230}
\definecolor{bgDIF}{RGB}{244,222,242}
\definecolor{cEnc}{RGB}{45,92,172}
\definecolor{cLat}{RGB}{92,52,162}
\definecolor{cDec}{RGB}{35,122,92}
\definecolor{cGen}{RGB}{198,92,22}
\definecolor{cDisc}{RGB}{148,32,62}
\definecolor{cFlow}{RGB}{38,122,158}
\definecolor{cDnet}{RGB}{108,45,155}
\definecolor{cArr}{RGB}{62,85,108}
\definecolor{cFwd}{RGB}{182,98,22}
\definecolor{cRev}{RGB}{82,42,142}
% --- Figure styles and pics ---
\tikzset{
  fT/.style={font=\sffamily\bfseries\large, text=inkF},
  fE/.style={font=\sffamily\small, text=mutF, align=center},
  fEq/.style={font=\sffamily\scriptsize, text=mutF, align=center},
  fL/.style={font=\sffamily\normalsize\itshape, text=mutF},
  fP/.style={rounded corners=10pt, draw=white!35!inkF, line width=0.3pt},
  fDv/.style={draw=white!45!inkF, line width=0.4pt, dashed},
  fB/.style={rounded corners=5pt, minimum height=11mm, minimum width=30mm, align=center, font=\sffamily\normalsize\bfseries, text=white},
  fS/.style={rounded corners=4pt, minimum height=8.5mm, minimum width=16mm, align=center, font=\sffamily\small\bfseries, text=white},
  fArr/.style={-{Stealth[length=2.8mm,width=1.9mm]}, line width=1.1pt, draw=cArr,
    line cap=round, line join=round},
  fBi/.style={{Stealth[length=2.8mm,width=1.9mm]}-{Stealth[length=2.8mm,width=1.9mm]},
    line width=1.1pt, draw=cArr, line cap=round, line join=round},
  fFw/.style={-{Stealth[length=2.8mm,width=1.9mm]}, line width=1.1pt, draw=cFwd,
    line cap=round, line join=round},
  fRv/.style={-{Stealth[length=2.8mm,width=1.9mm]}, line width=1.1pt, draw=cRev,
    line cap=round, line join=round},
  crystal/.pic={
    \draw[gray!52,line width=0.55pt]
      (0,0)--(0.54,0.27) (0,0)--(0.20,-0.36)
      (0.54,0.27)--(1.06,0.11) (0.54,0.27)--(0.74,-0.10)
      (0.20,-0.36)--(0.74,-0.10) (1.06,0.11)--(0.74,-0.10);
    \fill[blue!72!white]   (0,0)        circle (3.5pt);
    \fill[red!68!white]    (0.54,0.27)  circle (3.0pt);
    \fill[green!52!black]  (1.06,0.11)  circle (3.2pt);
    \fill[orange!78!black] (0.20,-0.36) circle (2.8pt);
    \fill[purple!62!white] (0.74,-0.10) circle (2.5pt);
  },
  noiseblob/.pic={
    \foreach \nx/\ny/\nr in {0/0/3.2, 0.22/0.14/2.4, -0.20/0.09/2.7, 0.14/-0.16/2.5, -0.14/-0.22/2.1, 0.32/-0.10/1.9, -0.28/0.22/2.2}{
      \fill[cRev!52,opacity=0.40] (\nx,\ny) circle (\nr pt);
    }
  }
}
\resizebox{\linewidth}{!}{%
\begin{tikzpicture}[x=1cm,y=1cm, >={Stealth[length=3.2mm,width=2.2mm]}]

\def\gp{0.65}
\def\pW{10.40}
\def\pH{7.35}

\coordinate (A) at (0,\pH+\gp);
\coordinate (B) at (\pW+\gp,\pH+\gp);
\coordinate (C) at (0,0);
\coordinate (D) at (\pW+\gp,0);

% ================================================================
% PANEL (a): VAE
% ================================================================
\path[fP,shade,top color=bgVAE,bottom color=bgVAE!38]
  (A) rectangle ++(\pW,\pH);

\node[fT,anchor=west] at ($(A)+(0.45,\pH-0.55)$)
  {(a)\enspace Variational Autoencoders (VAE)};

\node[fEq,anchor=west,text width=9.3cm] at ($(A)+(0.55,\pH-1.12)$)
  {$\mathcal{L}_{\mathrm{ELBO}}
  = \mathbb{E}_{q_\phi}[\log p_\theta(\mathbf{x}\mid\mathbf{z})]
  - D_{\mathrm{KL}}\!\left[q_\phi(\mathbf{z}\mid\mathbf{x})\;\|\;\mathcal{N}(\mathbf{0},\mathbf{I})\right]$};

\draw[fDv] ($(A)+(0.32,\pH-1.52)$)--($(A)+(\pW-0.32,\pH-1.52)$);

\node[fL,anchor=west] at ($(A)+(0.45,\pH-1.92)$) {Encode:};

\pic[scale=0.88] at ($(A)+(0.92,\pH-2.95)$) {crystal};
\node[fE] at ($(A)+(1.23,\pH-3.48)$) {$\mathbf{x}$};

\node[fB,fill=cEnc,minimum width=29mm] (ENC) at ($(A)+(3.78,\pH-2.62)$)
  {Encoder\\[2pt]$q_\phi(\mathbf{z}\mid\mathbf{x})$};

\node[fS,fill=cLat] (MU)  at ($(A)+(6.50,\pH-2.10)$) {$\boldsymbol{\mu}$};
\node[fS,fill=cLat] (SIG) at ($(A)+(6.50,\pH-3.12)$) {$\boldsymbol{\sigma}^2$};

\node[fB,fill=cLat,minimum width=23mm] (LAT) at ($(A)+(9.00,\pH-2.62)$)
  {$\mathbf{z}\sim q_\phi$};

\draw[fArr] ($(A)+(1.88,\pH-2.65)$)--(ENC.west);
\draw[fArr] ($(ENC.east)+(0,0.18)$)--($(MU.west)+(-0.02,0)$);
\draw[fArr] ($(ENC.east)+(0,-0.18)$)--($(SIG.west)+(-0.02,0)$);
\coordinate (LATmu)  at ($(LAT.west)+(0,0.52)$);
\coordinate (LATsig) at ($(LAT.west)+(0,-0.50)$);
\draw[fArr] (MU.east)--(LATmu);
\draw[fArr] (SIG.east)--(LATsig);

\draw[fDv] ($(A)+(0.32,\pH-3.78)$)--($(A)+(\pW-0.32,\pH-3.78)$);

\node[fL,anchor=west] at ($(A)+(0.45,\pH-4.20)$) {Generate:};

\node[fB,fill=cLat!80!black,minimum width=30mm] (PRI) at ($(A)+(2.65,\pH-5.05)$)
  {$\mathbf{z}\sim\mathcal{N}(\mathbf{0},\mathbf{I})$};

\node[fB,fill=cDec,minimum width=30mm] (DEC) at ($(A)+(6.55,\pH-5.05)$)
  {Decoder\\[2pt]$p_\theta(\mathbf{x}\mid\mathbf{z})$};

\pic[scale=0.88] at ($(A)+(9.00,\pH-5.48)$) {crystal};
\node[fE] at ($(A)+(9.30,\pH-6.02)$) {$\hat{\mathbf{x}}$};

\draw[fArr] (PRI.east)--(DEC.west);
\draw[fArr] (DEC.east)--($(A)+(8.82,\pH-5.05)$);

% ================================================================
% PANEL (b): GAN
% ================================================================
\path[fP,shade,top color=bgGAN,bottom color=bgGAN!38]
  (B) rectangle ++(\pW,\pH);
\begin{scope}
\clip (B) rectangle ++(\pW,\pH);

\node[fT,anchor=west] at ($(B)+(0.45,\pH-0.55)$)
  {(b)\enspace Generative Adversarial Networks (GAN)};

\node[fEq,anchor=west,text width=9.3cm] at ($(B)+(0.55,\pH-1.12)$)
  {$\min_\theta \max_\phi\;
  \mathbb{E}_{\mathbf{x}\sim p_{\mathrm{data}}}\!\big[\log D_\phi(\mathbf{x})\big]
  + \mathbb{E}_{\mathbf{z}\sim p(\mathbf{z})}\!\big[\log(1-D_\phi(G_\theta(\mathbf{z})))\big]$};

\draw[fDv] ($(B)+(0.32,\pH-1.52)$)--($(B)+(\pW-0.32,\pH-1.52)$);

\node[fL,anchor=west] at ($(B)+(0.45,\pH-1.92)$) {Generator branch:};

\node[fB,fill=cGen!80!black,minimum width=28mm] (ZIN) at ($(B)+(1.85,\pH-2.70)$)
  {$\mathbf{z}\sim\mathcal{N}(\mathbf{0},\mathbf{I})$};

\node[fB,fill=cGen,minimum width=30mm] (GEN) at ($(B)+(5.25,\pH-2.70)$)
  {Generator $G_\theta$};

%% fake crystal — shifted right; label placed ABOVE crystal for clarity
\pic[scale=0.72] at ($(B)+(7.35,\pH-2.68)$) {crystal};
\node[fE,fill=bgGAN!88,inner sep=2pt,rounded corners=2pt]
  at ($(B)+(7.63,\pH-2.18)$)
  {fake $\tilde{\mathbf{x}}$};

\node[fB,fill=cDisc,minimum width=23mm,minimum height=19mm] (DISC)
  at ($(B)+(9.18,\pH-4.20)$)
  {Discrim-\\inator $D_\phi$\\[2pt]$\to P(\mathrm{real})$};

%% z → Generator → fake crystal → Discriminator
\draw[fArr] (ZIN.east)--(GEN.west);
\draw[fArr] (GEN.east)--($(B)+(7.13,\pH-2.68)$);
\draw[fArr] ($(B)+(8.13,\pH-2.78)$) -| (DISC.north);

\draw[fDv] ($(B)+(0.32,\pH-3.88)$)--($(B)+(7.65,\pH-3.88)$);

\node[fL,anchor=west] at ($(B)+(0.45,\pH-4.24)$) {Real data branch:};

%% real crystal moved left to use available panel space
\pic[scale=1.05] at ($(B)+(2.60,\pH-5.10)$) {crystal};
\node[fE,anchor=north,fill=bgGAN!88,inner sep=2pt,rounded corners=2pt]
  at ($(B)+(2.60,\pH-5.90)$)
  {real $\mathbf{x}$};

\draw[fArr]
  ($(B)+(4.00,\pH-5.10)$)
  --($(DISC.west)+(0,-0.83)$);

%% adversarial feedback — shifted down to give clear vertical gap from fake-x label
\draw[cDisc,line width=1.1pt,dashed,-{Stealth[length=2.8mm,width=1.9mm]}]
  ($(DISC.west)+(-0.06,-0.28)$)--+(-4.50,0)
  node[midway,above=3pt,fE,text=cDisc,fill=bgGAN!80,inner sep=2pt,rounded corners=2pt]
  {adversarial loss updates $G_\theta$};
\end{scope}

% ================================================================
% PANEL (c): Normalizing Flows
% ================================================================
\path[fP,shade,top color=bgNF,bottom color=bgNF!38]
  (C) rectangle ++(\pW,\pH);

\node[fT,anchor=west] at ($(C)+(0.45,\pH-0.55)$)
  {(c)\enspace Normalizing Flows (NF)};

\node[fEq,anchor=west,text width=9.3cm] at ($(C)+(0.55,\pH-1.12)$)
  {$\log p(\mathbf{x}) = \log p(\mathbf{z})
  + \sum_{k=1}^{K}\log\!\left|\det \frac{\partial f_k}{\partial \mathbf{z}_{k-1}}\right|$\\[-1pt]
  (exact likelihood)};

\draw[fDv] ($(C)+(0.32,\pH-1.52)$)--($(C)+(\pW-0.32,\pH-1.52)$);

\begin{scope}[shift={($(C)+(1.30,3.42)$)},xscale=0.62,yscale=0.58]
  \fill[cLat!20]
    plot[domain=-1.5:1.5,samples=38,smooth]
    (\x,{0.95*exp(-1.2*\x*\x)}) -- (1.5,0) -- (-1.5,0) -- cycle;
  \draw[cLat!72,line width=1.0pt]
    plot[domain=-1.5:1.5,samples=38,smooth]
    (\x,{0.95*exp(-1.2*\x*\x)});
  \draw[draw=cArr,-Stealth,line width=0.55pt] (-1.75,0)--(1.75,0);
  \draw[draw=cArr,-Stealth,line width=0.55pt] (0,-0.08)--(0,1.15);
\end{scope}

\node[fE] at ($(C)+(1.30,\pH-1.92)$) {\textbf{Base}};
\node[fE] at ($(C)+(1.30,2.68)$) {$p(\mathbf{z})=\mathcal{N}(\mathbf{0},\mathbf{I})$};

\node[fB,fill=cFlow,minimum width=22mm] (F1) at ($(C)+(3.95,3.28)$)
  {$f_1$\\[1pt]\small bijective};

\node[fB,fill=cFlow!82!black,minimum width=22mm] (F2) at ($(C)+(6.75,3.28)$)
  {$f_2$\\[1pt]\small bijective};

\node[fE] at ($(C)+(8.28,3.62)$) {$\cdots\; f_K$};

\pic[scale=0.95] at ($(C)+(9.05,2.88)$) {crystal};
\node[fE] at ($(C)+(9.35,2.25)$) {$p(\mathbf{x})$};
\node[fE] at ($(C)+(9.35,\pH-1.92)$) {\textbf{Data}};

\draw[fFw]
  ($(C)+(1.95,4.10)$)--($(C)+(8.45,4.10)$)
  node[midway,above,fE,text=cFwd] {Generation ($\mathbf{z}\to\mathbf{x}$)};

\draw[fRv]
  ($(C)+(8.45,2.22)$)--($(C)+(1.95,2.22)$)
  node[midway,below,yshift=-4pt,fE,text=cRev,fill=bgNF!72,inner sep=1pt]
  {Inference ($\mathbf{x}\to\mathbf{z}$)};

\draw[fArr] ($(C)+(2.40,3.28)$)--(F1.west);
\draw[fArr] (F1.east)--(F2.west);
\draw[fArr] (F2.east)--($(C)+(8.25,3.28)$);
\draw[fArr] ($(C)+(8.45,3.28)$)--($(C)+(8.85,3.02)$);

% ================================================================
% PANEL (d): Diffusion / DDPM
% ================================================================
\path[fP,shade,top color=bgDIF,bottom color=bgDIF!38]
  (D) rectangle ++(\pW,\pH);
\begin{scope}
\clip (D) rectangle ++(\pW,\pH);

\node[anchor=west,font=\sffamily\bfseries\normalsize,text=inkF] at ($(D)+(0.45,\pH-0.55)$)
  {(d)\enspace Denoising Diffusion Probabilistic Models (DDPMs)};

\node[fEq,anchor=north west,text width=9.15cm] at ($(D)+(0.55,\pH-0.78)$)
  {Forward: $q(\mathbf{x}_t\mid\mathbf{x}_{t-1})
   = \mathcal{N}(\sqrt{1-\beta_t}\,\mathbf{x}_{t-1},\,\beta_t\mathbf{I})$\\[5pt]
   Reverse: $p_\theta(\mathbf{x}_{t-1}\mid\mathbf{x}_t)
   = \mathcal{N}(\boldsymbol{\mu}_\theta(\mathbf{x}_t,t),\,\sigma_t^2\mathbf{I})$};

\draw[fDv] ($(D)+(0.32,\pH-2.08)$)--($(D)+(\pW-0.32,\pH-2.08)$);

\node[fL,anchor=west] at ($(D)+(0.45,\pH-2.48)$)
  {Forward process (add noise):};

\draw[fFw]
  ($(D)+(0.38,\pH-3.13)$)--($(D)+(\pW-0.38,\pH-3.13)$)
  node[midway,above=2pt,fE,text=cFwd,fill=bgDIF!72,inner sep=1pt]
  {$q(\mathbf{x}_t\mid\mathbf{x}_{t-1})$};

\pic[scale=0.82] at ($(D)+(0.78,\pH-3.98)$) {crystal};
\node[fE] at ($(D)+(1.10,\pH-4.48)$) {$\mathbf{x}_0$};

\draw[fArr] ($(D)+(1.82,\pH-3.88)$)--($(D)+(2.32,\pH-3.88)$);

\pic[scale=1.40] at ($(D)+(2.95,\pH-3.88)$) {noiseblob};
\node[fE] at ($(D)+(2.95,\pH-4.48)$) {$\mathbf{x}_{t_1}$};

\draw[fArr] ($(D)+(3.62,\pH-3.88)$)--($(D)+(4.55,\pH-3.88)$);

\pic[scale=1.95] at ($(D)+(5.30,\pH-3.88)$) {noiseblob};
\node[fE] at ($(D)+(5.30,\pH-4.48)$) {$\mathbf{x}_{t_2}$};

\node[font=\large\bfseries,text=cArr] at ($(D)+(6.82,\pH-3.88)$) {$\cdots$};

\draw[fArr] ($(D)+(7.05,\pH-3.88)$)--($(D)+(7.62,\pH-3.88)$);

\pic[scale=2.65] at ($(D)+(8.35,\pH-3.88)$) {noiseblob};
\node[fE] at ($(D)+(8.35,\pH-4.48)$) {$\mathbf{x}_{T}$};

\draw[fDv] ($(D)+(0.32,\pH-4.82)$)--($(D)+(\pW-0.32,\pH-4.82)$);

\node[fL,anchor=west] at ($(D)+(0.45,\pH-5.20)$)
  {Reverse process (denoise):};

\draw[fRv]
  ($(D)+(\pW-0.38,\pH-5.90)$)--($(D)+(0.38,\pH-5.90)$)
  node[midway,above=2pt,fE,text=cRev,fill=bgDIF!72,inner sep=1pt]
  {$p_\theta(\mathbf{x}_{t-1}\mid\mathbf{x}_t)$};

\node[fB,fill=cDnet,minimum width=33mm,minimum height=8mm]
  at ($(D)+(5.45,\pH-6.52)$) {score net $\varepsilon_\theta(\mathbf{x}_t,t)$};

\pic[scale=2.40] at ($(D)+(8.75,\pH-6.42)$) {noiseblob};
\node[fE] at ($(D)+(8.75,\pH-6.90)$) {$\mathbf{x}_{T}$};

\pic[scale=1.25] at ($(D)+(2.35,\pH-6.42)$) {noiseblob};
\node[fE] at ($(D)+(2.35,\pH-6.90)$) {$\mathbf{x}_{t_1}$};

\pic[scale=0.82] at ($(D)+(0.78,\pH-6.45)$) {crystal};
\node[fE] at ($(D)+(1.12,\pH-6.92)$) {$\hat{\mathbf{x}}_0$};

%% arrows in reverse-process bottom row (added per reviewer request)
\node[font=\large\bfseries,text=cArr] at ($(D)+(5.70,\pH-6.42)$) {$\cdots$};
\draw[fArr] ($(D)+(8.05,\pH-6.42)$)--($(D)+(7.42,\pH-6.42)$);
\draw[fArr] ($(D)+(3.62,\pH-6.42)$)--($(D)+(2.82,\pH-6.42)$);
\draw[fArr] ($(D)+(1.98,\pH-6.42)$)--($(D)+(1.62,\pH-6.42)$);
\end{scope}

\end{tikzpicture}%
}
\caption{\textbf{Overview of the four principal crystal structure generation model families.}
(\textbf{a})~\textit{Variational Autoencoder (VAE):}
The encoder $q_\phi(\mathbf{z}|\mathbf{x})$, parameterized by $\phi$, maps a crystal $\mathbf{x}$ to a posterior over a latent code $\mathbf{z}$, characterized by mean $\boldsymbol{\mu}$ and variance $\boldsymbol{\sigma}^2$.
A latent sample $\mathbf{z}\!\sim\!q_\phi$ is passed to the decoder $p_\theta(\mathbf{x}|\mathbf{z})$, parameterized by $\theta$, to reconstruct $\hat{\mathbf{x}}$.
Training maximizes the Evidence Lower BOund (ELBO): $\mathcal{L}_\text{ELBO}=\mathbb{E}_{q_\phi}[\log p_\theta(\mathbf{x}|\mathbf{z})]-D_\mathrm{KL}[q_\phi(\mathbf{z}|\mathbf{x})\|\mathcal{N}(\mathbf{0},\mathbf{I})]$,
where $D_\mathrm{KL}$ is the Kullback-Leibler divergence regularizing the posterior toward the standard Gaussian prior $\mathcal{N}(\mathbf{0},\mathbf{I})$.
Generation involves sampling $\mathbf{z}\!\sim\!\mathcal{N}(\mathbf{0},\mathbf{I})$ and decoding via $p_\theta$.
(\textbf{b})~\textit{Generative Adversarial Network (GAN):}
A generator $G_\theta$ transforms latent noise $\mathbf{z}\!\sim\!p(\mathbf{z})$ into candidate structures, while a discriminator $D_\phi$ estimates $P(\mathrm{real})$, the probability that a sample originates from the true crystal distribution $p_\mathrm{data}$.
Both networks are optimized jointly through the minimax objective shown at the top of the panel;
the discriminator's adversarial loss signal propagates back to update $G_\theta$.
(\textbf{c})~\textit{Normalizing Flow (NF):}
A sequence of $K$ learnable invertible bijections $\{f_1,\dots,f_K\}$ maps a Gaussian base distribution $p(\mathbf{z}){=}\mathcal{N}(\mathbf{0},\mathbf{I})$ to the crystal distribution $p(\mathbf{x})$ via the change-of-variables relation: $\log p(\mathbf{x})=\log p(\mathbf{z})+\sum_{k=1}^K\log|\det\partial f_k/\partial\mathbf{z}_{k-1}|$, where each Jacobian term accounts for the volume change at step $k$.
Due to the invertibility of each $f_k$, this architecture supports both generation (latent to data, left-to-right) and exact inference (data-to-latent, right-to-left).
(\textbf{d})~\textit{Denoising Diffusion Probabilistic Model (DDPM):}
The forward (noising) process $q(\mathbf{x}_t|\mathbf{x}_{t-1})=\mathcal{N}(\sqrt{1{-}\beta_t}\,\mathbf{x}_{t-1},\,\beta_t\mathbf{I})$ gradually corrupts a crystal $\mathbf{x}_0$ into Gaussian noise $\mathbf{x}_T$ over $T$ steps with a noise schedule $\{\beta_t\}$.
The learned reverse (denoising) process $p_\theta(\mathbf{x}_{t-1}|\mathbf{x}_t)=\mathcal{N}(\boldsymbol{\mu}_\theta(\mathbf{x}_t,t),\,\sigma_t^2\mathbf{I})$ is parameterized by a score network $\varepsilon_\theta(\mathbf{x}_t,t)$ that predicts the noise component at each step, enabling iterative recovery of a crystal proposal $\hat{\mathbf{x}}_0$ from noise.
In all panels, $\phi$ and $\theta$ denote trainable network parameters.}
\label{fig:overview_rev}
\end{figure}

In this review, we highlight recent progress at the intersection of (i) generative crystal-structure modeling, (ii) multimodal representation learning, and (iii) inverse-design workflows for automated materials discovery. The discussion focuses on the core requirements of inverse design: steerability, constraint satisfaction, and verifiable objectives. It then compares the major families of generative approaches and how they incorporate physical and chemical constraints, examines multimodal alignment strategies for target specification and controllable inverse design, and outlines the evaluation principles and open challenges that will determine whether these systems can reliably accelerate materials discovery instead of simply generating plausible samples. \revised{As an organizing thread, we introduce a discovery-credibility ladder in the review (Section~\ref{sec:ladder}, Table~\ref{tab:ladder}), which fixes what evidence a discovery claim requires at each stage, and we return to it as a minimum reporting standard (Section~\ref{sec:ledger}) and a quantitative audit of what published generators actually report. Where the published evidence permits, we state verdicts as well as descriptions: which model family currently leads on which benchmark, at what reported rate, under which protocol, and which claims remain unvalidated (Sections~\ref{sec:gen_families}, \ref{sec:materials_behavior}, \ref{sec:search_strategies} and~\ref{sec:ledger}; Tables~\ref{tab:ladder}, \ref{tab:family_behaviour} and~\ref{tab:quant_B}).} Figure~\ref{fig:overview_rev} provides an overview of the major crystal-structure generation model families. 

\section{Inverse Materials Design as a Closed-Loop Framework}\label{sec:closedloop}

Inverse materials design redefines materials discovery from predicting properties of given candidates to searching for candidates that satisfy a design intent under constraints. In forward modeling, one typically learns a mapping $\hat{y}(x)$ or a distribution $p(y \mid x)$, where $x$ denotes a candidate material (composition and structure) and $y$ denotes target attributes (functional properties, stability margins, cost, or other design criteria). In inverse design, the objective shifts toward searching for candidate materials $x$ that satisfy desired attributes while remaining feasible [\ref{cit:Lookman2019}, \ref{cit:Chen2021}-\ref{cit:Jha2018}].

This process can be formally framed as targeted optimization under feasibility:
\begin{equation}
x^\ast \in \arg\max_{x \in \mathcal{X} \cap \mathcal{C}} U(x)
\label{eq:main_optimization}
\end{equation}
where $\mathcal{X}$ denotes the candidate design space, $\mathcal{C}$ represents the feasible manifold, constrained by rigorous physical, chemical, and synthetic requirements, including thermodynamical stability and structural integrity, $x^\ast$ denotes an optimal candidate, and $U(x)$ is a utility function that encodes the desired single- or multi-objective design goal. Here, feasibility may include geometric validity, charge balance, stability criteria, symmetry, density, and synthesizability proxies~[\ref{cit:Lookman2019}, \ref{cit:DeBreuck2025}].

\subsection{Proposal-Evaluation-Feedback as the Operational Loop}\label{sec:loop_op}

Operationally, inverse design follows an iterative loop with three main steps [\ref{cit:Lookman2019}, \ref{cit:Stach2021}, \ref{cit:Fudge2024}-\ref{cit:Rank2024}]:

\begin{enumerate}[leftmargin=*]
\item \textbf{Proposal}: Generate candidate materials using a proposal mechanism such as conditional generation, latent sampling, heuristic enumeration, or physics-driven global optimization [\ref{cit:DeBreuck2025}].
\item \textbf{Evaluation}: Score candidates using a hierarchy of models and filters, typically moving from inexpensive validity checks to surrogate predictors and then to high-fidelity validation including relaxations, DFT, advanced simulation, or targeted experiments [\ref{cit:Lookman2019}].
\item \textbf{Feedback}: Update proposal and evaluation based on outcomes, including failures, so that subsequent iterations concentrate effort on promising regions while maintaining adequate diversity and calibrated uncertainty [\ref{cit:Lookman2019}, \ref{cit:Wagner2026}].
\end{enumerate}
These steps define the operational loop for automated materials discovery, whose goal is to minimize manual intervention throughout the discovery process.

\subsection{What Makes Inverse Design Difficult}\label{sec:difficulty}
Inverse design is challenging because it is not simply a prediction problem; it is a constrained search problem carried out in a large, structured, and often poorly behaved design space. Operationally, success depends on finding candidates with desirable predicted properties and, equally, on ensuring that those candidates remain physically plausible, chemically valid, and worth the cost of downstream verification. The reversal creates several failure modes absent from forward modeling:

\textit{Constraint dominance}: Many feasibility constraints are hard: small violations can yield unphysical structures or invalid chemistries. Geometric sanity (such as no overlaps, realistic interatomic distances, reasonable density), chemical plausibility, charge neutrality, and symmetry consistency often determine whether downstream evaluation is meaningful at all. Constraint handling therefore defines the effective search space~[\ref{cit:DeBreuck2025}]. In Eq.~\eqref{eq:main_optimization}, this corresponds to restricting the candidate space from $\mathcal{X}$ to the feasible region $\mathcal{X}\cap\mathcal{C}$.

\textit{Optimization-induced failure modes}: During inverse design, the search process can exploit weaknesses in predictive models. As candidate generation becomes more aggressive, the method may propose materials that move away from the kinds of examples seen during training and into regions where model predictions are less reliable. The outcome is apparent improvement that collapses under downstream verification. To reduce this risk, inverse design workflows require constraint-aware objectives, uncertainty-aware candidate selection, and verification steps embedded directly into the loop [\ref{cit:Lookman2019}, \ref{cit:DeBreuck2025}]. \revised{This failure mode is stated here only as a property of the inverse problem; Section~\ref{sec:sampling_challenges} gives the mechanism common to every search strategy, and Section~\ref{sec:failure_modes} the consequences for evaluation.}

\textit{Verification bottleneck}: Generative models can generate candidates cheaply, often with stability-aware objectives or filters built in. However, discovery is still limited by the throughput of higher-fidelity validation, such as structural relaxations, rigorous stability evaluation, and, where relevant, synthesis and characterization. Inverse design is therefore also a budgeting problem: it must allocate scarce validation calls to candidates that are simultaneously promising, feasible, and informative, without collapsing diversity [\ref{cit:Stach2021}, \ref{cit:Wagner2026}].

\subsection{Algorithms for Targeted Search: A Roadmap}\label{sec:roadmap}

 The proposal-evaluation-feedback loop can be steered by several algorithmic frameworks that trade off controllability, diversity, and sample efficiency under expensive validation. In this review, we focus on latent-space optimization, guided diffusion sampling, Bayesian optimization in learned embeddings, reinforcement-learning policy search, and active learning with generative loops, which we analyze in detail (Sections~\ref{sec:latent_opt}-\ref{sec:active_learning}) in terms of their control knobs, constraint handling, and characteristic failure modes under optimization pressure [\ref{cit:Lookman2019}-\ref{cit:Hase2021}, \ref{cit:Wagner2026}, \ref{cit:Han2025}].

\subsection{Where Proposal, Multimodality, and Optimization Enter the Loop}\label{sec:where_enter}

Each component of the review maps onto a distinct role in the closed-loop inverse-design frameworks. Crystal structure generators (Section~\ref{sec:generation}) serve as the \emph{proposal engine}: they convert design intent into candidate structures by sampling from a learned prior over plausible crystals, steered by conditioning signals. Multimodal learning (Section~\ref{sec:multimodal_section}) serves as the \emph{design interface}: it aligns heterogeneous evidence (structure, spectra, text, and processing context) into a shared representational space from which targets can be specified, candidates retrieved, and generation conditioned on richer forms of intent. Optimization frameworks (Section~\ref{sec:sampling_opt}) are the \emph{search strategies} that decide how to allocate validation budget, update the proposal policy, and converge toward high-utility feasible candidates. Evaluation and deployment (Section~\ref{sec:failure_modes}) provides the \emph{ground truth feedback} that prevents surrogate exploitation, ensures novelty and stability claims are trustworthy, and ultimately determines whether proposals translate into experimentally realizable materials. Understanding how these four roles interact and where each can fail is the central theme of this review. \revised{Placed alongside the credibility ladder introduced next (Section~\ref{sec:ladder}), these four roles are the machinery that moves a candidate up the ladder: generators produce a raw sample (rung~L0), feasibility screening carries it to geometric and chemical validity (L1), the machine-learned-interatomic-potential tier (Section~\ref{sec:mlip_oracle}) to L2, selective DFT to L3-L5, and the laboratory to L7-L8.}

\revised{\subsection{A Discovery-Credibility Ladder for Inverse Design}\label{sec:ladder}
Because inverse-design systems optimize aggressively, they exploit any evaluation gap, and a claim of \emph{discovery} is only as strong as the highest rung of evidence it has actually cleared. We therefore fix, before discussing any method, an explicit ladder of evidence that the rest of the review uses as a common yardstick [\ref{cit:DeBreuck2025}-\ref{cit:Dunn2020}, \ref{cit:Riebesell2025}]. The ladder has nine rungs: an AI-generated sample (L0); a geometrically and chemically valid structure (L1); one that survives MLIP relaxation (L2); one that survives DFT relaxation (L3); low energy above a declared convex hull, judged by a chemistry-calibrated threshold (L4) [\ref{cit:Sun2016}-\ref{cit:Aykol2018}]; dynamical stability confirmed by phonons (L5) [\ref{cit:Gouvea2026}, \ref{cit:Loew2025}]; the target property confirmed at high fidelity (L6); a synthesized compound (L7); and a functional property measured under application-relevant conditions (L8). Table~\ref{tab:ladder} states, for each rung, what it certifies, the reported survival numbers that locate the field on it, and the claim language it licenses.

Two principles govern the ladder's use. Under \emph{evidential weight}, a rung's weight is set by how hard it is to pass by accident or by exploitation: operationally, by the probability that a candidate which is \emph{not} a real material passes it under optimization pressure. Because modern generators pass validity (L1) at $83$-$100\%$ wherever it is reported (Tables~\ref{tab:quant_A}-\ref{tab:quant_B}), L1 carries almost no evidence, whereas phonon stability, high-fidelity property confirmation, and synthesis fail routinely and therefore carry much. Under \emph{monotone claims}, a discovery claim of level~$k$ asserts that the candidate passed \emph{every} rung at or below~$k$ under declared protocols (database snapshot, matching tolerances, functional, correction scheme) [\ref{cit:Riebesell2025}, \ref{cit:Negishi2025}]; skipping a rung leaves the claim unsupported. The ladder orders \emph{evidence} and not execution; a pipeline may run MLIP phonons before DFT relaxation for cost reasons, but the claim is graded by rungs passed. Accordingly, a study should report the rung distribution of \emph{all} generated candidates, where the median candidate dies, and not only where the best survivor peaked, and the staged pass-rate ledger of Section~\ref{sec:ledger} is the reporting instrument for that distribution.}

\revised{\begin{table}[htbp]\color{revred}
\centering
\footnotesize
\caption{The discovery-credibility ladder for inverse materials design. Each rung is an entry condition for the next; a discovery claim of level~$k$ asserts passage of every rung at or below~$k$ under declared protocols. Survival exemplars are the figures reported in the cited studies.}
\label{tab:ladder}
\setlength{\tabcolsep}{5pt}
\renewcommand{\arraystretch}{1.30}
\begin{tabularx}{\textwidth}{
  >{\raggedright\arraybackslash}p{0.06\textwidth}
  >{\raggedright\arraybackslash}p{0.29\textwidth}
  >{\raggedright\arraybackslash}Y
  >{\raggedright\arraybackslash}p{0.24\textwidth}}
\toprule
\textbf{Rung} & \textbf{What it certifies} & \textbf{Reported survival exemplar} & \textbf{Claim language it supports} \\
\midrule
L0 & A sample from a learned distribution & 100\% by construction & \emph{generated samples} \\
L1 & Geometric and chemical plausibility & 83-100\% across the models of Table~\ref{tab:quant_B} that report it & \emph{valid candidate structures} \\
L2 & Mechanical integrity under approximate physics & Passes MLIP relaxation (Table~\ref{tab:mlip}) & \emph{MLIP-screened candidates} \\
L3 & Survives reference-fidelity relaxation & 95\% within RMSD 0.076~\AA\ of the DFT minimum [\ref{cit:Zeni2025}] & \emph{DFT-relaxed candidates} \\
L4 & Thermodynamic competitiveness & 78\% below 0.1~eV/atom [\ref{cit:Zeni2025}]; 20/102 [\ref{cit:Antunes2024}] & \emph{computationally (meta)stable hypotheses} \\
L5 & Dynamical stability (phonons) & Reported for two of the seven models of Table~\ref{tab:quant_B} [\ref{cit:Park2025}, \ref{cit:Babu2026}] & \emph{dynamically stable predicted compounds} \\
L6 & Function as designed, at the fidelity the design claim requires & Cheap for single-number electronic properties (e.g.\ band gap); seldom reported for transport, response or device-level targets & \emph{predicted materials with confirmed computed property} \\
L7 & Physical existence & 41/58 targets [\ref{cit:Szymanski2023}]; one candidate [\ref{cit:Zeni2025}] & \emph{synthesized compounds}; \emph{discovery} begins here \\
L8 & Function as made & Property within 20\% of target, one case [\ref{cit:Zeni2025}] & \emph{demonstrated functional materials} \\
\bottomrule
\end{tabularx}
\end{table}}

\revised{Figure~\ref{fig:pipeline_funnel} assembles these ideas into a single operational picture that the rest of the review refers back to. Candidates enter as a dense population sampled from the generative proposal manifold $p_\theta(x\mid c)$ (left) and pass through five stages: generation, surrogate scoring, machine-learned-interatomic-potential (MLIP) relaxation, DFT validation, and reporting, each of which removes candidates, so the surviving set (the funnel population beneath each stage) contracts monotonically. Two quantities travel with the funnel: the \emph{survival counts} $N_0 \ge N_{\mathrm{chem}} \ge \dots \ge N_{\mathrm{prop}}$ and their ratios $r_\bullet = N_\bullet/N_0$, which together form the staged pass-rate ledger of Section~\ref{sec:ledger}. Each stage carries a characteristic \emph{failure mode} (top, red): validity drift, surrogate bias, MLIP out-of-distribution false stability, reference dependence, and novelty leakage, and each is checked by an explicit diagnostic (surrogate-DFT disagreement, ranking consistency, and calibration) under a per-stage decision rule. The right-hand \emph{credibility ladder} (Table~\ref{tab:ladder}) records what each surviving candidate has earned the right to claim, from an unchecked sample (rung~L0) to a synthesized, property-verified material (rung~L8), which the figure groups into six milestone levels, Level~0 to Level~5. The remainder of the review develops each element of this picture: the generators (Section~\ref{sec:generation}), the shared representations that condition them (Section~\ref{sec:multimodal_section}), the search strategies that drive the loop (Section~\ref{sec:sampling_opt}), and the evaluation instruments (ledger, MLIP-oracle tier, reporting standard, and failure-mode diagnostics) that keep its claims honest (Section~\ref{sec:failure_modes}).}

\revised{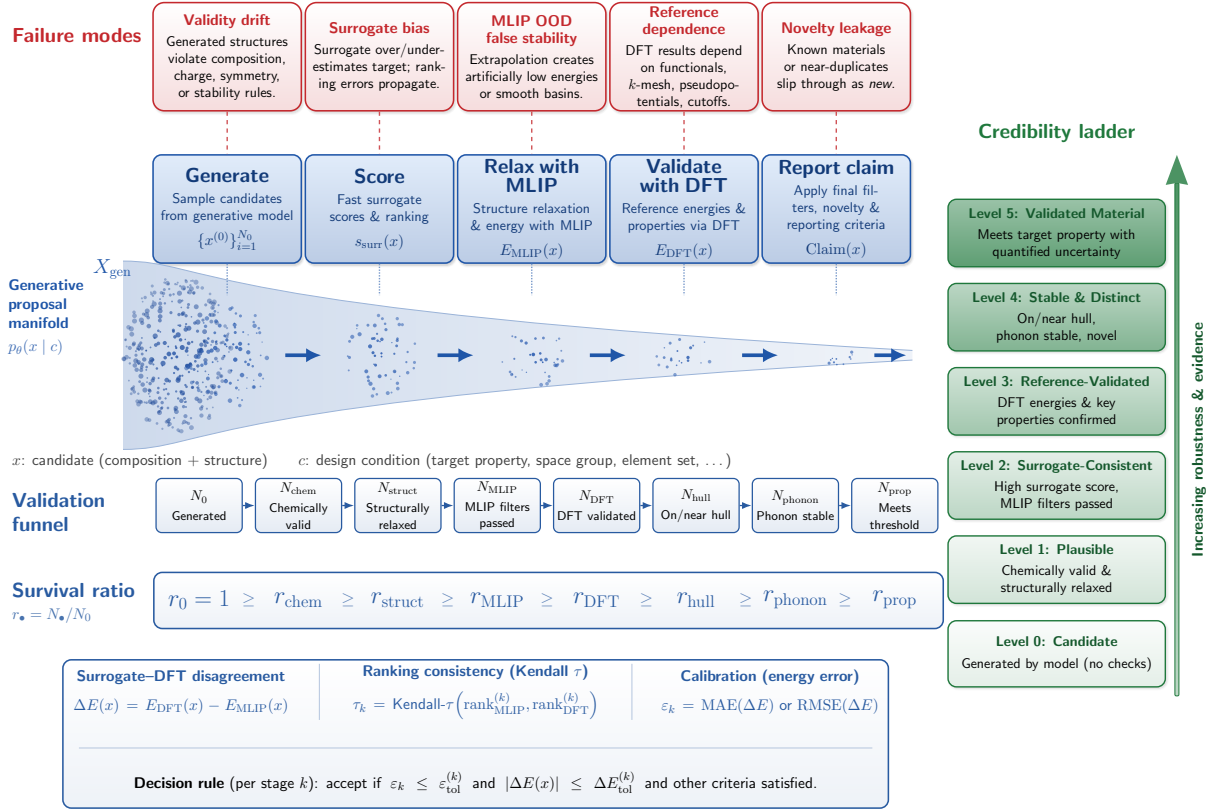
\begin{figure*}[htbp]
\centering
\resizebox{\textwidth}{!}{%
\begin{tikzpicture}[
  font=\sffamily,
  >=Latex,
  stagecard/.style={rounded corners=6pt, draw=pipeblue, line width=1.1pt,
                    top color=pipelt, bottom color=pipeblue!30,
                    text=pipedk, align=center, inner sep=0pt, text width=3.4cm,
                    minimum height=2.7cm, minimum width=3.7cm, blur shadow={shadow blur steps=6,shadow xshift=1pt,shadow yshift=-1.5pt,shadow opacity=25}},
  failcard/.style={rounded corners=8pt, draw=failred, line width=1.1pt,
                   top color=white, bottom color=failred!16,
                   align=center, inner sep=0pt, text width=3.4cm, minimum height=2.7cm,
                   minimum width=3.7cm, font=\sffamily\small, text=black,
                   blur shadow={shadow blur steps=6,shadow xshift=1pt,shadow yshift=-1.5pt,shadow opacity=20}},
  evcard/.style={rounded corners=6pt, draw=evgreen, line width=1.0pt,
                 align=center, inner sep=6pt, text width=5.0cm, minimum height=1.65cm,
                 font=\sffamily\small, blur shadow={shadow blur steps=5,shadow xshift=1pt,shadow yshift=-1pt,shadow opacity=18}},
  metricbox/.style={rounded corners=4pt, draw=metricblue, line width=0.9pt,
                    top color=white, bottom color=metricblue!12,
                    align=center, inner sep=0pt, text width=2.05cm, minimum height=1.35cm,
                    minimum width=2.15cm, font=\sffamily\small},
  flow/.style={-{Latex[length=3.8mm,width=3.6mm]}, line width=2.6pt, draw=pipeblue},
  faildash/.style={dashed, line width=1.0pt, draw=failred!80},
]

% ============================================================ geometry
\def\SXa{5.7}\def\SXb{9.5}\def\SXc{13.3}\def\SXd{17.1}\def\SXe{20.9}
\def\labelx{0.2}

% ============================================================ FAILURE MODES (top red row)
\def\fy{11.55}
\node[failcard] (f1) at (\SXa,\fy) {\textbf{\textcolor{failred}{\normalsize Validity drift}}\\[3pt]Generated structures violate composition, charge, symmetry, or stability rules.};
\node[failcard] (f2) at (\SXb,\fy) {\textbf{\textcolor{failred}{\normalsize Surrogate bias}}\\[3pt]Surrogate over/under-estimates target; ranking errors propagate.};
\node[failcard] (f3) at (\SXc,\fy) {\textbf{\textcolor{failred}{\normalsize MLIP OOD false stability}}\\[3pt]Extrapolation creates artificially low energies or smooth basins.};
\node[failcard] (f4) at (\SXd,\fy) {\textbf{\textcolor{failred}{\normalsize Reference dependence}}\\[3pt]DFT results depend on functionals, $k$-mesh, pseudopotentials, cutoffs.};
\node[failcard] (f5) at (\SXe,\fy) {\textbf{\textcolor{failred}{\normalsize Novelty leakage}}\\[3pt]Known materials or near-duplicates slip through as \emph{new}.};
\node[align=left, font=\sffamily\bfseries\Large, text=failred, anchor=west] at (\labelx,\fy+0.55) {Failure modes};

% ============================================================ PIPELINE STAGE CARDS
\def\py{7.75}
\node[stagecard] (p1) at (\SXa,\py) {\textbf{\Large Generate}\\[3pt]\small Sample candidates from generative model\\[4pt]\normalsize$\{x^{(0)}\}_{i=1}^{N_0}$};
\node[stagecard] (p2) at (\SXb,\py) {\textbf{\Large Score}\\[3pt]\small Fast surrogate scores \& ranking\\[4pt]\normalsize$s_{\mathrm{surr}}(x)$};
\node[stagecard] (p3) at (\SXc,\py) {\textbf{\Large Relax with MLIP}\\[3pt]\small Structure relaxation \& energy with MLIP\\[4pt]\normalsize$E_{\mathrm{MLIP}}(x)$};
\node[stagecard] (p4) at (\SXd,\py) {\textbf{\Large Validate with DFT}\\[3pt]\small Reference energies \& properties via DFT\\[4pt]\normalsize$E_{\mathrm{DFT}}(x)$};
\node[stagecard] (p5) at (\SXe,\py) {\textbf{\Large Report claim}\\[3pt]\small Apply final filters, novelty \& reporting criteria\\[4pt]\normalsize$\mathrm{Claim}(x)$};

\foreach \i in {1,...,5}{ \draw[faildash] (f\i.south) -- (p\i.north); }

% ============================================================ FUNNEL (cone + per-stage populations aligned to cards)
\def\fcy{4.1}
\def\mouthx{3.10}
\def\tipx{22.9}
\def\mouthh{2.35}
\def\tiph{0.12}
% local cone half-height helper values at each stage center (linear taper)
\begin{scope}
  % smooth cone body
  \shade[left color=pipeblue!22, right color=pipeblue!7]
    (\mouthx,\fcy+\mouthh)
    .. controls (\mouthx+2.2,\fcy+\mouthh) and (\SXa-1.0,\fcy+1.15) ..
    (\SXe+1.9,\fcy+\tiph)
    -- (\SXe+1.9,\fcy-\tiph)
    .. controls (\SXa-1.0,\fcy-1.15) and (\mouthx+2.2,\fcy-\mouthh) ..
    (\mouthx,\fcy-\mouthh) -- cycle;
  \draw[pipeblue!45, line width=0.7pt]
    (\mouthx,\fcy+\mouthh) .. controls (\mouthx+2.2,\fcy+\mouthh) and (\SXa-1.0,\fcy+1.15) .. (\SXe+1.9,\fcy+\tiph);
  \draw[pipeblue!45, line width=0.7pt]
    (\mouthx,\fcy-\mouthh) .. controls (\mouthx+2.2,\fcy-\mouthh) and (\SXa-1.0,\fcy-1.15) .. (\SXe+1.9,\fcy-\tiph);

  % ---- dense generative blob at the mouth (left of Generate) ----
  \foreach \n in {1,...,340}{
    \pgfmathsetmacro\rr{sqrt(rnd)}
    \pgfmathsetmacro\th{360*rnd}
    \pgfmathsetmacro\bx{4.45+\rr*cos(\th)*1.35}
    \pgfmathsetmacro\by{\fcy+\rr*sin(\th)*1.95}
    \pgfmathsetmacro\sz{0.020+0.045*rnd}
    \pgfmathsetmacro\op{0.35+0.5*(1-\rr)}
    \pgfmathsetmacro\cmix{40+40*(1-\rr)}
    \fill[dotblue!\cmix!dotdk, opacity=\op] (\bx,\by) circle (\sz);
  }
  % ---- per-stage populations: one tight, distinct cluster centered UNDER each card ----
  \foreach \cx/\cnt/\clh/\clw/\dsz in {\SXa/90/1.35/1.05/0.042, \SXb/58/1.05/0.95/0.038, \SXc/36/0.80/0.85/0.034, \SXd/21/0.55/0.72/0.030, \SXe/11/0.36/0.58/0.026}{
    \foreach \n in {1,...,\cnt}{
      \pgfmathsetmacro\ang{360*rnd}
      \pgfmathsetmacro\rad{sqrt(rnd)}
      \pgfmathsetmacro\jx{\cx+\rad*cos(\ang)*\clw}
      \pgfmathsetmacro\jy{\fcy+\rad*sin(\ang)*\clh}
      \pgfmathsetmacro\sz{0.015+\dsz*rnd}
      \pgfmathsetmacro\op{0.5+0.4*(1-\rad)}
      \fill[dotblue, opacity=\op] (\jx,\jy) circle (\sz);
    }
  }
\end{scope}

% ---- clean vertical stems: each stage card -> its funnel population ----
\foreach \i/\cx in {1/\SXa, 2/\SXb, 3/\SXc, 4/\SXd, 5/\SXe}{
  \draw[pipeblue!60, line width=1.0pt, densely dotted]
     (p\i.south) -- (\cx,\fcy+1.5);
}

\node[text=pipeblue, font=\sffamily\bfseries\Large, anchor=east] at (3.45,\fcy+2.15) {$X_{\mathrm{gen}}$};
\node[align=left, anchor=west, inner sep=1pt]
     at (\labelx,\fcy+0.95)
     {\color{pipeblue}\sffamily\bfseries Generative\\\color{pipeblue}\sffamily\bfseries proposal\\\color{pipeblue}\sffamily\bfseries manifold\\[3pt]
      \color{pipeblue!85}\sffamily $p_\theta(x\mid c)$};

% flow arrows in the gaps between populations
\foreach \xa/\xb in {\SXa/\SXb, \SXb/\SXc, \SXc/\SXd, \SXd/\SXe}{
  \pgfmathsetmacro\amid{(\xa+\xb)/2}
  \draw[flow] (\amid-0.45,\fcy) -- (\amid+0.45,\fcy);
}
\draw[flow] (\SXe+0.95,\fcy) -- (\SXe+1.75,\fcy);

\node[align=left, font=\sffamily\normalsize, text=black!75, anchor=west] at (\labelx,\fcy-2.65)
  {$x$: candidate (composition + structure)\qquad
   $c$: design condition (target property, space group, element set, \dots)};

% ============================================================ METRIC ROW
\def\my{0.35}
\def\mx{5.0}
\def\dxm{2.48}
\foreach \i/\lab/\sub in {0/{$N_0$}/Generated, 1/{$N_{\mathrm{chem}}$}/{Chemically valid},
   2/{$N_{\mathrm{struct}}$}/{Structurally relaxed}, 3/{$N_{\mathrm{MLIP}}$}/{MLIP filters passed},
   4/{$N_{\mathrm{DFT}}$}/{DFT validated}, 5/{$N_{\mathrm{hull}}$}/{On/near hull},
   6/{$N_{\mathrm{phonon}}$}/{Phonon stable}, 7/{$N_{\mathrm{prop}}$}/{Meets threshold}}{
   \pgfmathsetmacro\mmx{\mx+\i*\dxm}
   \node[metricbox] (m\i) at (\mmx,\my) {\lab\\[2pt]\footnotesize \sub};
}
\foreach \i in {0,...,6}{ \pgfmathtruncatemacro\j{\i+1} \draw[-{Latex[length=2.4mm]}, draw=metricblue, line width=0.9pt] (m\i) -- (m\j); }
\node[align=left, text=metricblue, font=\sffamily\bfseries\Large, anchor=west] at (\labelx,\my+0.15) {Validation};
\node[align=left, text=metricblue, font=\sffamily\bfseries\Large, anchor=west] at (\labelx,\my-0.45) {funnel};

% ============================================================ SURVIVAL RATIOS (enlarged, panel)
\def\ry{-2.0}
\node[align=left, text=metricblue, font=\sffamily\bfseries\Large, anchor=west] at (\labelx,\ry+0.25) {Survival ratio};
\node[align=left, text=metricblue!85, font=\sffamily\normalsize, anchor=west] at (\labelx,\ry-0.38) {$r_\bullet = N_\bullet/N_0$};
\node[rounded corners=6pt, draw=metricblue, line width=1.1pt,
      top color=white, bottom color=metricblue!8, inner sep=0pt,
      minimum width=19.7cm, minimum height=1.40cm, anchor=north west] at (3.85,\ry+0.70) {};
\node[font=\sffamily\LARGE, text=metricblue] at (\mx,\ry) {$r_0=1$};
\foreach \i/\rr in {1/{$r_{\mathrm{chem}}$},
   2/{$r_{\mathrm{struct}}$}, 3/{$r_{\mathrm{MLIP}}$},
   4/{$r_{\mathrm{DFT}}$}, 5/{$r_{\mathrm{hull}}$},
   6/{$r_{\mathrm{phonon}}$}, 7/{$r_{\mathrm{prop}}$}}{
   \pgfmathsetmacro\mmx{\mx+\i*\dxm}
   \node[font=\sffamily\LARGE, text=metricblue] at (\mmx,\ry) {\rr};
   \pgfmathsetmacro\gex{\mx+(\i-0.5)*\dxm}
   \node[font=\sffamily\large, text=metricblue!75] at (\gex,\ry) {$\ge$};
}

% ============================================================ DIAGNOSTIC BOX
\def\dgy{-5.35}
\node[rounded corners=6pt, draw=metricblue, line width=1.1pt,
      top color=white, bottom color=metricblue!8,
      inner sep=0pt, minimum width=20.6cm, minimum height=3.7cm] (diagbg) at (11.9,\dgy) {};
\node[align=center, text width=6.3cm, font=\sffamily\normalsize] at (4.55,\dgy+1.05)
  {\color{metricblue}\textbf{Surrogate--DFT disagreement}\\[7pt]
   $\Delta E(x)=E_{\mathrm{DFT}}(x)-E_{\mathrm{MLIP}}(x)$};
\node[align=center, text width=7.1cm, font=\sffamily\normalsize] at (11.9,\dgy+1.05)
  {\color{metricblue}\textbf{Ranking consistency (Kendall $\tau$)}\\[7pt]
   $\tau_k=\text{Kendall-}\tau\!\left(\mathrm{rank}^{(k)}_{\mathrm{MLIP}},\mathrm{rank}^{(k)}_{\mathrm{DFT}}\right)$};
\node[align=center, text width=6.3cm, font=\sffamily\normalsize] at (19.25,\dgy+1.05)
  {\color{metricblue}\textbf{Calibration (energy error)}\\[7pt]
   $\varepsilon_k=\mathrm{MAE}(\Delta E)\ \text{or}\ \mathrm{RMSE}(\Delta E)$};
\draw[draw=metricblue!35, line width=0.6pt] (8.0,\dgy+0.35) -- (8.0,\dgy+1.75);
\draw[draw=metricblue!35, line width=0.6pt] (15.8,\dgy+0.35) -- (15.8,\dgy+1.75);
\draw[draw=metricblue!35, line width=0.7pt] (2.0,\dgy-0.35) -- (21.8,\dgy-0.35);
\node[align=center, text width=19.5cm, font=\sffamily\normalsize] at (11.9,\dgy-1.2)
  {\textbf{Decision rule} (per stage $k$): accept if\;
   $\varepsilon_k \le \varepsilon^{(k)}_{\mathrm{tol}}$\; and\; $|\Delta E(x)| \le \Delta E^{(k)}_{\mathrm{tol}}$\;
   and other criteria satisfied.};

% ============================================================ CREDIBILITY LADDER
\def\lx{26.4}
\foreach \i/\ttl/\sub/\opa/\opb in {
   0/{Level 0: Candidate}/{Generated by model (no checks)}/2/12,
   1/{Level 1: Plausible}/{Chemically valid \& structurally relaxed}/6/20,
   2/{Level 2: Surrogate-Consistent}/{High surrogate score, MLIP filters passed}/12/30,
   3/{Level 3: Reference-Validated}/{DFT energies \& key properties confirmed}/20/42,
   4/{Level 4: Stable \& Distinct}/{On/near hull, phonon stable, novel}/30/56,
   5/{Level 5: Validated Material}/{Meets target property with quantified uncertainty}/45/72}{
   \pgfmathsetmacro\ly{-3.35+\i*2.1}
   \node[evcard, top color=evgreen!\opa, bottom color=evgreen!\opb] (lev\i) at (\lx,\ly)
         {\textbf{\textcolor{evdk}{\ttl}}\\[3pt]\small \sub};
}
\node[align=center, text=evgreen, font=\sffamily\bfseries\Large] at (\lx,9.65) {Credibility ladder};
\shade[top color=evgreen!80, bottom color=evgreen!30,
       draw=evgreen, line width=0.5pt, rounded corners=1pt]
   ($(\lx+2.95,-4.4)$) rectangle ($(\lx+3.05,8.4)$);
\fill[evgreen!80] (\lx+3.0,8.4) -- ++(-0.28,-0.05) -- ++(0.28,0.75) -- ++(0.28,-0.75) -- cycle;
\node[rotate=90, text=evdk, font=\sffamily\normalsize\bfseries] at (\lx+3.55,2.0)
   {Increasing robustness \& evidence};

\end{tikzpicture}%
}
\caption{\textbf{The inverse-design pipeline as a funnel feeding a discovery-credibility ladder.} A dense population of candidates sampled from the generative proposal manifold $p_\theta(x\mid c)$ (left) flows through five sequential stages: \emph{Generate}, \emph{Score}, \emph{Relax with MLIP}, \emph{Validate with DFT}, and \emph{Report claim}, each associated with a governing equation and a characteristic failure mode (top, red). The funnel population beneath each stage contracts monotonically, giving the survival counts $N_0 \ge N_{\mathrm{chem}} \ge N_{\mathrm{struct}} \ge N_{\mathrm{MLIP}} \ge N_{\mathrm{DFT}} \ge N_{\mathrm{hull}} \ge N_{\mathrm{phonon}} \ge N_{\mathrm{prop}}$ and the survival ratios $r_\bullet = N_\bullet/N_0$ that constitute the staged pass-rate ledger (Section~\ref{sec:ledger}). Stage-level diagnostics (surrogate-DFT energy disagreement $\Delta E$, ranking consistency (Kendall $\tau$), and calibration ($\varepsilon_k$)) feed a per-stage decision rule. The credibility ladder (right) condenses the full nine-rung ladder of Table~\ref{tab:ladder} (L0-L8) into six milestone levels, from an unchecked sample (Level~0) to a synthesized, property-verified material (Level~5); the level-to-rung correspondence follows the reporting standard of Table~\ref{tab:report}, and a discovery claim at any level requires passing every rung at or below it under declared protocols.}
\label{fig:pipeline_funnel}
\end{figure*}}

\section{Crystal Structure Generation as the Materials Proposal Engine}\label{sec:generation}

Generative crystal modeling for inverse materials design is critical, transforming candidate generation from a combinatorial bottleneck into a controllable, data-driven operation. Instead of enumerating candidates and ranking them afterward, generative models learn a distribution over plausible periodic structures and can bias sampling toward target objectives and feasibility constraints. This matters most when chemical space is too large for exhaustive screening, when multiple constraints must be satisfied simultaneously, and when validation budgets limit the number of candidates that can be tested [\ref{cit:Luo2024}-\ref{cit:Luo2025}, \ref{cit:Ren2022}, \ref{cit:DeBreuck2025}, \ref{cit:Antunes2024}, \ref{cit:Chenebuah2024}].

\subsection{What It Means to Generate a Valid Crystal}\label{sec:valid_crystal}
A periodic crystal can be formalized as a tuple
\begin{equation}
x \;=\; \big(\mathbf{L},\, \mathbf{Z},\, \mathbf{S}\big)
\label{eq:crystal_tuple}
\end{equation}
where $\mathbf{L}\in\mathbb{R}^{3\times 3}$ is the lattice matrix, $\mathbf{Z}=(Z_1,\dots,Z_N)\in\mathcal{A}^N$ denotes the atomic species (e.g., atomic numbers) on $N$ crystallographic sites, and $\mathbf{S}=[\mathbf{s}_1,\dots,\mathbf{s}_N]^\top\in[0,1)^{N\times 3}$ are the fractional coordinates.
Real-space positions follow as
\begin{equation}
\mathbf{r}_i \;=\; \mathbf{L}\,\mathbf{s}_i,\qquad i=1,\dots,N
\label{eq:realspace_coords}
\end{equation}
with periodicity enforced by identifying $\mathbf{s}_i \equiv \mathbf{s}_i + \mathbf{n}$ for any $\mathbf{n}\in\mathbb{Z}^3$.

In this representation, crystal generation amounts to learning a distribution over such tuples,
\begin{equation}
x \sim p_{\theta}(x),
\qquad\text{or (for inverse design)}\qquad
x \sim p_{\theta}(x\mid y)
\label{eq:gen_objective}
\end{equation}
where $y$ encodes design intent (target properties, composition, symmetry cues, or other constraints).
Symmetry can be incorporated by restricting $(\mathbf{L},\mathbf{S})$ to a space-group-consistent parameterization (e.g., Wyckoff degrees of freedom), while chemical plausibility can be enforced through charge neutrality or oxidation-state constraints during sampling and screening~[\ref{cit:DeBreuck2025}, \ref{cit:Davies2019}].

This structure makes crystal generation qualitatively different from standard molecular generation: periodic boundary conditions couple local coordination to global cell parameters, and feasibility constraints are frequently hard (e.g., overlaps, unrealistic densities, inconsistent coordination patterns). As a result, the representation used by a generator strongly determines (i) which constraints can be imposed efficiently and (ii) which failure modes appear under conditioning and optimization [\ref{cit:DeBreuck2025}]. The following steps are currently adopted:

\begin{enumerate}[leftmargin=*]
\item \textbf{Continuous geometric representation}: Here, lattice parameters and fractional atomic coordinates are parametrized directly. This representation is natural for gradient-based steering and diffusion-enabled sampling, but it requires careful handling of invariances and physical validity [\ref{cit:Xie2021}-\ref{cit:Jiao2023}, \ref{cit:Qiu2025}, \ref{cit:Ren2022}].
\item \textbf{Discrete representation}: Structures are serialized as sequences, for example as CIF-like tokens. Token sequences suit autoregressive modeling and integrate naturally with text, but physical validity must be enforced through constrained decoding and post-generation screening, such as symmetry checks, chemical-validity filters, and MLIP or DFT-based relaxation [\ref{cit:Antunes2024}, \ref{cit:Wang2021}-\ref{cit:Goodall2020}].
\item \textbf{Symmetry-aware representation}: Crystals are encoded through space groups and Wyckoff positions, which enforce global structural consistency by construction but may limit flexibility when disorder, defects, or low-symmetry targets are central [\ref{cit:DeBreuck2025}].
\end{enumerate}

\subsection{Families of Generative Models and Their Inverse-Design Trade-Offs}\label{sec:gen_families}

Multiple families of generative models have been adapted to periodic crystals. At the most basic level, a generative model learns a probability distribution over crystal candidates,
\begin{equation}
x \sim p_{\theta}(x)
\label{eq:generative_model}
\end{equation}
where $x$ denotes a crystal representation and $\theta$ denotes the model parameters. For inverse design, we usually want candidates that satisfy the predefined condition $c$, such as a target composition, symmetry class, or desired property range. The problem therefore becomes conditional generation,
\begin{equation}
x \sim p_{\theta}(x \mid c)
\label{eq:conditional_generation}
\end{equation}
or in other words, guided generation in which samples are steered toward higher design utility while remaining feasible.

For inverse design, a model earns its place by generating plausible crystals and then doing three further things: accepting steering toward desired targets, absorbing physical and chemical constraints, and staying reliable beyond the training distribution [\ref{cit:Luo2024}-\ref{cit:Jiao2023}, \ref{cit:Zeni2025}-\ref{cit:Luo2025}, \ref{cit:DeBreuck2025}, \ref{cit:Antunes2024}].

\paragraph{Early 3D grid, voxel, and GAN generators:}
Some of the earliest crystal generators borrowed ideas from computer vision, representing crystals on 3D grids or tensors. They established that crystal generation could be posed as a learning problem, displacing brute-force enumeration, but \revised{their inductive biases, developed for pictures instead of periodic solids, left periodicity, exact stoichiometry, atom identity, and rotational invariance unrepresented, demanding heavy post-processing to recover a valid periodic structure; Section~\ref{sec:symmetry_axis} analyzes this failure in architectural terms}\cut{ in practice such models struggled with the specific structure of crystalline matter, and substantial post-processing was required to recover periodic structure and basic chemical validity} [\ref{cit:Court2020}-\ref{cit:Kim2020}].

\paragraph{Latent-variable generators:}
Latent-variable models introduce a continuous hidden representation $z$ for each crystal:
\begin{equation}
z \sim p(z), \qquad x \sim p_{\theta}(x \mid z)
\label{eq:latent_variable_model}
\end{equation}
where $z$ is a latent code sampled from prior $p(z)$, and $p_{\theta}(x \mid z)$ is the conditional decoder parameterized by $\theta$.
Crucially, proximity in the latent space reflects structural similarity between crystals, facilitating interpolation, controlled perturbation, and optimization. For inverse design, one often searches in latent space, not directly in crystal space:
\begin{equation}
z^{\ast} \in \arg\max_{z} \; U\!\big(g_{\theta}(z)\big)
\label{eq:latent_optimization}
\end{equation}
where $g_{\theta}$ is the decoder and $U(\cdot)$ is a target-dependent utility function (Eq.~\ref{eq:latent_optimization} is used throughout Section~\ref{sec:latent_opt}). Latent-variable models are therefore attractive for proposal-and-refine workflows; diffusion-decoder VAEs strengthen the decoder to improve sample quality. Their main limitation is that latent-space optimization can push decoded structures off the physically plausible manifold unless priors, feasibility filters, or explicit constraints are imposed [\ref{cit:Xie2021}, \ref{cit:Ren2022}].

\paragraph{Normalizing flows:}
Flow-based models learn an invertible mapping between a simple base distribution and the crystal representation:
\begin{equation}
{\color{revred}
z \sim \mathcal{N}(0,I), \qquad (\mathbf{L},\mathbf{S}) = f_{\theta}\!\left(z \mid \mathbf{Z}, c\right), \qquad
\log p_{\theta}\!\left(\mathbf{L},\mathbf{S} \mid \mathbf{Z}, c\right) = \log p(z) + \log \left| \det \frac{\partial f_{\theta}^{-1}}{\partial (\mathbf{L},\mathbf{S})} \right|}
\label{eq:normalizing_flow}
\end{equation}
where $f_{\theta}$ is a learnable invertible bijection (flow), $p(z) = \mathcal{N}(0,I)$ is the Gaussian base distribution with $I$ denoting the identity covariance matrix, and the Jacobian term accounts for the change of volume under the transformation. \revised{Here $\mathbf{L}$, $\mathbf{Z}$ and $\mathbf{S}$ are the lattice matrix, the species list and the fractional coordinates of Eq.~\ref{eq:crystal_tuple}, and $c$ is the target condition (composition, symmetry class, a property value, or a text prompt). The bijection transports only the continuous variables $\mathbf{L}$ and $\mathbf{S}$ at a species assignment $\mathbf{Z}$ set by conditioning, since discrete species admit no diffeomorphism onto a Gaussian base.}
Because the transformation is invertible, flows offer exact likelihoods, efficient sampling, and direct latent manipulation. \revised{The materials consequence of that invertibility is the family's defining trade-off: a tractable Jacobian constrains the architecture, so imposing symmetry, atomic discreteness, and hard feasibility conditions inside a fully invertible map is difficult, and such constraints are typically delegated to the representation or enforced outside the flow [\ref{cit:Luo2025}, \ref{cit:Ren2022}].}

\paragraph{Autoregressive sequence models:}
Autoregressive models treat a crystal as a sequence of tokens and factorize generation step by step:
\begin{equation}
{\color{revred}
p_{\theta}(x \mid c)=\prod_{t=1}^{T} p_{\theta}(\tau_t \mid \tau_{<t}, c),
\qquad
\tau_t \sim p_{\theta}(\,\cdot \mid \tau_{<t}, c)\big|_{\mathcal{M}_t},
\qquad
\mathcal{M}_t \subseteq \mathcal{V}_t}
\label{eq:autoregressive}
\end{equation}
where $\tau_t$ is the $t$-th token of the serialized crystal, $\tau_{<t}=(\tau_1,\dots,\tau_{t-1})$ denotes all tokens preceding position $t$, $c$ is optional conditioning information such as composition or symmetry cues, and $T$ is the total sequence length. \revised{The tokens $\tau_t$ are the fields of a CIF-like record (species symbols, lattice parameters, coordinates and space-group entries), $\mathcal{V}_t$ is the full vocabulary available at step $t$, and $\mathcal{M}_t \subseteq \mathcal{V}_t$ is the chemically admissible subset to which decoding is restricted, enforcing charge balance and space-group consistency (Section~\ref{sec:staged_validity}).}
In crystal applications, the tokens may correspond to CIF-like syntax, structural descriptors, or serialized geometry information. The main strength of this family is flexible conditioning: composition, symmetry cues, or text-derived prompts can be injected naturally into the sequence-generation process. Autoregressive models therefore suit design interfaces built from discrete specifications or when one wants direct coupling with scientific language. Their weakness is equally important: syntactic correctness is not the same as physical correctness. A valid-looking sequence does not guarantee geometric plausibility, charge balance, or low-energy structure. In practice, these models often depend on decoders, rules, or post-generation screening to ensure that syntactically valid outputs are also scientifically meaningful [\ref{cit:Antunes2024}].

\paragraph{Diffusion and score-based generators:}
Diffusion models have become prominent because they can model complex high-dimensional distributions while also supporting gradual, controllable generation. In simplified form, they learn to reverse a noising process,
\begin{equation}
{\color{revred}
(\mathbf{L}_T,\mathbf{Z}_T,\mathbf{S}_T) \rightarrow \cdots \rightarrow (\mathbf{L}_0,\mathbf{Z}_0,\mathbf{S}_0),
\qquad
\mathbf{S}_t \equiv \mathbf{S}_t + \mathbf{n}, \quad \mathbf{n} \in \mathbb{Z}^{3}}
\label{eq:diffusion_reverse_chain}
\end{equation}
where $(\mathbf{L}_T,\mathbf{Z}_T,\mathbf{S}_T)$ is the fully noised state, $(\mathbf{L}_0,\mathbf{Z}_0,\mathbf{S}_0)$ is the clean crystal candidate, and the intermediate states represent progressively denoised structures along the reverse trajectory. \revised{The three channels of Eq.~\ref{eq:crystal_tuple} are denoised jointly, and $\mathbf{n} \in \mathbb{Z}^3$ is an integer lattice translation, so the coordinate channel is diffused on the periodic torus and periodicity holds by construction at every step (Section~\ref{sec:symmetry_axis}).}
One starts from noise and iteratively denoises toward a crystal candidate. Inverse design enters through conditional denoising or guidance during sampling. For example, the reverse process can be steered using a target score. The advantage of using the diffusion model is to generate candidates under target-property, symmetry, or composition constraints while preserving diversity. The central strength of diffusion models is therefore controllable sampling under complex constraints. However, they are susceptible to guidance-induced brittleness; overly aggressive signals from imperfect predictors can produce candidates that appear optimal in silico but fail during external validation [\ref{cit:Jiao2023}, \ref{cit:Gao2025}-\ref{cit:Karras2022}].

These model families differ in architecture, representation, and optimization style, but the practical lesson is broader. No single family solves inverse design on its own. A useful proposal engine must generate candidates and stay compatible with feasibility control, uncertainty-aware ranking, and downstream verification. The relevant comparison is therefore which family can be embedded most effectively into a closed-loop discovery workflow, and only secondarily which samples the most realistic crystals. In a closed loop, candidate proposal is followed by guidance and conditioning, feasibility screening, high-fidelity verification, and iterative model feedback.
The best choice therefore depends on the scientific regime: whether the problem demands strict controllability, easy integration of hard constraints, efficient latent optimization, discrete prompt conditioning, or robust guided sampling under expensive validation [\ref{cit:DeBreuck2025}-\ref{cit:Dunn2020}].

Inverse design is useful only when target-conditioned candidates remain physically feasible and verifiable after generation. Therefore, the practical value of conditioning should be assessed on four counts together: whether the model moves samples toward the requested objective, whether those samples satisfy the imposed constraints, whether they retain chemical and structural diversity, and whether they survive downstream relaxation or validation such as geometry optimization, stability screening, or higher-fidelity calculation [\ref{cit:Shahriari2016}, \ref{cit:DeBreuck2025}, \ref{cit:Dunn2020}].

\subsubsection*{\revised{Verdict: which family leads, and on what evidence}}
\revised{On the figures reported in the original studies, and subject to the limits of self-reported evaluation discussed in Section~\ref{sec:failure_modes}: equivariant diffusion is the current default proposal engine. It leads the CSP benchmarks (MP-20 match rate 51.5\% for DiffCSP against 33.9\% for CDVAE at $k{=}1$ [\ref{cit:Xie2021}-\ref{cit:Jiao2023}]), leads de novo generation (78\% of MatterGen samples within 0.1~eV/atom of the convex hull after DFT relaxation, with stable-unique-novel rates more than twice those of earlier baselines [\ref{cit:Zeni2025}]), and is so far the only family reporting an experimentally synthesized candidate whose designed property was confirmed [\ref{cit:Zeni2025}]. Autoregressive CIF models are the choice when the design interface is language or composition and training data are limited: CrystaLLM matches 88.1\% of held-out compositions within three attempts, yet only 20 of its 102 DFT-relaxed novel candidates fell within the declared hull threshold [\ref{cit:Antunes2024}], at the structural price of mandatory post-generation screening (Section~\ref{sec:symmetry_axis}). Within that family, a further distinction matters for inverse design. Most of the models above denoise in structure space, with the lattice and fractional coordinates carrying the noise directly [\ref{cit:Jiao2023}, \ref{cit:Zeni2025}]. A second branch encodes the crystal into a learned latent space with a VAE and runs the denoising there: Chemeleon-2 uses a diffusion transformer over VAE latents [\ref{cit:Park2026}], and the same construction allows a policy-gradient method to be attached to the sampler, because a latent trajectory gives the tractable score that policy-gradient optimization requires [\ref{cit:Park2026}]. Latent diffusion's advantage is that denoising in a smooth, low-dimensional latent space gives the sampler a tractable score, so that reinforcement-learning fine-tuning and property conditioning can be applied to the sampler itself instead of to its output (Sections~\ref{sec:latent_opt}-\ref{sec:guided_diffusion}, \ref{sec:rl}). Flows trade a small match-rate deficit for exact likelihoods and cheaper sampling (51.9\% at $k{=}1$ for CrystalFlow [\ref{cit:Luo2025}]); latent-variable models are no longer only components of hybrid systems, since the VAE in a latent-diffusion generator is doing exactly this work [\ref{cit:Xie2021}, \ref{cit:Ren2022}, \ref{cit:Raccuglia2016}]; and voxel-based GANs are superseded for periodic solids [\ref{cit:Court2020}-\ref{cit:Kim2020}]. Table~\ref{tab:family_behaviour} and Tables~\ref{tab:quant_A}-\ref{tab:quant_B} carry the underlying numbers together with the protocols that qualify them.}

\revised{\subsection{Symmetry as an Architectural Axis: How Generators Respect or Repair Crystal Invariances}\label{sec:symmetry_axis}
The family taxonomy above answers \emph{how} a model learns a distribution, but it misses the axis that most sharply separates early failures from current successes: how the architecture treats the symmetries of crystalline matter. The relevant symmetry group is larger than is often acknowledged. The energy and identity of a crystal are invariant under (i)~the Euclidean group $E(3)$ of global rotations, reflections, and translations; (ii)~permutations of identical atoms; (iii)~periodic translations of fractional coordinates, which wrap around the cell; and (iv)~the discrete group of cell re-parameterizations: the same crystal is described by infinitely many equivalent cells related by integer changes of basis, conventionally fixed by Niggli reduction. Formally, a distribution $p(x)$ is \emph{invariant} when $p(g\!\cdot\!x)=p(x)$ for every symmetry operation $g$, and a network $f$ is \emph{equivariant} when $f(g\!\cdot\!x)=g\!\cdot\!f(x)$. A generator handles these invariances in one of three ways: build them in by construction, learn them approximately from data at a cost in capacity and examples, or violate them and repair the damage by post-processing. The history of crystal generation is largely the move from the third strategy to the first.

\subsubsection*{\revised{Three ways to handle the crystal invariances}}
The theoretical foundation for building them in is well established: if the base (noise) distribution is invariant and every learned transformation or transition kernel is equivariant, the resulting model density is invariant by construction [\ref{cit:Kohler2020}-\ref{cit:Xu2022}]. Equivariance and diffusion are natural partners for this reason: the score of an invariant density is an equivariant vector field, so an equivariant denoiser guarantees an exactly symmetric learned distribution, effectively learning each training structure together with all its rotated, translated, and re-indexed copies. Two idioms dominate equivariant message passing: scalarization networks [\ref{cit:Satorras2021}] that communicate through invariant scalars (distances, angles), and irreducible-representation networks (the design underlying modern universal interatomic potentials [\ref{cit:Batatia2022}]) that carry higher-order tensor features from spherical harmonics and tensor products. DiffCSP [\ref{cit:Jiao2023}] brought this machinery to crystals, jointly denoising the lattice and fractional coordinates with a periodic $E(3)$-equivariant network, and foundation-scale generators such as MatterGen [\ref{cit:Zeni2025}] adopt the same design.

A second mechanism sidesteps equivariance by describing the structure entirely in already-invariant quantities: fractional coordinates, interatomic distances and angles, or invertible invariant crystallographic representations [\ref{cit:Ren2022}, \ref{cit:Xiao2023}]. The trade-off is decodability: strongly invariant descriptor sets are not always uniquely invertible, and reconstruction ambiguity becomes a new source of invalid outputs. Autoregressive CIF-token models [\ref{cit:Antunes2024}] sit at a weaker point on the same axis: canonicalized cell settings impose invariance through data normalization, leaving the architecture agnostic, which is why post-generation screening remains indispensable for that family.

The third mechanism enforces global symmetry by construction. Inorganic crystals are not uniformly distributed over symmetry groups: experimentally known structures concentrate in nontrivial space groups, whereas unconstrained generators drift toward $P1$, producing structures that are formally valid but crystallographically atypical [\ref{cit:Jiao2024}-\ref{cit:Cao2024}]. Wyckoff- and space-group-constrained generation selects a space group and generates only the compatible Wyckoff degrees of freedom: space-group-constrained diffusion projects the denoising trajectory onto the symmetry-consistent subspace [\ref{cit:Jiao2024}], and space-group-informed autoregressive generation factorizes the structure over Wyckoff positions [\ref{cit:Cao2024}]. Beyond guaranteeing symmetry, this sharply reduces the effective search dimensionality, from the $3N+9$ continuous degrees of freedom of an unconstrained cell to a handful of free Wyckoff parameters, concentrating capacity in the high-symmetry regions where stable materials live.

\subsubsection*{\revised{What ignoring symmetry costs, and what respecting it buys}}
The cost of ignoring this axis is visible in the first generation of generators, and the benefit of respecting it is quantifiable. Voxel- and image-style GANs [\ref{cit:Court2020}-\ref{cit:Kim2020}] inherited biases developed for pictures: translation was handled by convolution, but rotational invariance, atom indistinguishability, and periodic boundary conditions were not represented, so such models spend data rediscovering symmetry through augmentation and produce spurious symmetry breaking (small triclinic distortions of higher-symmetry structures). On the metrics reported in the original studies (subject to the limits of self-reported evaluation discussed in Section~\ref{sec:failure_modes}), the shift to equivariant architectures produced discontinuous jumps in performance: replacing a non-equivariant decoding pipeline with joint equivariant diffusion raised the MP-20 structure-match rate from roughly $34\%$ for CDVAE to over $51\%$ for DiffCSP [\ref{cit:Xie2021}-\ref{cit:Jiao2023}], and MatterGen reports generated structures more than twice as likely to be simultaneously novel and stable, and more than an order of magnitude closer to their DFT local energy minimum, than earlier generative baselines [\ref{cit:Zeni2025}]. Equivariance therefore converts invariances from statistical regularities a model may fail to learn into structural guarantees, improving data efficiency and raising downstream validity and relaxation-survival rates.

This axis is a design decision with its own cost: hard symmetry constraints restrict expressiveness where disorder, defects, or low-symmetry ground states are central, and equivariant architectures are more expensive per parameter. Treated as a decision axis, symmetry handling gives concrete guidance: space-group/Wyckoff constraints [\ref{cit:Jiao2024}-\ref{cit:Cao2024}] for ordered high-symmetry targets (the common case in functional-materials discovery); unconstrained equivariant diffusion [\ref{cit:Jiao2023}, \ref{cit:Zeni2025}] where disorder or low-symmetry ground states matter; and autoregressive models [\ref{cit:Antunes2024}] where the interface is language- or composition-driven, for which the Stage~D screening of Section~\ref{sec:staged_validity} is a structural requirement. The symmetry-handling strategy of a generator should therefore be reported alongside its family, and Table~\ref{tab:gen_families} records both as a per-family audit of which invariances each architecture enforces \emph{by construction} and which residual failure modes are left to \emph{post-processing}.}

\subsection{Where Constraints Enter: A Staged View of Validity and Feasibility}\label{sec:staged_validity}

A central practical question is when physical constraints are imposed. For materials discovery-oriented workflows, it is useful to treat constraint handling as a staged process.

\textit{Stage A: Representation-level constraints.}
Representations can enforce invariances or global structure (including symmetry-aware parameterizations), reducing invalid degrees of freedom at the source [\ref{cit:DeBreuck2025}].

\textit{Stage B: Training-time constraints.}
Feasibility can be encouraged through losses, priors, or regularizers that penalize implausible geometry or chemistry, improving average sample quality~[\ref{cit:DeBreuck2025}, \ref{cit:Davies2019}, \ref{cit:Gao2025}].

\textit{Stage C: Sampling-time constraints.}
Constraints can be enforced during sampling (guided diffusion, constrained decoding rules), which is often where inverse-design controllability is achieved in practice.

\textit{Stage D: Post-generation screening and verification.} After generation, candidate structures must pass hard feasibility gates, including distance and density checks, charge-neutrality or oxidation-state plausibility where applicable, and basic bonding sanity checks. These filters are followed by relaxation and stability assessment, for example using MLIP, force-field, or DFT-based validation, including vibrational stability screening to identify dynamically unstable candidates, before any new discovery claim can be made [\ref{cit:DeBreuck2025}, \ref{cit:Gouvea2026}].

\revised{\subsection{Materials-Specific Behavior: Periodicity, Stoichiometry, Variable Cell Content, and Survival Under Relaxation}\label{sec:materials_behavior}
The learning objectives above are common to generative modeling at large; what distinguishes \emph{crystal} generation is how each family answers four materials questions the formalism leaves open: how it represents periodicity, whether it can enforce exact stoichiometry, whether it admits a variable cell size and atom count, and how often its samples survive physical relaxation. Table~\ref{tab:family_behaviour} answers these four questions family by family. Two points deserve emphasis. First, fixed-dimension generators (diffusion, latent-variable, flows) must choose the atom count before denoising and typically draw it from the training distribution of cell sizes, so chemistries whose stable phases require large or low-symmetry cells are structurally under-sampled, and no downstream screening recovers candidates that were never generated. Second, survival under relaxation, the first stage at which physics instead of statistics judges a sample, is reported thinly across the literature, which is why Table~\ref{tab:report} asks for it explicitly. The best available self-reported numbers indicate both real progress and remaining distance: for CSP, the MP-20 match rate rose from roughly $34\%$ (CDVAE) to over $51\%$ (DiffCSP) [\ref{cit:Xie2021}-\ref{cit:Jiao2023}]; for de novo generation, MatterGen reports $78\%$ of generated structures below $0.1$~eV/atom above the hull after DFT relaxation and $95\%$ relaxing to within RMSD $0.076$~\AA\ of their generated geometry [\ref{cit:Zeni2025}]. Until relaxation survival is reported as routinely as validity, headline generation rates remain uninterpretable.}

\revised{\begin{table}[htbp]\color{revred}
\centering
\footnotesize
\caption{\revised{How each generative family answers the four materials-specific questions that the learning objective leaves open. \emph{CSP} denotes the crystal-structure-prediction setting (fixed composition); \emph{de novo} denotes joint generation of composition and structure.}}
\label{tab:family_behaviour}
\setlength{\tabcolsep}{4pt}
\renewcommand{\arraystretch}{1.3}
\begin{tabularx}{\textwidth}{
  >{\raggedright\arraybackslash}p{0.15\textwidth}
  >{\raggedright\arraybackslash}Y
  >{\raggedright\arraybackslash}Y
  >{\raggedright\arraybackslash}Y
  >{\raggedright\arraybackslash}p{0.19\textwidth}}
\toprule
\textbf{Family} & \textbf{Periodicity} & \textbf{Exact stoichiometry} & \textbf{Variable cell / atom count} & \textbf{Survival under relaxation (reported)} \\
\midrule
Diffusion / score-based [\ref{cit:Xie2021}-\ref{cit:Jiao2023}, \ref{cit:Zeni2025}]
  & Native: periodic graph; lattice denoised jointly with fractional coordinates
  & Exact in CSP (types fixed); drifts in de novo unless masked/conditioned (adapters)
  & Fixed: $N$ chosen before denoising, biased to training cell sizes
  & MatterGen: 78\% below 0.1~eV/atom; 95\% within RMSD 0.076~\AA\ [\ref{cit:Zeni2025}]; DiffCSP MP-20 match 51\% [\ref{cit:Jiao2023}] \\
Latent-variable / VAE [\ref{cit:Xie2021}]
  & Native: periodic multi-graph encoder/decoder
  & Soft, conditionable preference
  & Fixed dimensionality per model
  & CDVAE MP-20 match 34\% (re-benchmarked) [\ref{cit:Xie2021}-\ref{cit:Jiao2023}] \\
Autoregressive (CIF tokens) [\ref{cit:Park2025}, \ref{cit:Antunes2024}]
  & None geometric: lattice is tokens, checked only at screening
  & Exact via constrained decoding
  & Free: sequence length is unconstrained
  & CrystaLLM: 20/102 novel candidates within hull threshold after DFT [\ref{cit:Antunes2024}] \\
Normalizing flows [\ref{cit:Luo2025}]
  & Via invariant representation
  & Via conditioning
  & Most constrained: invertibility fixes dimensionality
  & Reported on DFT subsets; rate not headline [\ref{cit:Luo2025}] \\
GANs (voxel/grid) [\ref{cit:Court2020}-\ref{cit:Kim2020}]
  & Not exact on finite grids
  & Occupancy-thresholded
  & Implicit in voxel density
  & Not systematically reported \\
\bottomrule
\end{tabularx}
\end{table}}

Tables~\ref{tab:gen_families} and~\ref{tab:gen_families_tradeoffs} consolidate the major crystal-generation algorithm families across all four stages: Table~\ref{tab:gen_families} surveys their statistical learning objectives\revised{, structural representations, and a per-family symmetry audit,} while Table~\ref{tab:gen_families_tradeoffs} profiles their practical strengths, failure modes\revised{, and materials-specific behavior} relevant to inverse-design deployment.

\begin{table}[htbp]
\centering
\footnotesize
\caption{Algorithm families for crystal structure generation: learning paradigms, structural representations, \revised{and a symmetry audit: the invariances each architecture guarantees \emph{by construction} (fourth column) versus the failure modes left for \emph{post-generation repair} (fifth column).}}
\label{tab:gen_families}
\setlength{\tabcolsep}{4pt}
\renewcommand{\arraystretch}{1.25}
\begin{tabularx}{\textwidth}{
  >{\raggedright\arraybackslash}p{0.14\textwidth}
  >{\raggedright\arraybackslash}p{0.15\textwidth}
  >{\raggedright\arraybackslash}p{0.15\textwidth}
  >{\raggedright\arraybackslash\leavevmode}Y
  >{\raggedright\arraybackslash\leavevmode}Y}
\toprule
\textbf{Family}
  & \textbf{Core learning idea}
  & \textbf{Typical representation}
  & \revised{\textbf{Symmetry by construction}}
  & \revised{\textbf{Left to post-processing}} \\
\midrule
VAE and latent-variable models\allowbreak\textsuperscript{[\ref{cit:Xie2021}, \ref{cit:Ren2022}]}
  & ELBO maximization; latent traversal for inverse design
  & Lattice, coordinates, and species as a continuous latent code
  & \revised{$E(3)$/permutation in graph encoder; periodicity via periodic multi-graphs}
  & \revised{Latent traversal can decode off-manifold, broken-symmetry structures} \\[3pt]
Auto\-regressive Transformer\allowbreak\textsuperscript{[\ref{cit:Park2025}, \ref{cit:Antunes2024}]}
  & Autoregressive factorization of $p(\mathbf{x})$ over discrete tokens
  & CIF or structure tokens (atom types, coordinates, symmetry)
  & \revised{None for $E(3)$; only through space-group tokens or Wyckoff factorization [\ref{cit:Cao2024}]}
  & \revised{Symmetry and geometry checks, dedup, MLIP/DFT relaxation are structural} \\[3pt]
Normalizing flows\allowbreak\textsuperscript{[\ref{cit:Luo2025}]}
  & Exact likelihood via invertible bijection
  & Continuous crystallographic descriptors under invertible transforms
  & \revised{Via invariant descriptor inputs [\ref{cit:Ren2022}, \ref{cit:Xiao2023}]}
  & \revised{Hard constraints conflict with invertibility; feasibility enforced outside the flow} \\[3pt]
Diffusion and score-based models\allowbreak\textsuperscript{[\ref{cit:Jiao2023}, \ref{cit:Zeni2025}, \ref{cit:Ho2020}-\ref{cit:Song2020}]}
  & Score network reverses a forward noising process
  & Atoms, lattice, and species under joint diffusion
  & \revised{An $E(3)$-equivariant score guarantees an invariant density [\ref{cit:Kohler2020}-\ref{cit:Xu2022}]; space-group projection [\ref{cit:Jiao2024}]}
  & \revised{Strong guidance can re-break symmetry (Sec.~5.4)} \\[3pt]
GANs\allowbreak\textsuperscript{[\ref{cit:Court2020}-\ref{cit:Kim2020}]}
  & Adversarial training; no explicit likelihood
  & Voxel grids or unit-cell latent codes
  & \revised{Translation only (convolution)}
  & \revised{Atom extraction, stoichiometry and periodicity restoration: the burden that motivated the move to equivariant models} \\[3pt]
Energy-based and multimodal\slash RL hybrids\allowbreak\textsuperscript{[\ref{cit:Babu2026}, \ref{cit:DeBreuck2025}, \ref{cit:Karpovich2024}]}
  & Physics- or target-informed reward guides a generative prior
  & Diffusion\slash equivariant backbone with a surrogate or target model
  & \revised{Inherits backbone equivariance; energy/target reward is a symmetry invariant}
  & \revised{Reward hacking can exploit any invariance the backbone misses} \\
\bottomrule
\end{tabularx}
\end{table}

\begin{table}[htbp]
\centering
\footnotesize
\caption{Practical trade-offs and inverse-design roles of crystal structure generation families\revised{. The final two columns state how each family handles the two constraints that distinguish crystals from molecules: the periodic boundary conditions, and the composition and size of the unit cell.}}
\label{tab:gen_families_tradeoffs}
\setlength{\tabcolsep}{3.5pt}
\renewcommand{\arraystretch}{1.30}
\begin{tabularx}{\textwidth}{
  >{\raggedright\arraybackslash}p{0.140\textwidth}
  >{\raggedright\arraybackslash}p{0.135\textwidth}
  >{\raggedright\arraybackslash}p{0.135\textwidth}
  >{\raggedright\arraybackslash}p{0.115\textwidth}
  >{\raggedright\arraybackslash\leavevmode}Y
  >{\raggedright\arraybackslash\leavevmode}Y}
\toprule
\textbf{Family}
  & \textbf{Key strengths}
  & \textbf{Common failure modes}
  & \textbf{Inverse-design task}
  & \revised{\textbf{Periodicity handling}}
  & \revised{\textbf{Composition and cell size}} \\
\midrule
VAE and latent-variable models
  & Smooth latent space; gradient property steering
  & Validity drift; posterior collapse
  & Latent space optimization
  & \revised{Built in, through a periodic multi-graph encoder and decoder}
  & \revised{Composition can be steered but only softly; the atom count is fixed per model and biased toward training cell sizes} \\[3pt]
Autoregressive Transformer
  & Flexible conditioning on composition or text
  & Long-range consistency fails; post-generation screening needed
  & Conditional generation from composition or text
  & \revised{Not represented: the lattice is emitted as tokens and checked only at screening}
  & \revised{Composition is exact under constrained decoding, and sequence length is unconstrained, so cell size is free} \\[3pt]
Normalizing flows
  & Exact likelihood; direct latent manipulation
  & Hard constraints nontrivial in invertible architectures
  & Likelihood scoring; posterior-guided refinement
  & \revised{Inherited from an invariant crystallographic representation}
  & \revised{Composition is set by conditioning; invertibility fixes the dimensionality, so the atom count cannot vary} \\[3pt]
Diffusion and score-based models
  & High quality and diversity; CFG guidance; equivariant symmetry
  & Expensive sampling; poor scaling to large cells
  & De~novo generation with property guidance
  & \revised{Built in: the lattice is denoised jointly with the fractional coordinates}
  & \revised{Composition is exact when types are fixed but drifts in de novo generation; the atom count is chosen before denoising} \\[3pt]
GANs
  & Fast single-pass sampling
  & Mode collapse; validity not enforced
  & High-throughput candidate pre-screening
  & \revised{Approximate only, since a finite voxel grid cannot represent it exactly}
  & \revised{Composition follows from an occupancy threshold, and cell content is implicit in the voxel density} \\[3pt]
Energy-based and RL hybrids
  & Targets objectives with physics-guided selection
  & Reward hacking; sparse feedback
  & Targeted sampling under property constraints
  & \revised{Inherited from whichever backbone generator is used}
  & \revised{Composition is preserved by the action space, and the atom count can change as the candidate is built} \\
\bottomrule
\end{tabularx}
\end{table}

\section{\revised{Shared Multimodal Materials Representations for Inverse Design}}
\label{sec:multimodal_section}

Multimodal learning aims to learn jointly from different forms of data within a shared representation space. In inverse design, the challenge is the scale of chemical space compounded by a second difficulty: evidence about materials is scattered across many modalities, including crystal structures, compositions, measured and computed properties, microscopy, spectroscopy, processing conditions, and scientific text. Multimodal learning addresses this fragmentation by building a common representational space, a shared language in which these heterogeneous descriptions become comparable and interoperable [\ref{cit:Wu2025}-\ref{cit:Moro2025}, \ref{cit:Kononova2019}-\ref{cit:Venugopal2024}, \ref{cit:PyzerKnapp2025}, \ref{cit:Isayev2017}].

\revised{As an ideal, one can conceive of a \emph{universal multimodal chemical space}: a representation in which every description of the same material coincides and all physically meaningful relationships between materials are preserved as geometric ones. This ideal is well-defined because every experimental and computational modality is ultimately a partial measurement of the same underlying object, an atomic configuration together with its processing history, so a universal space would be the representational counterpart of the structure-property map itself; the observation that large models trained on different modalities converge toward increasingly similar internal representations (the platonic-representation hypothesis [\ref{cit:Huh2024}]) suggests the limit is directionally meaningful. But an ideal is a direction of travel, not a claim about current systems. What the field has demonstrated are \emph{shared multimodal materials representations}: alignments of specific modality pairs, under specific pairing assumptions, with measurable failure modes. Throughout this review, \emph{universal multimodal chemical space} refers only to the labeled ideal, and every claim about existing systems uses \emph{shared multimodal materials representation.}}

\revised{Such a representation can be assessed against three operational criteria, each with a measurable test and a held-out protocol. The first is consistency: paired descriptions of the same material should map to nearby embeddings, measured by cross-modal retrieval (Recall@$K$, mean reciprocal rank) on held-out pairs against a declared distractor-pool size, and with compositional-generalization splits, which separate alignment from memorization of training pairs [\ref{cit:Wu2025}, \ref{cit:Olivetti2020}, \ref{cit:Suzuki2025}-\ref{cit:Ozawa2024}]. The second is property-awareness: frozen multimodal embeddings should predict held-out properties through a linear probe at accuracy competitive with supervised single-modality baselines, and any loss of accuracy relative to the best single modality should be reported, since it indicates that alignment has been bought at the price of information [\ref{cit:Moro2025}, \ref{cit:Dunn2020}-\ref{cit:PyzerKnapp2025}, \ref{cit:Batra2021}]. The third is design usability: conditioning a generator on the shared representation should raise pass rates through the ledger of Section~\ref{sec:ledger} above unconditional generation at a matched sampling budget, since a shared space that steers generation without improving ledger survival adds little [\ref{cit:Park2025}-\ref{cit:Ren2022}, \ref{cit:Wu2025}-\ref{cit:Moro2025}, \ref{cit:Chenebuah2024}]. Progress toward a universal space is therefore a matter of measured movement on these three axes.}

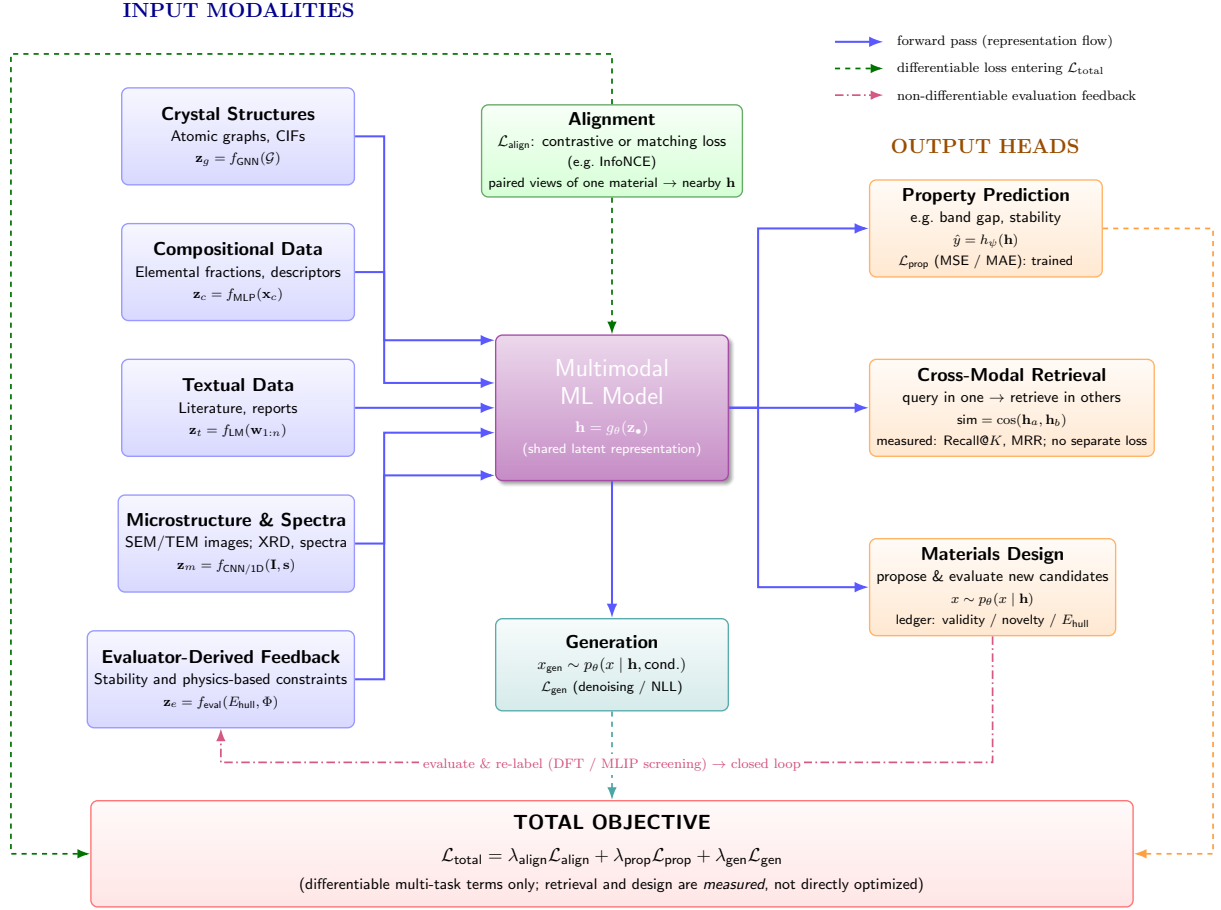
\begin{figure}[h!]
\centering
\resizebox{\linewidth}{!}{%
\begin{tikzpicture}[
    font=\sffamily,
    >=Latex,
    box/.style={
        rounded corners=4pt,
        inner sep=4pt,
        minimum width=4.8cm,
        minimum height=2.0cm,
        align=center,
        draw,
        thick
    },
    inputbox/.style={box, top color=blue!3, bottom color=blue!15, draw=blue!40},
    outputbox/.style={box, top color=orange!3, bottom color=orange!18, draw=orange!60},
    centralbox/.style={
        box, minimum height=3.0cm,
        top color=violet!15, bottom color=violet!45,
        text=white, draw=violet!70,
        blur shadow={shadow xshift=2pt,shadow yshift=-2pt}
    },
    alignbox/.style={box, minimum height=1.9cm, top color=green!3, bottom color=green!16, draw=green!45!black!60},
    genbox/.style={box, minimum height=1.9cm, top color=teal!3, bottom color=teal!16, draw=teal!65},
    totalbox/.style={
        rounded corners=4pt, inner sep=8pt,
        minimum width=21.5cm, minimum height=2.0cm,
        align=center, draw=red!55, thick,
        top color=red!2, bottom color=red!10
    },
    flow/.style={->,line width=1.2pt,draw=blue!65},
    alignflow/.style={->,line width=1.0pt,draw=green!45!black,dashed},
    genflow/.style={->,line width=1.0pt,draw=teal!70,dashed},
    proploss/.style={->,line width=1.0pt,draw=orange!75,dashed},
    feedback/.style={->,line width=1.0pt,draw=purple!65,dash pattern=on 5pt off 2pt on 1pt off 2pt},
    grouplabel/.style={font=\bfseries\large}
]

% ============================ column headings ==============================
\node[grouplabel,text=blue!55!black]   at (-7.7, 8.2) {INPUT MODALITIES};
\node[grouplabel,text=orange!60!black] at ( 7.7, 5.4) {OUTPUT HEADS};

% ================================= legend ==================================
\begin{scope}[shift={(4.6,6.6)}]
  \draw[flow]      (0,0.95) -- (1.0,0.95);
  \node[anchor=west,font=\scriptsize] at (1.15,0.95) {forward pass (representation flow)};
  \draw[alignflow] (0,0.40) -- (1.0,0.40);
  \node[anchor=west,font=\scriptsize] at (1.15,0.40) {differentiable loss entering $\mathcal{L}_{\text{total}}$};
  \draw[feedback]  (0,-0.15) -- (1.0,-0.15);
  \node[anchor=west,font=\scriptsize] at (1.15,-0.15) {non-differentiable evaluation feedback};
\end{scope}

% ============================ input modalities =============================
\node[inputbox, anchor=east] (crystal) at (-5.3, 5.6)
    {\textbf{Crystal Structures}\\[-1pt]\footnotesize Atomic graphs, CIFs\\[-1pt]{\scriptsize$\mathbf{z}_g=f_{\text{GNN}}(\mathcal{G})$}};

\node[inputbox, anchor=east] (comp)    at (-5.3, 2.8)
    {\textbf{Compositional Data}\\[-1pt]\footnotesize Elemental fractions, descriptors\\[-1pt]{\scriptsize$\mathbf{z}_c=f_{\text{MLP}}(\mathbf{x}_c)$}};

\node[inputbox, anchor=east] (text)    at (-5.3, 0)
    {\textbf{Textual Data}\\[-1pt]\footnotesize Literature, reports\\[-1pt]{\scriptsize$\mathbf{z}_t=f_{\text{LM}}(\mathbf{w}_{1:n})$}};

\node[inputbox, anchor=east] (micro)   at (-5.3,-2.8)
    {\textbf{Microstructure \& Spectra}\\[-1pt]\footnotesize SEM/TEM images; XRD, spectra\\[-1pt]{\scriptsize$\mathbf{z}_m=f_{\text{CNN/1D}}(\mathbf{I},\mathbf{s})$}};

\node[inputbox, anchor=east] (evalb)   at (-5.3,-5.6)
    {\textbf{Evaluator-Derived Feedback}\\[-1pt]\footnotesize Stability and physics-based constraints\\[-1pt]{\scriptsize$\mathbf{z}_e=f_{\text{eval}}(E_{\text{hull}},\Phi)$}};

% ====================== alignment / central / generation ===================
\node[alignbox] (align) at (0, 5.3)
    {\textbf{Alignment}\\[-1pt]{\footnotesize$\mathcal{L}_{\text{align}}$: contrastive or matching loss}\\[-2pt]{\scriptsize(e.g.\ InfoNCE)}\\[-1pt]{\scriptsize paired views of one material $\rightarrow$ nearby $\mathbf{h}$}};

\node[centralbox] (central) at (0, 0)
    {\Large Multimodal\\[1mm] \Large ML Model\\[1mm]{\footnotesize$\mathbf{h}=g_\theta(\mathbf{z}_\bullet)$}\\[-1pt]{\scriptsize (shared latent representation)}};

\node[genbox] (gen) at (0,-5.3)
    {\textbf{Generation}\\[-1pt]{\footnotesize$x_{\text{gen}}\sim p_\theta(x\mid\mathbf{h},\text{cond.})$}\\[-1pt]{\scriptsize$\mathcal{L}_{\text{gen}}$ (denoising / NLL)}};

% =============================== output heads ==============================
\node[outputbox, anchor=west] (prop)   at (5.3, 3.7)
    {\textbf{Property Prediction}\\[-1pt]\footnotesize e.g.\ band gap, stability\\[-1pt]{\scriptsize$\hat{y}=h_\psi(\mathbf{h})$}\\[-1pt]{\scriptsize$\mathcal{L}_{\text{prop}}$ (MSE / MAE): trained}};

\node[outputbox, anchor=west] (retr)   at (5.3, 0)
    {\textbf{Cross-Modal Retrieval}\\[-1pt]\footnotesize query in one $\rightarrow$ retrieve in others\\[-1pt]{\scriptsize$\text{sim}=\cos(\mathbf{h}_a,\mathbf{h}_b)$}\\[-1pt]{\scriptsize measured: Recall@$K$, MRR; no separate loss}};

\node[outputbox, anchor=west] (design) at (5.3,-3.7)
    {\textbf{Materials Design}\\[-1pt]\footnotesize propose \& evaluate new candidates\\[-1pt]{\scriptsize$x\sim p_\theta(x\mid\mathbf{h})$}\\[-1pt]{\scriptsize ledger: validity / novelty / $E_{\text{hull}}$}};

% ============================= total objective =============================
\node[totalbox] (total) at (0,-9.2)
    {\textbf{\large TOTAL OBJECTIVE}\\[2mm]
     $\mathcal{L}_{\text{total}}=\lambda_{\text{align}}\mathcal{L}_{\text{align}}
       +\lambda_{\text{prop}}\mathcal{L}_{\text{prop}}
       +\lambda_{\text{gen}}\mathcal{L}_{\text{gen}}$\\[1mm]
     {\footnotesize (differentiable multi-task terms only; retrieval and design are \emph{measured}, not directly optimized)}};

% ============================== input arrows ===============================
\draw[flow] (crystal.east) -- ++(0.6,0) |- (central.150);
\draw[flow] (comp.east)    -- ++(0.6,0) |- (central.168);
\draw[flow] (text.east)    -- ++(0.6,0) -- (central.west);
\draw[flow] (micro.east)   -- ++(0.6,0) |- (central.192);
\draw[flow] (evalb.east)   -- ++(0.6,0) |- (central.210);

% ===================== alignment in, generation out ========================
\draw[alignflow] (align.south)   -- (central.north);
\draw[flow]      (central.south) -- (gen.north);

% ============================== output arrows ==============================
\draw[flow] (central.east) -- ++(0.6,0) |- (prop.west);
\draw[flow] (central.east) --             (retr.west);
\draw[flow] (central.east) -- ++(0.6,0) |- (design.west);

% ============ loss contributions to the total objective ====================
% alignment: routed ABOVE the crystal box, in the gap below the column heading
\draw[alignflow] (align.north) -- (0,7.3) -- (-12.4,7.3) -- (-12.4,-9.2) -- (total.west);
\draw[proploss]  (prop.east)   -- (12.4,3.7) -- (12.4,-9.2) -- (total.east);
\draw[genflow]   (gen.south)   -- (total.north);

% ==================== closed loop: design -> evaluator ======================
\draw[feedback] (design.south) -- ++(0,-2.6) -| (evalb.south);
\node[font=\scriptsize,text=purple!70,fill=white,inner sep=2pt] at (0,-7.35)
    {evaluate \& re-label (DFT / MLIP screening) $\rightarrow$ closed loop};

\end{tikzpicture}%
}
\caption{\textbf{Schematic of a multimodal materials machine learning workflow.}
The framework encodes and fuses five distinct input modalities into a \revised{shared multimodal representation model (an approximation, under stated pairing assumptions, of the universal-space ideal of Section~\ref{sec:multimodal_section})}. It comprises \textit{(i) crystal structures} encoded as atomic graphs or crystallographic information files (CIFs);
\textit{(ii) compositional data}, including elemental fractions and derived descriptors;
\textit{(iii) microstructure images} such as scanning electron microscopy (SEM), transmission electron microscopy (TEM), and optical micrographs;
\textit{(iv) textual data} extracted from scientific literature and experimental reports; and
\textit{(v) physics-based signals} from equivariant graph neural network embeddings and structural stability assessments.
From this integrated space, three downstream tasks emerge:
\textit{property prediction} (e.g., band gap, thermodynamical stability);
\textit{cross-modal retrieval} (querying across structures, images, and text to identify chemically similar candidates); and
\textit{materials design} via large-scale foundation-model generation conditioned on the shared latent space.
The central model serves as the key enabling component: by aligning these five heterogeneous modalities in a common representation, it supports target specification, candidate ranking, and constrained generation within a single unified framework. \revised{The figure is generic: the encoders and the alignment loss are drawn as slots, with labels naming only the commonest current choices ($f_{\text{GNN}}$ for structures, $f_{\text{MLP}}$ for compositions, $f_{\text{LM}}$ for text, $f_{\text{CNN/1D}}$ for images and spectra). A shared projection $g_\theta$ maps the embeddings $\mathbf{z}_\bullet$ into a common space, where an alignment term $\mathcal{L}_{\text{align}}$ pulls paired descriptions of the same material together. Any contrastive or matching objective can occupy that slot; InfoNCE (the noise-contrastive estimation objective, in CLIP-style form [\ref{cit:Radford2021}]) is the usual choice in the systems surveyed here [\ref{cit:Park2025}, \ref{cit:Wu2025}, \ref{cit:Suzuki2025}-\ref{cit:Ozawa2024}], while triplet, sigmoid-pairwise, and optimal-transport objectives are in use elsewhere and differ in their false-negative behavior, which matters for the reason given in Section~\ref{sec:mm_limits}. The evaluator channel supplies a physics-derived embedding $\mathbf{z}_e=f_{\text{eval}}(E_{\text{hull}},\Phi)$. The aligned embedding $\mathbf{h}$ drives a property head $\hat{y}=h_\psi(\mathbf{h})$, cosine-similarity retrieval, and conditional generation $x\sim p_\theta(x\mid\mathbf{h})$. Only three differentiable terms enter the joint objective $\mathcal{L}_{\text{total}}=\lambda_{\text{align}}\mathcal{L}_{\text{align}}+\lambda_{\text{prop}}\mathcal{L}_{\text{prop}}+\lambda_{\text{gen}}\mathcal{L}_{\text{gen}}$; retrieval is not separately trained but \emph{measured} (Recall@$K$, mean reciprocal rank on held-out pairs), and design quality (validity, novelty, distance to the hull) is a non-differentiable ledger (Section~\ref{sec:ledger}) that re-enters training only through the closed loop at the bottom, in which screened candidates are relabeled by DFT or MLIP screening and returned as evaluator-derived feedback. Consistent with Section~\ref{sec:multimodal_section}, the figure depicts a \emph{shared} representation under stated pairing assumptions, not a demonstrated universal space.}}
\label{fig:proposal_engine_rev}
\end{figure}

Figure~\ref{fig:proposal_engine_rev} illustrates the multimodal materials ML workflow: diverse input modalities (crystal structures, composition, microstructure images, textual data) are aligned in a shared model that supports property prediction, cross-modal retrieval, and materials design.

In practice, no single modality captures the full state of a material. Structure, properties, processing history, experimental, and computational characterization, each provide only a partial view, and it is their combination that yields a more complete basis for understanding and design. Multimodal learning is therefore valuable because it can bring these distributed forms of evidence into a shared representational space, allowing information from one modality to inform, complement, and constrain another. For inverse design, this is crucial because promising candidates often need to be judged from several partially informative signals, no one of which suffices on its own, and because many high-value modalities remain sparse, costly, or unavailable [\ref{cit:Wu2025}, \ref{cit:Kononova2019}, \ref{cit:Olivetti2020}-\ref{cit:Venugopal2024}, \ref{cit:PyzerKnapp2025}, \ref{cit:Rank2024}, \ref{cit:Isayev2017}].

\subsection{What the \revised{shared} space does for inverse design: reason, search, create}\label{sec:shared_space}

We can anchor multimodality to the three discovery operations it enables:
\begin{enumerate}[leftmargin=*]
\item \textbf{Reason}: Multimodal models can infer properties from whatever evidence is available and reconcile signals that would otherwise remain siloed, for example by linking processing conditions to microstructure and, in turn, to properties. In this way, they support more coherent reasoning even when the available evidence is incomplete or noisy [\ref{cit:Wu2025}-\ref{cit:Moro2025}, \ref{cit:Olivetti2020}, \ref{cit:Das2023}].
\item \textbf{Search}: Once heterogeneous evidence is placed in a shared space, multimodal learning enables cross-modal retrieval: one can query with text, a partial measurement, or a structure and retrieve materials that are nearby under the model’s learned notion of chemical similarity. Such embeddings support analogical discovery and the ranking of candidates under limited experimental or computational budgets [\ref{cit:Wu2025}-\ref{cit:Moro2025}, \ref{cit:Tshitoyan2019}, \ref{cit:PyzerKnapp2025}, \ref{cit:Suzuki2025}-\ref{cit:Ozawa2024}].
\item \textbf{Create}: A \revised{shared} space becomes a design tool when it is coupled to a materials proposal framework, either by conditioning a generator on the shared representation to produce structures consistent with desired attributes or by moving through that space toward target behaviors and then decoding candidate materials. Text-guided generation and multimodal conditional generation provide concrete examples of this direction. [\ref{cit:Park2025}, \ref{cit:Wu2025}, \ref{cit:Babu2026}]
\end{enumerate}

Table~\ref{tab:model_overview_generation} provides a system-level reference for the crystal structure generators that underpin each of these three roles, mapping their core representations and internal mechanisms.

\begin{table}[htbp]
\centering
\footnotesize
\caption{Representative crystal structure generation models and their roles in the \revised{shared multimodal materials representation}: structural representations and core mechanisms. Models are ordered by family type and are referenced in the text as examples of generation platforms supporting the Reason-Search-Create operations described above.}
\label{tab:model_overview_generation}
\setlength{\tabcolsep}{5pt}
\renewcommand{\arraystretch}{1.30}
\begin{tabularx}{\textwidth}{
  >{\raggedright\arraybackslash}p{0.24\textwidth}
  >{\raggedright\arraybackslash}p{0.24\textwidth}
  >{\raggedright\arraybackslash}p{0.42\textwidth}}
\toprule
\textbf{Model}
  & \textbf{Representation}
  & \textbf{Core mechanism} \\
\midrule
3D GAN generators\textsuperscript{[\ref{cit:Court2020}-\ref{cit:Kim2020}]}
  & 3D grids or discretized unit cells
  & Adversarial training with periodicity post-processing \\[3pt]
CDVAE\textsuperscript{[\ref{cit:Xie2021}]}
  & Periodic graphs, fractional coordinates, lattice
  & VAE with diffusion decoder; periodic invariances enforced \\[3pt]
DiffCSP\textsuperscript{[\ref{cit:Jiao2023}]}
  & Fractional coordinates, lattice
  & Equivariant denoising of lattice and atomic positions \\[3pt]
Self-supervised generative platform\textsuperscript{[\ref{cit:DeBreuck2025}]}
  & SE(3)-equivariant Transformer latent space
  & Self-supervised training with adversarial reliability signal \\[3pt]
CrystaLLM\textsuperscript{[\ref{cit:Antunes2024}]}
  & CIF syntax as discrete tokens
  & Autoregressive CIF model; outputs steerable by reranking \\[3pt]
Chemeleon\textsuperscript{[\ref{cit:Park2025}]}
  & Aligned text and crystal embeddings
  & Contrastive text-structure alignment; text-conditioned diffusion \\[3pt]
\revised{Chemeleon-2\textsuperscript{[\ref{cit:Park2026}]}}
  & \revised{VAE latent space over crystals; no text channel}
  & \revised{Latent denoising diffusion (diffusion transformer over VAE latents); GRPO fine-tuning against verifiable multi-objective rewards} \\[3pt]
\revised{MEIDNet\textsuperscript{[\ref{cit:Babu2026}]}}
  & \revised{Equivariant structure encoder; shared multimodal (structure-target) embedding}
  & \revised{Multimodal alignment with objective-driven generation and downstream feasibility checks} \\[3pt]
MatterGen\textsuperscript{[\ref{cit:Zeni2025}]}
  & Atoms, coordinates, lattice under joint diffusion
  & Foundation diffusion model; adapter fine-tuning; CFG at inference \\[3pt]
CrystalFlow\textsuperscript{[\ref{cit:Luo2025}]}
  & Invertible map over periodic crystal space
  & Exact-likelihood flow; direct latent manipulation \\[3pt]
Active learning loops\textsuperscript{[\ref{cit:Settles2009}, \ref{cit:Han2025}]}
  & Diffusion generator with active learning
  & Iterative generate-score-label-relax-fine-tune cycle \\[3pt]
Transformer-based diffusion\textsuperscript{[\ref{cit:Zeni2025}, \ref{cit:Ho2020}-\ref{cit:Song2020}]}
  & Transformer denoising over crystal tokens
  & Joint lattice and atomic modeling; scales with Transformer capacity \\
\bottomrule
\end{tabularx}
\end{table}

\subsection{Multimodal Models Serving Various Functional Roles}\label{sec:mm_roles}

Multimodal learning is becoming increasingly relevant to inverse materials design because the information needed for materials discovery is seldom contained in a single modality. From the perspective of inverse design, the key question is whether combining modalities strengthens a specific operational role: \emph{evaluation}, \emph{retrieval}, \emph{text-conditioned proposal}. \revised{These roles are often conflated, yet they rest on different data, losses, and validation needs; Table~\ref{tab:mm_threads} separates the six threads, and the paragraphs below group them by the role each plays in the loop, naming in each case the objective that is actually trained and the evidence a claim about that thread requires. \revised{Three points apply across all six threads. First, the loss families are not interchangeable. Supervised fusion learns a labeled map and is judged by predictive error; contrastive objectives learn only relative geometry and are judged by ranking; and conditional generative objectives learn a sampler that can only be judged by whether its samples survive. A claim proved under one of these does not transfer to another. Second, the validation need follows from the loss, not from the modality. A fusion model owes an ablation against its best single-modality baseline, because the entire claim is that the added modality contributes something; a contrastive model owes a held-out retrieval protocol with a declared distractor-pool size and declared pair granularity, because Recall@$K$ is meaningless without knowing how many distractors it was measured against and what counted as a positive; a text-conditioned generator owes the full ledger of Section~\ref{sec:ledger} against an unconditional baseline at matched sampling budget, because a prompt that steers samples without improving what survives has changed the output distribution without improving the science. Third, retrieval is the one thread with no loss of its own: it is the capability the contrastive term buys, measured after the fact, which is why it appears in Table~\ref{tab:mm_threads} with an empty loss column and a full validation column.}}

\revised{\begin{table}[htbp]\color{revred}
\centering
\footnotesize
\caption{\revised{Six problems often grouped under multimodal materials learning, separated by data requirement, loss family, and validation need. Conflating threads with different validation needs is how shared-representation claims become unfalsifiable.}}
\label{tab:mm_threads}
\setlength{\tabcolsep}{4pt}
\renewcommand{\arraystretch}{1.25}
\begin{tabularx}{\textwidth}{
  >{\raggedright\arraybackslash}p{0.17\textwidth}
  >{\raggedright\arraybackslash}p{0.19\textwidth}
  >{\raggedright\arraybackslash}p{0.15\textwidth}
  >{\raggedright\arraybackslash}Y
  >{\raggedright\arraybackslash}p{0.13\textwidth}}
\toprule
\textbf{Thread} & \textbf{Data requirement} & \textbf{Loss family} & \textbf{Validation need} & \textbf{Example systems} \\
\midrule
Multimodal property prediction & Paired structure and text or descriptors, with labels & Supervised fusion & Ablation vs.\ single-modality baselines; label-leakage audit of the text channel & CrysMMNet [\ref{cit:Das2023}]; MultiMat [\ref{cit:Moro2025}] \\
Text-structure contrastive learning & Corpora of structure-description pairs & InfoNCE / CLIP-style [\ref{cit:Radford2021}] & Held-out Recall@$K$; false-negative control; declared pair granularity & CLaSP [\ref{cit:Suzuki2025}]; CLICS [\ref{cit:Ozawa2024}] \\
Cross-modal retrieval & Shared embedding from contrastive training & No additional training objective; retrieval reuses the contrastive embedding and is scored, not optimized & Declared distractor-pool size; compositional-generalization splits & CLaSP [\ref{cit:Suzuki2025}] \\
Text / literature mining & Corpus with annotation schema & Information-extraction / LM objectives & Extraction precision/recall vs.\ annotated gold sets & Kononova [\ref{cit:Kononova2019}]; Tshitoyan [\ref{cit:Tshitoyan2019}]; Olivetti [\ref{cit:Olivetti2020}]; MatKG [\ref{cit:Venugopal2024}] \\
Image / spectroscopy fusion & Micrographs or spectra with process metadata & Contrastive or supervised fusion & Pair-definition audit (specimen vs.\ material); imaging-noise robustness & MatMCL [\ref{cit:Wu2025}] \\
Text-conditioned generation & Prompt-structure pairs & Conditional generative loss, with guidance & Full ledger pass rates vs.\ unconditional (Section~\ref{sec:ledger}); hallucination protocol & Chemeleon [\ref{cit:Park2025}]; MEIDNet [\ref{cit:Babu2026}] \\
\bottomrule
\end{tabularx}
\end{table}}

\paragraph{Multimodality for evaluation:}
Some models primarily strengthen the \emph{evaluation} side of the loop by improving property prediction or candidate ranking through fusion of complementary information sources. CrysMMNet~[\ref{cit:Das2023}] combines structural encoders with curated textual descriptors for multimodal fusion in property prediction. The reported improvements suggest that text can supply global or domain-level cues that may be difficult to extract robustly from crystal structure alone, including symmetry-related information accumulated from prior scientific understanding. If multimodal fusion yields better or more stable property predictors, then candidate ranking and prioritization become more reliable, a form of multimodality that contributes to inverse-design performance without generating structures itself. \revised{The evidence such a claim owes is an ablation against the strongest single-modality baseline trained on the same split, together with an audit of the text channel for label leakage (Section~\ref{sec:mm_limits}), since the improvement is only attributable to fusion if the added modality is doing work that the structure encoder could not do alone.} MultiMat~[\ref{cit:Moro2025}, \ref{cit:Butler2018}-\ref{cit:Schmidt2019}] similarly learns transferable joint representations from large corpora in which structure/composition and multiple computed properties are co-observed, embedding materials and multi-property targets together within a shared latent space. A design objective can then be represented as a profile of several properties at once, supporting similarity-based discovery around known high-performing examples.

\paragraph{Multimodality for retrieval:}
A key function of multimodal learning is cross-modal retrieval. By mapping heterogeneous evidence into a shared embedding space, a model can use a text description, a partial measurement, or a reference structure as a query to retrieve materials that are nearby under its learned representation of chemical and structural similarity. CLaSP~[\ref{cit:Suzuki2025}] and related contrastive approaches (e.g.\ CLICS~[\ref{cit:Ozawa2024}]) align structural representations with scientific text, improving the robustness and semantic organization of structure embeddings while enabling text-to-structure retrieval. That capability matters for inverse design, because design intent is usually voiced first in qualitative scientific language well before it can be written as a fully specified numerical target vector [\ref{cit:Olivetti2020}]. \revised{Retrieval claims are trained by a contrastive objective and validated by ranking on held-out pairs, so three things must be declared for a reported Recall@$K$ or mean reciprocal rank to be interpretable: the size of the distractor pool the query was ranked against, the granularity at which a positive pair was defined (specimen, material, or phase), and whether the split was random or compositionally grouped. Without the last of these, retrieval of memorized pairs is indistinguishable from alignment; without the first, the number is not comparable across papers.}

\paragraph{Multimodality for text-conditioned proposal:}
The magnetocaloric workflow reported by Weston \emph{et al.}~[\ref{cit:Weston2021}] illustrates the operational application of multimodality in which literature text is mined into structured training data and integrated with prediction, candidate generation, and DFT-based validation. Text thereby becomes an actionable design resource that feeds directly into a proposal-evaluation-verification cycle [\ref{cit:Kononova2019}-\ref{cit:Tshitoyan2019}]. Chemeleon~[\ref{cit:Park2025}] extends this direction by aligning language and crystal representations and conditioning crystal generation on textual prompts. Instead of requiring every target condition to be expressed as a fixed numerical label, the model allows the specification of design directions via natural language and then generates candidate crystal structures consistent with that prompt.

\paragraph{Multimodality for integrated inverse-design systems:}
Beyond simple retrieval or property prediction, multimodal learning can place alignment directly within the proposal mechanism. MatMCL~[\ref{cit:Wu2025}] links processing parameters, microstructure images (e.g.\ SEM), and measured properties in a shared embedding space, making the process-structure-property chain operational within a unified representation. It enables cross-modal tasks such as retrieving similar morphologies from process conditions and generating plausible microstructures conditioned on synthesis variables, a capability that matters in experimentally grounded problems where processing history is inseparable from the resulting functional response [\ref{cit:Ren2022}]. MEIDNet~[\ref{cit:Babu2026}] presents multimodality as an integrated inverse-design system by combining equivariant structure encoding, multimodal alignment between structure and target information, objective-driven generation, and downstream checks. The approach turns the shared representation into an operational design space, where targets, structures, and objectives can be related, navigated, and optimized jointly. It therefore points toward integrated workflows in which multimodal understanding, candidate generation, and verification-aware decision-making are treated as connected parts of the same inverse-design loop [\ref{cit:Chitturi2024}]. \revised{Because their architectures natively support complex conditioning, latent denoising-diffusion models like Chemeleon-2 [\ref{cit:Park2026}]  represent a highly promising avenue for multimodal inverse design as well. By navigating and decoding from a shared latent representation (Section~\ref{sec:latent_opt}), these models provide a unified space where diverse modalities can jointly guide the generation process.}

\revised{Fusion-based property prediction and contrastive retrieval are demonstrated capabilities, with held-out gains reported on the benchmarks used in the original studies [\ref{cit:Wu2025}-\ref{cit:Moro2025}, \ref{cit:Suzuki2025}-\ref{cit:Ozawa2024}, \ref{cit:Das2023}], though, per Section~\ref{sec:mm_limits}, text-augmented gains remain unaudited for label leakage. Text-conditioned generation is demonstrated at the level of composition- and symmetry-class prompts, with Chemeleon the only system reporting full DFT validation campaigns for its prompted candidates (e.g.\ Ti-O: 549 generated structures reduced to 122 unique metastable ones; Table~\ref{tab:quant_B}) [\ref{cit:Park2025}]. No multimodal system has yet reported an experimentally realized material attributable to the multimodal conditioning itself. The claims of this section therefore reach L4 routinely and L6 in the limited sense that a cheap computed property can be checked at DFT fidelity, while L7 and L8 remain unreached. Table~\ref{tab:mm_threads} records the validation each thread still owes.}

\subsection{Limitations and Open Challenges}\label{sec:mm_limits}

The field is still in its early stages, and most existing systems remain specialized to specific modalities. Furthermore, several limitations are common to all such functional roles. \emph{Incomplete modalities} are the norm in practice: many useful signals are expensive or sparsely available, requiring models to function even when evidence is partial. In addition, \emph{weak grounding} remains a persistent challenge, as semantic alignment between text and structure does not guarantee that generated candidates are physically valid or energetically stable. Finally, \emph{dataset mismatch} introduces distribution shift when models trained on curated computational databases are applied to experimental settings dominated by noise, missing data, and process variability. To address these issues, robust evaluation of multimodal inverse-design systems should test whether performance is maintained when modalities are missing, whether cross-modal retrieval remains reliable under distribution shift, and whether multimodal conditioning improves downstream success rates. The staged evaluation ledger described in Section~\ref{sec:failure_modes} provides one way to make these checks explicit.

\revised{Beneath these lies the harder question of what, exactly, is being aligned. A crystal graph is an idealized description of an infinite periodic solid; a micrograph is a lossy 2D projection of one finite, defective specimen; recipe text is an ambiguous account of intent. Aligning them is aligning views of \emph{different objects}, the ideal phase and an actual sample, related by an unobserved processing chain. Three consequences follow. Contrastive pair definition is a modeling decision, not a given: positives can be declared at the level of specimen, material, or phase, each injecting a different label noise (one structure maps to many micrographs; one micrograph may contain several phases), and the chosen granularity should be reported. The discrete-continuous gap (atomic coordinates vs.\ pixels, tokens vs.\ spectra) is bridged by modality-specific encoders meeting in a CLIP-style projection space [\ref{cit:Radford2021}], so the shared space is only as physical as its weakest encoder. Furthermore, false negatives are endemic: in any large batch, chemically near-identical materials appear as negative pairs, forcing the contrastive loss to push apart representations that physics says belong together. The standard recipe for contrastive learning was built for internet images and captions, where near-duplicates are rare, an assumption that materials data violate.}

\revised{Four questions then determine whether a shared representation earns its keep. \emph{What does text add beyond structure and composition?} Real but modest property-prediction gains where text supplies context expensive to compute: synthesis conditions, morphology, application context [\ref{cit:Moro2025}, \ref{cit:Das2023}], although one qualification is seldom stated. Text about a material frequently contains the property value itself, so any text-augmented gain must be audited for label leakage before it is credited to multimodal synergy. \emph{How often does text-conditioning yield physically valid candidates?} Encouraging for composition- and symmetry-level prompts [\ref{cit:Park2025}], but systematically under-reported; text-conditioned generators should be scored on the same ledger as any other generator (Section~\ref{sec:ledger}), with conditioning-strength sweeps (Section~\ref{sec:guided_diffusion}), since a prompt is just another conditioning signal and inherits every pathology of Section~\ref{sec:sampling_challenges}. \emph{How are hallucinated associations detected?} The term \emph{hallucination protocol} is used here for a bundle of four testable checks, which Table~\ref{tab:mm_threads} asks of any text-conditioned generator. (i) \emph{Cycle consistency}: generate a structure from a prompt, caption or classify that structure with an independent model, and measure agreement with the original prompt. (ii) \emph{Retrieval controls}: verify that a generated structure lies closer in the shared space to its own prompt than to randomly drawn prompts, which detects a generator that is ignoring its conditioning. (iii) \emph{Negative-prompt controls}: submit physically impossible or self-contradictory prompts and measure the abstention rate, since a model that returns a structure for every prompt, including impossible ones, provides no evidence of grounding. (iv) \emph{Database-grounding audit}: check whether the associations a prompt evokes are supported by the training corpus or are artifacts of surface co-occurrence in text. Reported together, the four allow grounding to be assessed quantitatively. \emph{What benchmarks exist?} Almost none that are shared: current evaluations are per-paper with incompatible pair definitions, distractor pools, and splits, so a leakage-controlled, compositionally split community benchmark for materials retrieval and text-conditioned generation, with generation scored on the ledger, is among the highest-leverage infrastructure investments the field could make (Section~\ref{sec:repro}).}

\section{Inverse Design and Sampling Optimization}\label{sec:sampling_opt}

Having established in Sections \ref{sec:generation}-\ref{sec:multimodal_section} that crystal generators can serve as scalable proposal engines and that multimodal learning can help construct a \revised{shared chemical representation space}, we turn to the operational challenge of inverse design: how can we search reliably when objectives are conflicting, constraints are hard, uncertainty is unavoidable, and high-fidelity validation is expensive?

Inverse design is better understood as an uncertainty-aware pipeline than as a single monolithic model that outputs the final answer, one that integrates (a) proposal generation, (b) constraint handling, (c) surrogate-based evaluation, and (d) selective high-fidelity validation, with feedback linked to the proposal policy. Table~\ref{tab:inverse_design_paradigms} summarizes the main inverse-design paradigms and their practical limitations under realistic constraints and validation budgets.

\begin{table}[H]
\centering
\footnotesize
\caption{Principal inverse-design paradigms for inorganic crystal materials: proposal strategy, constraint handling, and key limitations.}
\label{tab:inverse_design_paradigms}
\setlength{\tabcolsep}{5pt}
\renewcommand{\arraystretch}{1.30}
\begin{tabularx}{\textwidth}{
  >{\raggedright\arraybackslash}p{0.20\textwidth}
  >{\raggedright\arraybackslash}p{0.22\textwidth}
  >{\raggedright\arraybackslash}p{0.22\textwidth}
  >{\raggedright\arraybackslash}p{0.25\textwidth}}
\toprule
\textbf{Paradigm}
  & \makecell[tl]{\textbf{Optimization}\\\textbf{strategy}}
  & \makecell[tl]{\textbf{Constraint}\\\textbf{handling}}
  & \makecell[tl]{\textbf{Key}\\\textbf{limitations}} \\
\midrule
Property-conditioned diffusion\textsuperscript{[\ref{cit:Luo2024}, \ref{cit:Zeni2025}]}
  & Conditional sampling toward property targets
  & CFG; symmetry and chemistry conditioning
  & Surrogate bias; DFT screening required \\[3pt]
Text-guided design\textsuperscript{[\ref{cit:Park2025}, \ref{cit:Tshitoyan2019}]}
  & Generate from natural-language descriptions
  & Contrastive text-crystal alignment; text-conditioned diffusion
  & Weak structural grounding from text \\[3pt]
\revised{Multimodal integrated design\textsuperscript{[\ref{cit:Babu2026}]}}
  & \revised{Objective-driven generation in a shared multimodal embedding}
  & \revised{Equivariant encoding; structure-target alignment; downstream feasibility checks}
  & \revised{Weak grounding; validation still required; early-stage} \\[3pt]
Flow-based design\textsuperscript{[\ref{cit:Luo2025}]}
  & Exact-likelihood sampling or latent manipulation
  & Conditioning networks; gradient-based latent editing
  & Hard constraints handled outside the flow \\[3pt]
Latent-space optimization\textsuperscript{[\ref{cit:Xie2021}, \ref{cit:Ren2022}, \ref{cit:Chen2021}]}
  & Gradient-based navigation in continuous latent space
  & Differentiable property heads; validity regularizers
  & Collapse under strong constraints \\[3pt]
Reinforcement learning\textsuperscript{[\ref{cit:Karpovich2024}]}
  & Policy maximizing stability and property reward
  & Reward shaping; action-space constraints
  & Reward hacking; costly without surrogates \\[3pt]
Bayesian optimization\textsuperscript{[\ref{cit:Shahriari2016}, \ref{cit:Diwale2022}-\ref{cit:Tian2025}]}
  & Sequential querying under probabilistic surrogate
  & Uncertainty-aware acquisition; constrained penalization
  & Struggles in high-dimensional spaces \\[3pt]
Active learning loops\textsuperscript{[\ref{cit:Settles2009}, \ref{cit:Han2025}]}
  & Iterative generate-screen-label-fine-tune cycle
  & Diversity filters; novelty thresholds
  & Diversity collapse; selection bias \\[3pt]
LLM-guided search\textsuperscript{[\ref{cit:Antunes2024}]}
  & Large pool generation with reranking or tree search
  & Energy and property predictors; syntax filtering
  & Predictor-dependent; physics confirmation needed \\
\bottomrule
\end{tabularx}
\end{table}

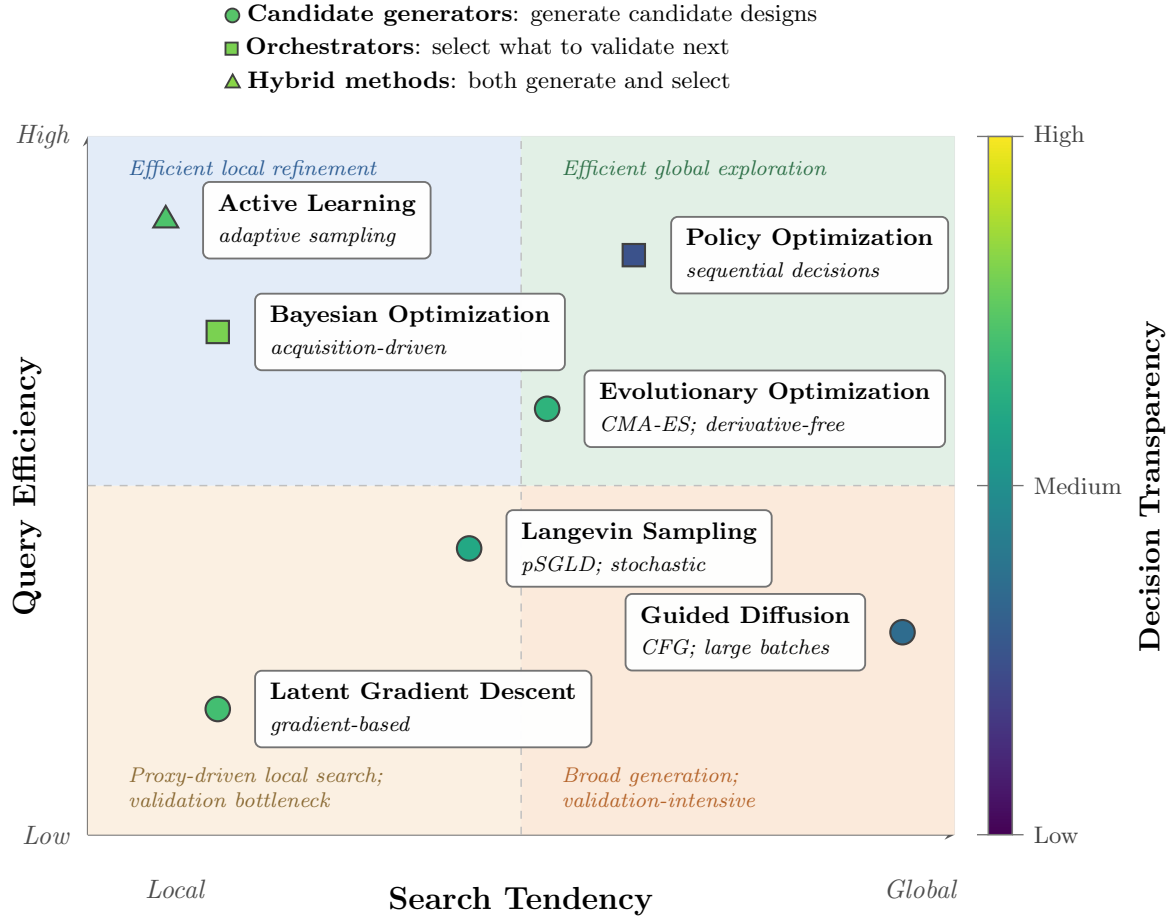
\begin{figure}[ht]
\centering
\begin{tikzpicture}

\tikzset{
  algtag/.style={
    font=\footnotesize,
    align=left,
    fill=white,
    fill opacity=0.97,
    text opacity=1,
    rounded corners=2.5pt,
    inner xsep=5.5pt,
    inner ysep=4.5pt,
    draw=black!55,
    line width=0.7pt,
    text=black,
  },
}

\begin{axis}[
    width=0.72\linewidth,
    height=0.58\linewidth,
    scale only axis,
    xmin=0, xmax=10,
    ymin=0, ymax=10,
    axis lines=left,
    axis line style={line width=0.65pt, color=black!65},
    tick style={draw=none},
    grid=both,
    grid style={line width=0.18pt, draw=black!8},
    major grid style={line width=0.28pt, draw=black!12},
    xtick={0,2,4,6,8,10},
    ytick={0,2,4,6,8,10},
    xticklabels={,,,,,},
    yticklabels={,,,,,},
    xlabel={\large\textbf{Search Tendency}},
    ylabel={\large\textbf{Query Efficiency}},
    xlabel style={font=\normalsize, yshift=-6pt},
    ylabel style={font=\normalsize, yshift=4pt},
    axis background/.style={fill=gray!3},
    clip=false,
    colormap name=viridis,
    point meta min=0, point meta max=1,
    colorbar,
    colorbar style={
      width=9pt,
      axis line style={draw=black!55, line width=0.6pt},
      ytick={0, 0.5, 1.0},
      yticklabels={Low, Medium, High},
      yticklabel style={font=\footnotesize, text=black!75},
      tick style={draw=black!55, line width=0.5pt},
      ytick align=outside,
      ylabel={\normalsize\textbf{Decision Transparency}},
      ylabel style={font=\normalsize, yshift=1pt},
    },
]

%% Quadrant background fills (cool = efficient use of validation; warm = validation-hungry)
\addplot[draw=none, fill=qLL, forget plot]
  coordinates {(0,0)(5,0)(5,5)(0,5)} \closedcycle;
\addplot[draw=none, fill=qUL, forget plot]
  coordinates {(0,5)(5,5)(5,10)(0,10)} \closedcycle;
\addplot[draw=none, fill=qLR, forget plot]
  coordinates {(5,0)(10,0)(10,5)(5,5)} \closedcycle;
\addplot[draw=none, fill=qUR, forget plot]
  coordinates {(5,5)(10,5)(10,10)(5,10)} \closedcycle;

%% Quadrant divides
\addplot[draw=black!25, dashed, line width=0.55pt, forget plot]
  coordinates {(5,0)(5,10)};
\addplot[draw=black!25, dashed, line width=0.55pt, forget plot]
  coordinates {(0,5)(10,5)};

%% ---- Candidate generators (filled circles) -------------------------------
\addplot[only marks, scatter, scatter src=explicit,
         scatter/use mapped color={draw=black!75, fill=mapped color},
         line width=0.8pt, forget plot, mark=*, mark size=4.6pt]
  coordinates {(1.50,1.80) [0.70] (4.40,4.10) [0.60] (9.40,2.90) [0.35] (5.30,6.10) [0.65]};

%% ---- Orchestrators (filled squares) --------------------------------------
\addplot[only marks, scatter, scatter src=explicit,
         scatter/use mapped color={draw=black!75, fill=mapped color},
         line width=0.8pt, forget plot, mark=square*, mark size=4.2pt]
  coordinates {(1.50,7.20) [0.80] (6.30,8.30) [0.25]};

%% ---- Hybrid methods (filled triangle) ------------------------------------
\addplot[only marks, scatter, scatter src=explicit,
         scatter/use mapped color={draw=black!75, fill=mapped color},
         line width=0.8pt, forget plot, mark=triangle*, mark size=5.4pt]
  coordinates {(0.90,8.80) [0.72]};

%% ---- Label boxes: every box vertically centred on, and butted against,
%% ---- its own marker (placement verified collision-free by explicit solve)
\node[algtag, anchor=west] (labAL)   at (axis cs:1.32, 8.80) {%
  {\bfseries Active Learning}\\[1pt]{\scriptsize\itshape adaptive sampling}};

\node[algtag, anchor=west] (labBO)   at (axis cs:1.92, 7.20) {%
  {\bfseries Bayesian Optimization}\\[1pt]{\scriptsize\itshape acquisition-driven}};

\node[algtag, anchor=west] (labRL)   at (axis cs:6.72, 8.30) {%
  {\bfseries Policy Optimization}\\[1pt]{\scriptsize\itshape sequential decisions}};

\node[algtag, anchor=west] (labEvo)  at (axis cs:5.72, 6.10) {%
  {\bfseries Evolutionary Optimization}\\[1pt]{\scriptsize\itshape CMA-ES; derivative-free}};

\node[algtag, anchor=west] (labLang) at (axis cs:4.82, 4.10) {%
  {\bfseries Langevin Sampling}\\[1pt]{\scriptsize\itshape pSGLD; stochastic}};

\node[algtag, anchor=east] (labCFG)  at (axis cs:8.98, 2.90) {%
  {\bfseries Guided Diffusion}\\[1pt]{\scriptsize\itshape CFG; large batches}};

\node[algtag, anchor=west] (labGD)   at (axis cs:1.92, 1.80) {%
  {\bfseries Latent Gradient Descent}\\[1pt]{\scriptsize\itshape gradient-based}};

%% ---- Quadrant descriptors -------------------------------------------------
\node[font=\scriptsize\itshape, text=qULtxt, anchor=north west]
  at (axis cs:0.35,9.80) {Efficient local refinement};
\node[font=\scriptsize\itshape, text=qURtxt, anchor=north west]
  at (axis cs:5.35,9.80) {Efficient global exploration};
\node[font=\scriptsize\itshape, text=qLLtxt, anchor=south west, align=left]
  at (axis cs:0.35,0.25) {Proxy-driven local search;\\validation bottleneck};
\node[font=\scriptsize\itshape, text=qLRtxt, anchor=south west, align=left]
  at (axis cs:5.35,0.25) {Broad generation;\\validation-intensive};

%% ---- Axis endpoint labels -------------------------------------------------
\node[font=\small\itshape, text=black!70, anchor=east]
  at (axis description cs:-0.01,1.0) {High};
\node[font=\small\itshape, text=black!70, anchor=east]
  at (axis description cs:-0.01,0.0) {Low};
\node[font=\small\itshape, text=black!70, anchor=north]
  at (axis description cs:0.10,-0.05) {Local};
\node[font=\small\itshape, text=black!70, anchor=north]
  at (axis description cs:0.96,-0.05) {Global};

%% ---- Class key, above the panel ------------------------------------------
\node[anchor=south, font=\footnotesize, align=left, inner sep=2pt]
  at (axis description cs:0.5, 1.055) {%
  \tikz[baseline=-0.55ex]{\draw[draw=black!75, fill=virc65, line width=0.7pt]
      (0,0) circle (2.6pt);}\;\textbf{Candidate generators}: generate candidate designs\\[1.5pt]
  \tikz[baseline=-0.55ex]{\draw[draw=black!75, fill=virc70, line width=0.7pt]
      (-2.3pt,-2.3pt) rectangle (2.3pt,2.3pt);}\;\textbf{Orchestrators}: select what to validate next\\[1.5pt]
  \tikz[baseline=-0.55ex]{\draw[draw=black!75, fill=virc72, line width=0.7pt]
      (-2.7pt,-2.1pt) -- (2.7pt,-2.1pt) -- (0,2.8pt) -- cycle;}\;\textbf{Hybrid methods}: both generate and select};

\end{axis}
\end{tikzpicture}
\caption{\textbf{\revised{Selection guide for inverse-design search strategies.}} \revised{Each marker places one search strategy with respect to two qualitative axes. The horizontal axis is \emph{search tendency}, running from local refinement of an existing candidate (left) to broad global exploration of the design space (right). The vertical axis is \emph{query efficiency}, meaning the information gained per high-fidelity validation, so that strategies near the top reach a given design target with fewer density-functional-theory calculations or experiments, and those near the bottom rely on generating and screening many candidates. Marker shape encodes the role a strategy plays in the closed loop of Section~\ref{sec:pipeline}: filled circles are \emph{candidate generators}, the proposal engines of Section~\ref{sec:generation}, which propose new designs; filled squares are \emph{orchestrators}, which propose nothing themselves and instead select which existing candidate to validate next and how the budget is spent; and the filled triangle marks a \emph{hybrid} method that does both, since an active-learning loop generates candidates and also chooses which of them to label. Marker color encodes \emph{decision transparency}, that is, the interpretability of the selection mechanism: the extent to which the mechanism that selects a candidate can be inspected and audited, which is low for a policy internalized in network weights and high for an explicit acquisition function. A generator and an orchestrator are complementary and are typically paired, which is the configuration recommended in Section~\ref{sec:search_strategies}. Each label box is placed against its own marker, on whichever side leaves room inside the panel; marker positions carry the information and are not moved to suit the labels. Abbreviations: CFG, classifier-free guidance; pSGLD, preconditioned stochastic gradient Langevin dynamics; CMA-ES, covariance matrix adaptation evolution strategy. Placements are qualitative and indicate typical behavior: the position of any particular implementation depends on the surrogate model, the acquisition strategy, the batch size, and the validation infrastructure available, so the figure is intended to guide method selection and not to rank methods.}}
\label{fig:pipeline_engines_rev}
\end{figure}

Figure~\ref{fig:pipeline_engines_rev} illustrates how various frameworks, including latent optimization, guided diffusion, BO, RL, and active learning, attach to this loop under budget constraints.

\subsection{Selecting Search Strategies for Inverse Design}\label{sec:search_strategies}

\revised{The methods described below are often listed together, although they differ in what they act on, what they require, and how they fail. Three questions organize the rest of this section. The first is \emph{where the method acts}. Latent-space optimization (Section~\ref{sec:latent_opt}) moves a point in a learned representation; guided diffusion (Section~\ref{sec:guided_diffusion}) bends a sampling trajectory; Bayesian optimization (Section~\ref{sec:bayesian_opt}) and active learning (Section~\ref{sec:active_learning}) act on neither, and instead choose which already-proposed candidate to spend a validation call on; reinforcement learning (Section~\ref{sec:rl}) sits in both camps, since a policy can either construct a candidate step by step or fine-tune the sampler that produces it. The second question is \emph{what the method requires}, which may be a differentiable surrogate and a continuous latent space, a conditioning signal, a calibrated posterior, or a reward that can be computed cheaply enough to be queried thousands of times. The third is \emph{what it costs when the requirement is not met}, which is the subject of Section~\ref{sec:sampling_challenges} and is the same in every case. This gives the practical division of labour used throughout the section and drawn in Figure~\ref{fig:pipeline_engines_rev}. \emph{Candidate generators} produce designs; \emph{orchestrators} produce none and instead decide which existing candidate to verify next; and \emph{hybrid} methods do both, active learning being the case in point, since its loop generates candidates and also chooses which of them to label. A working system needs the generating and the selecting function present, whether supplied by two components or by one. With that in place, s}electing an inverse-design search strategy depends on matching the search method to the practical structure of the design problem. In practice, the appropriate approach is dictated by gradient availability, surrogate reliability, requirements for structural diversity, and the computational budget available for high-quality validation within each cycle. These trade-offs are summarized in Figure~\ref{fig:pipeline_engines_rev}, which serves as a regime map. \revised{The horizontal axis represents search tendency, from local refinement to global exploration, and the vertical axis represents query efficiency, defined as the information gained per high-fidelity validation. Together they distinguish methods that reach a target with few expensive evaluations from those that reach it by generating and screening many candidates, which is the distinction that matters once a validation budget is fixed.}

\subsubsection*{\revised{Matching the method to the structure of the problem}}
When a continuous latent representation and a differentiable surrogate are available, latent gradient descent is often the most efficient option for local refinement [\ref{cit:Xiao2023}, \ref{cit:ref2018}]. While it can move rapidly toward a target, it is also highly vulnerable to surrogate error and representation bias. By contrast, Covariance Matrix Adaptation Evolution Strategy (CMA-ES) is better suited to rugged or poorly differentiable objectives, because it performs black-box population search and preserves diversity more naturally [\ref{cit:Hansen2016}]. Langevin-type updates occupy an intermediate regime: Stochastic Gradient Langevin Dynamics (SGLD) and its preconditioned variant pSGLD combine gradient information with controlled stochasticity, enabling both precise optimization and uncertainty-aware exploration of promising regions [\ref{cit:Welling2011}-\ref{cit:Li2016}].

Guided diffusion lies toward the globally exploratory end of the regime map. Classifier-free guidance (CFG) provides a practical mechanism for steering generation toward desired conditions while retaining stochastic sampling diversity [\ref{cit:Ho2022}]. Recent materials models have further shown that diffusion-based generators can produce stable and property-conditioned inorganic crystals at scale [\ref{cit:Zeni2025}]. In inverse design, this positions diffusion as an ideal high-throughput proposal engine: it can generate a vast candidate pool, but those candidates still require filtering, deduplication, and downstream validation.

\subsubsection*{\revised{What the experimental record actually supports}}
The policy-optimization regime becomes most relevant once inverse design is embedded in a sequential workflow. Here, the challenge shifts from mere candidate generation to deciding which experiment or simulation to perform next, in what order, and under which resource constraints. The family of Proximal Policy Optimization (PPO) reinforcement learning offers a standard formulation for such sequential decision problems [\ref{cit:Schulman2017}], and recent materials studies have begun applying RL directly to inverse inorganic design under simultaneous property and synthesis objectives [\ref{cit:Karpovich2024}, \ref{cit:Banik2023}]. Even so, Bayesian optimization and active learning remain the most established selection and orchestration frameworks in materials discovery [\ref{cit:Lookman2019}, \ref{cit:Kusne2020}]. Closed-loop platforms such as CAMEO and A-Lab make clear why: in real-world workflows, the value of an optimizer is determined less by proposal quality than by how efficiently it converts proposals into validated outcomes under experimental constraints [\ref{cit:Szymanski2023}, \ref{cit:Kusne2020}-\ref{cit:Volk2024}]. \revised{The experimental record sharpens this into a verdict. Bayesian active learning is, to date, the only search strategy with repeated closed-loop experimental wins: CAMEO's on-the-fly Bayesian loop located the optimal phase-change composition Ge$_4$Sb$_6$Te$_7$ in a single autonomous synchrotron campaign [\ref{cit:Kusne2020}], and the A-Lab's active-learning-driven platform realized 41 of 58 computationally proposed targets in 17 days of autonomous operation [\ref{cit:Szymanski2023}]. Among generative proposal strategies, guided diffusion is the only one with an experimentally confirmed designed-property material, and so far for a single candidate [\ref{cit:Zeni2025}]; materials RL [\ref{cit:Karpovich2024}, \ref{cit:Banik2023}] and latent-space optimization [\ref{cit:Xie2021}, \ref{cit:Ren2022}, \ref{cit:Xiao2023}, \ref{cit:ref2018}] remain validated only at the DFT level. Under a real experimental budget, the evidence-backed configuration is therefore a Bayesian or active-learning orchestrator allocating validation calls over candidates supplied by a (possibly diffusion-based) proposal engine, with the pure policy-search and latent-optimization strategies of Sections~\ref{sec:latent_opt} and~\ref{sec:rl} treated as promising but experimentally unproven. \revised{Figure~\ref{fig:pipeline_engines_rev} reflects this functional distinction by separating methods by marker shape. Rather than competing as alternative proposal mechanisms, Bayesian optimization and generative models (such as guided diffusion or latent gradient descent) serve complementary roles. An orchestrator like BO evaluates a given candidate pool to optimize the next validation call, meaning it can wrap around any generator. Active learning is depicted as a hybrid framework because its loop inherently couples candidate generation with selection, though this dual role means it inherits the failure modes of both (Section~\ref{sec:active_learning}). Ultimately, designing a closed-loop system relies on pairing an effective generator with an efficient selection framework, and not on choosing between them.}}

\subsection{A Pipeline Blueprint: Proposal, Filtering, Validation, and Update}\label{sec:pipeline}

An inverse-design workflow can be formulated as a constrained, budget-limited, multi-objective search problem. Let $x \in \mathcal{X}$ denote a candidate material, where $\mathcal{X}$ is the design space of allowable compositions and structures. Let $\mathbf{f}(x) = (f_1(x), \ldots, f_K(x))$ denote the vector of target properties or objectives, and let $\mathcal{C} \subseteq \mathcal{X}$ denote the feasible subset defined by physical, chemical, geometric, or processing constraints. The aim is to identify feasible candidates with high utility under a limited high-fidelity evaluation budget $B$:
\begin{equation}
\max_{x \in \mathcal{C}} \; U\!\left(\mathbf{f}(x)\right)
\quad
\text{s.t.}
\quad
\mathrm{cost}\!\left(\mathcal{E}(x)\right) \leq B.
\label{eq:budget_constrained_problem}
\end{equation}

Here, $\mathcal{X}$ denotes the full candidate space, $\mathcal{C}$ denotes the feasible region, $\mathbf{f}(x)$ is the objective vector, $U(\mathbf{f}(x))$ is a scalar utility function that maps multiple objectives to a ranking criterion, $\mathcal{E}(x)$ denotes an expensive high-fidelity evaluation such as density functional theory (DFT) or experiment, and $B$ is the total validation budget.

A practical inverse-design pipeline typically consists of the following stages:

\begin{enumerate}[leftmargin=*]

\item \textbf{Candidate proposal}:
Candidates are first drawn from a proposal mechanism, for example a conditional generative model,
\begin{equation}
x \sim q_{\theta}(x \mid c).
\label{eq:candidate_proposal}
\end{equation}

Here, $q_{\theta}(x \mid c)$ denotes a parameterized proposal distribution with parameters $\theta$, and $c$ represents the design intent, such as target properties, compositional restrictions, symmetry requirements, or text-based conditioning. This stage determines which regions of the design space are explored and therefore strongly influences both diversity and attainable discovery outcomes.

\item \textbf{Low-cost feasibility screening}:
Proposed candidates are then subjected to inexpensive hard constraints to eliminate obviously invalid structures before surrogate evaluation:
\begin{equation}
\mathbb{I}_{\mathcal{C}_{\mathrm{fast}}}(x) \in \{0,1\},
\qquad
x \text{ is retained only if } \mathbb{I}_{\mathcal{C}_{\mathrm{fast}}}(x)=1.
\label{eq:feasibility_screening}
\end{equation}

Here, $\mathbb{I}_{\mathcal{C}_{\mathrm{fast}}}(x)$ is an indicator function, and $\mathcal{C}_{\mathrm{fast}}$ denotes the subset of candidates satisfying fast feasibility checks. These checks may include the absence of atomic overlaps, acceptable interatomic distance ranges, density or volume bounds, charge neutrality, stoichiometric consistency, and other basic invariants required for meaningful downstream assessment.

\item \textbf{Surrogate-based scoring with uncertainty}:
The remaining candidates are evaluated using fast predictive models for objectives and soft constraints. For a multi-objective problem with $K$ objectives, the surrogate returns, for each objective $k=1,\ldots,K$, both a predictive mean and an uncertainty estimate:
\begin{equation}
\bigl(\mu_k(x), \sigma_k(x)\bigr),
\qquad k = 1,\ldots,K.
\label{eq:surrogate_outputs}
\end{equation}

Here, $\mu_k(x)$ denotes the predicted value of the $k$th objective and $\sigma_k(x)$ denotes the associated predictive uncertainty. Candidates can then be ranked using an uncertainty-aware scalarized utility that balances predicted performance, uncertainty-driven exploration, and constraint violation:
\begin{equation}
\hat{U}(x) =
\sum_{k=1}^{K} w_k
\left[
\mu_k(x) + \beta \sigma_k(x)
\right]
-\lambda\,\Phi(x).
\label{eq:surrogate_utility}
\end{equation}

Here, $\hat{U}(x)$ is the surrogate utility, $w_k$ is the weight assigned to the $k$th objective, $\beta$ controls how predictive uncertainty enters the decision rule, $\Phi(x)$ is a penalty term for predicted constraint violations, and $\lambda$ controls the strength of that penalty. A positive $\beta$ gives an exploration-oriented upper-confidence-bound form, favoring candidates that are both promising and uncertain, whereas a negative $\beta$ gives a more conservative risk-averse utility. More generally, $\hat{U}(x)$ may be replaced by an acquisition function $\alpha(x)$, such as expected improvement or constrained expected improvement, to balance exploitation, uncertainty-driven exploration, and feasibility.

\item \textbf{\revised{MLIP-based physics screening}}\revised{: Candidates surviving surrogate triage are relaxed with a universal machine-learned interatomic potential [\ref{cit:Batatia2022}, \ref{cit:Chen2022}-\ref{cit:Batatia2025}], which provides forces, relaxed geometries, energy-above-hull estimates and, for shortlisted candidates, phonon spectra [\ref{cit:Gouvea2026}], at three-to-five orders of magnitude below DFT cost. This tier converts \emph{scores well on a property head} into \emph{behaves like a material under approximate physics,} and is where most physically fragile candidates should be eliminated. Its verdicts must nevertheless be treated as conditionally trustworthy, because the potential and the generator are typically trained on the same databases and are therefore unreliable in the same places. Section~\ref{sec:mlip_oracle} sets out what this tier can and cannot be trusted to decide.}

\item \textbf{Selective high-fidelity validation}:
Only a small subset of candidates is advanced to expensive evaluation:
\begin{equation}
\mathcal{S} \subseteq \{x \in \mathcal{X} : \mathbb{I}_{\mathcal{C}_{\mathrm{fast}}}(x)=1\},
\qquad
|\mathcal{S}| \text{ determined by the available budget}.
\label{eq:selective_validation}
\end{equation}

Here, $\mathcal{S}$ denotes the batch selected for high-fidelity validation, and $|\mathcal{S}|$ is its size. At this stage, candidates are evaluated using rigorous methods to establish ground-truth objective values and feasibility assessments. Such checks limit the damage from surrogate model errors and keep predictive uncertainty meaningful.

\item \textbf{Feedback-driven model and dataset update}:
Validated outcomes are incorporated into the accumulated dataset:
\begin{equation}
\mathcal{D}
\leftarrow
\mathcal{D} \cup
\left\{
\bigl(x, \mathbf{f}(x), \mathrm{feasibility}, \mathrm{metadata}\bigr)
: x \in \mathcal{S}
\right\}.
\label{eq:dataset_update}
\end{equation}

Here, $\mathcal{D}$ denotes the current dataset, $\mathbf{f}(x)$ is the validated objective vector, and the additional terms record feasibility outcomes and auxiliary metadata such as relaxation status, synthesis conditions, provenance, or uncertainty estimates. The proposal model and surrogate predictors are then updated using the expanded dataset, for example through retraining, fine-tuning, recalibration, or policy adaptation.

To reduce collapse toward near-duplicate candidates, inverse-design pipelines often monitor novelty or diversity explicitly, for example through the minimum distance to previously observed structures:
\begin{equation}
\mathrm{Novelty}(x) = \min_{x' \in \mathcal{D}} d(x,x').
\label{eq:novelty}
\end{equation}

Here, $d(x,x')$ denotes how dissimilar two candidates are, from which $\mathrm{Novelty}(x)$ is derived to quantify how distinct the candidate $x$ is relative to the existing dataset. \revised{Continuous formulations of this metric should be preferred to binary match thresholds, and we state that as a recommendation. Negishi et al.\ [\ref{cit:Negishi2025}] build such formulations from the Element Mover's Distance between compositions and the $L_\infty$ distance between average-minimum-distance (AMD) structural fingerprints, reporting \emph{how far} a candidate sits from the nearest known material instead of whether it crossed a tolerance, which removes the dependence of the headline number on an arbitrary cutoff and forms the basis of continuous stable-unique-novel (SUN) metrics for benchmarking generative models [\ref{cit:Negishi2025}]. A binary rate cannot distinguish a genuinely new prototype from a strained or singly substituted copy of a known one; a distance distribution can, and it should be reported as a distribution.} Geographic-style navigation maps of materials space built on local novelty distance further enable real-time duplicate checking against large structural databases, even when candidates are disguised by unit-cell transformations or elemental substitutions [\ref{cit:Widdowson2025}]. Diversity-aware selection based on such measures helps maintain coverage of the design space and improves the long-term efficiency of closed-loop discovery.

\end{enumerate}

\subsection{Latent-Space Sampling Optimization: Fast Steering with Validity Control}\label{sec:latent_opt}

When a generator provides a continuous latent representation $z$, a natural strategy is to optimize $z$ to improve the predicted utility while regularizing toward plausibility. Formally, one solves
\begin{equation}
z^{\ast} \in \arg\max_{z} \; U\!\big(g_{\theta}(z)\big) \quad \text{s.t.} \quad z \in \mathcal{Z}_{\text{valid}}
\label{eq:latent_space_optimization}
\end{equation}
where $g_{\theta}$ is the decoder, $U(\cdot)$ is a utility function, and $\mathcal{Z}_{\text{valid}}$ is a feasibility-regularized region of latent space. In practice, gradient-based methods (Adam, Langevin dynamics) are used to iteratively update the latent variable $z$ within the feasible region $Z_{\mathrm{valid}}$, and the decoded crystal at each step is evaluated by fast property heads or surrogate models. \revised{
In latent-space diffusion models [\ref{cit:Park2026}], a learned diffusion prior constrains the underlying VAE latent space. This mitigates the primary weakness of standard latent optimization—decoding unphysical, off-manifold structures—by actively keeping $z$ within the trained distribution. Furthermore, this architecture naturally facilitates property-guided design, as conditioning is integrated directly into the latent denoiser instead of relying on a post-hoc property head.}

The main strengths of latent-space optimization are speed and compatibility with multi-objective scalarization. Because gradients flow through the decoder, the method can efficiently move toward a target region, and weighted sums can handle multiple objectives simultaneously.
One of the major failure modes is surrogate exploitation: if the property head is imperfect, gradient ascent can move $z$ into regions where the predictor is overconfident but the decoded structure is physically implausible or unstable. Countermeasures include (i) hard feasibility gates applied after decoding, (ii) uncertainty-aware utility terms that penalize overconfident extrapolation, (iii) re-seeding or re-anchoring the latent search using candidates that have passed higher-fidelity relaxation or stability checks, and (iv) diversity-aware selection to prevent collapse into a narrow family of structures [\ref{cit:Xie2021}, \ref{cit:Ren2022}, \ref{cit:Chen2021}].

\subsection{Guided Diffusion Sampling: Controllability Without Losing Diversity}\label{sec:guided_diffusion}

\revised{The term \emph{diffusion model} is too broad for this discussion, and two sub-families should be separated because they behave differently under guidance. In \emph{structure-space} (or coordinate-space) denoising diffusion, the noise is added to the lattice matrix, the fractional coordinates, and the atom types themselves, and the denoiser is an equivariant network acting on the periodic graph [\ref{cit:Jiao2023}, \ref{cit:Zeni2025}, \ref{cit:Park2025}]. In \emph{latent} denoising diffusion, a variational autoencoder first encodes the crystal, and the denoising runs on that latent representation, typically with a diffusion transformer [\ref{cit:Park2026}]. The first keeps every intermediate state interpretable as a structure, which allows hard geometric or space-group constraints to be projected onto the trajectory (Section~\ref{sec:symmetry_axis}); the second gives a smoother, lower-dimensional trajectory with a tractable score, which makes policy-gradient fine-tuning of the sampler practical (Section~\ref{sec:rl}). The guidance discussion below applies to both, and $x_t$ should be read as the state of whichever space the model denoises in.} Diffusion models generate candidates by starting from noise and iteratively denoising it into a structured sample. In the crystal-generation setting, the model learns a reverse process that maps a noisy representation $x_t$ toward a cleaner candidate $x_{t-1}$, eventually producing a crystal proposal $x_0$. In simplified form, this can be written as
\begin{equation}
{\color{revred}
x_T=(\mathbf{L}_T,\mathbf{Z}_T,\mathbf{S}_T) \sim p_{\mathrm{prior}}, \qquad x_{t-1} \sim p_{\theta}(x_{t-1}\mid x_t, c)}
\label{eq:diffusion_sampling}
\end{equation}
where $x_T$ is the fully noised crystal, $p_{\theta}(x_{t-1}\mid x_t, c)$ is the learned reverse transition, and $c$ denotes optional conditioning information such as composition, symmetry, text prompts, or target-property information. \revised{The starting distribution $p_{\mathrm{prior}}$ is defined channel by channel for a periodic crystal: Gaussian on the lattice, wrapped-Gaussian on the fractional coordinates because these live on the torus, and categorical on the species channel.} The main appeal of diffusion models for inverse design is that controllability can be built into the denoising process itself.

\subsubsection*{\revised{Steering the trajectory: sampling-time guidance}}
A key idea is \emph{sampling-time guidance}: instead of drawing samples only from the learned data prior, the reverse trajectory is biased toward candidates that better satisfy the desired condition. This guidance may come from explicit conditioning networks, property-prediction adapters, or classifier-free guidance. In classifier-free guidance, for example, the conditional and unconditional denoisers are combined as
\begin{equation}
\hat{\epsilon}_{\theta}(x_t,c)
=
\epsilon_{\theta}(x_t,\varnothing)
+
w\Big[
\epsilon_{\theta}(x_t,c)-\epsilon_{\theta}(x_t,\varnothing)
\Big]
\label{eq:cfg}
\end{equation}
where $\hat{\epsilon}_{\theta}(x_t,c)$ is the guided noise prediction, $\epsilon_{\theta}(x_t,c)$ and $\epsilon_{\theta}(x_t,\varnothing)$ are the conditional and unconditional denoiser outputs respectively, and $w \geq 0$ is the guidance strength. When $w=0$, sampling follows the unconditional model; as $w$ increases, samples are pushed more strongly toward the requested condition.

\subsubsection*{\revised{Guidance strength as a trade-off curve, not a tuned constant}}
Guided diffusion is appealing for inverse design because it can bias generation toward a target without forcing all samples into the same small region of design space. Unlike direct latent optimization, which can quickly overconcentrate the search around a few high-scoring candidates, diffusion keeps sampling within the model’s learned distribution of plausible crystals throughout the denoising trajectory. Doing so tends to preserve diversity while still enriching for the desired condition. The benefit, however, depends on guidance strength. When guidance is too weak, sampling stays close to the generic prior and the target is reached only occasionally. When it is well calibrated, the sampler produces candidates that are both plausible and relevant. When it is too strong, samples become overly similar, artifacts increase, and the model may exploit errors in the guidance predictor, yielding candidates that score well internally but fail external validation [\ref{cit:Zeni2025}, \ref{cit:Ren2022}-\ref{cit:Song2020}, \ref{cit:DeBreuck2025}, \ref{cit:Karras2022}, \ref{cit:Rombach2022}]. Its value lies in offering a controllable compromise between plausibility, diversity, and target-directed generation, which helps when downstream validation is expensive and candidate batches must be both enriched and reliable. \revised{How strongly to condition is a quantitative design decision. The guidance weight $w$ traces an explicit trade-off curve on which target-hit rate rises with $w$ while validity, stability, and diversity fall, because strong guidance pushes samples off the learned manifold into the regions where the guidance predictor is least reliable [\ref{cit:Zeni2025}, \ref{cit:Ren2022}-\ref{cit:Song2020}, \ref{cit:Karras2022}, \ref{cit:Ho2022}, \ref{cit:Rombach2022}]. \revised{The trade-off is general. It was characterized for classifier-free guidance in image diffusion, where raising the guidance weight improves adherence to the requested condition while reducing sample variety [\ref{cit:Ho2022}], and it transfers to materials generation with an added penalty, because the conditioning signal here is a learned property predictor whose errors grow off the data manifold, so strong guidance buys target adherence partly by steering into the region where the predictor is least trustworthy (Section~\ref{sec:sampling_challenges}).} A model evaluated at a single tuned $w$ reports the peak of a curve, not the behavior a practitioner meets across targets. Reporting practice should therefore include conditioning-strength sweeps (property-hit rate, validity, uniqueness, and post-relaxation stability as functions of $w$), which are cheap relative to a single DFT validation campaign and convert \emph{"our model is controllable"} from a claim into a curve.}

\subsection{Bayesian Optimization: Sample Efficiency Under Expensive Validation}\label{sec:bayesian_opt}

Bayesian optimization (BO) is well suited to inverse-design problems in which each high-fidelity evaluation is expensive, for example when validation requires DFT relaxation, phase-stability calculations, experimental synthesis and characterization. Instead of evaluating large numbers of candidates indiscriminately, BO builds a surrogate model of the objective and uses an acquisition function to decide which candidate should be tested next. If the true objective for a candidate $x$ is denoted by $f(x)$, BO replaces direct global optimization of $f$ with a sequential decision process,
\begin{equation}
f(x) \approx \hat{f}(x), \qquad x_{t+1} = \arg\max_{x \in \mathcal{X}} \alpha_t(x)
\label{eq:bo_main}
\end{equation}
where $\hat{f}(x)$ is the surrogate model and $\alpha_t(x)$ is the acquisition function at iteration $t$. The central idea is to spend a limited validation budget where it is most useful: on candidates that appear promising, informative, or both [\ref{cit:Lookman2019}-\ref{cit:Frazier2018}, \ref{cit:Hase2018}-\ref{cit:Hase2021}, \ref{cit:Diwale2022}].

In BO, the surrogate provides both an expected value and an uncertainty estimate,
\begin{equation}
\hat{f}(x) = \big(\mu(x), \sigma(x)\big)
\label{eq:bo_surrogate}
\end{equation}
where $\mu(x)$ estimates performance and $\sigma(x)$ reflects how uncertain the model is at that point in the search space. Acquisition functions are then constructed from these two terms. For example, the upper confidence bound (UCB),
\begin{equation}
\alpha_{\mathrm{UCB}}(x) = \mu(x) + \kappa \sigma(x)
\label{eq:ucb}
\end{equation}
where $\kappa > 0$ is a parameter controlling the exploration-exploitation trade-off. This acquisition
favors candidates that are predicted to perform well and/or lie in regions where the model remains uncertain. Expected improvement (EI),
\begin{equation}
\alpha_{\mathrm{EI}}(x)
=
\mathbb{E}\!\left[\max\big(f(x)-f^{+},0\big)\right]
\label{eq:ei}
\end{equation}
where $f(x)$ is the objective value at candidate $x$ and $f^{+}$ is the current best observed value.
This criterion prefers candidates that are likely to improve on the best observation obtained so far. In both cases, the next experiment is chosen for what it teaches the search, and not for its expected score alone. \revised{Here $\mu(x)$ and $\sigma(x)$ are the surrogate mean and predictive uncertainty for a physical target evaluated on the crystal tuple of Eq.~\ref{eq:crystal_tuple}, $\kappa$ is the exploration weight, and $f^{+}$ is the best objective value confirmed so far at the declared validation fidelity and not the best surrogate score seen (Section~\ref{sec:sampling_challenges}).}

This weighs heavily in inverse design, where promising candidates must also remain feasible. If feasibility is described by a constraint $g(x)\leq 0$, the acquisition can be modified as
\begin{equation}
{\color{revred}
\alpha_{\mathrm{cons}}(x)=\alpha(x)\,
\underbrace{P\!\big(g(x)\leq 0\big)}_{\text{chemical feasibility}}\,
\underbrace{P_{\mathrm{relax}}(x)}_{\text{survives MLIP/DFT relaxation}}}
\label{eq:constrained_acquisition}
\end{equation}
where $\alpha(x)$ is the base acquisition function (e.g., UCB or EI), $g(x) \leq 0$ encodes a feasibility constraint, and $P(g(x) \leq 0)$ is the predicted probability of constraint satisfaction. \revised{The first factor is the probability that the chemical feasibility constraint $g(x)\leq 0$ holds; the second is materials-specific, since a candidate can be chemically admissible yet collapse once forces are minimized, so the acquisition weights it by $P_{\mathrm{relax}}(x)$, its estimated probability of surviving the MLIP or DFT relaxation stage of Section~\ref{sec:pipeline}.}
This prevents the search from repeatedly selecting candidates that look attractive according to the surrogate objective but are unlikely to be structurally valid, chemically plausible, or experimentally accessible.

\subsubsection*{\revised{Why BO needs a compact, chemically meaningful search space}}
However, the success of BO in crystal design depends strongly on representation. The search space for materials is not usually a simple continuous domain; it is typically high-dimensional, structured, and mixed in type, combining discrete variables such as composition or space group with continuous variables such as lattice parameters or latent coordinates. Standard BO methods degrade in such spaces, because a Gaussian-process surrogate with a stationary kernel requires a metric under which nearby points behave similarly, and acquisition optimization itself becomes hard as dimensionality grows [\ref{cit:Shahriari2016}-\ref{cit:Frazier2018}]. \revised{For this reason, BO is usually most effective when the search is carried out in a compact and chemically meaningful embedding,}
\begin{equation}
z = \phi(x), \qquad z \in \mathbb{R}^{d}, \quad d \ll \dim(x)
\label{eq:embedding}
\end{equation}
where $\phi: \mathcal{X} \to \mathbb{R}^d$ is a learned embedding function, $z$ is the compact low-dimensional representation, and $d \ll \dim(x)$ indicates the embedding is substantially lower-dimensional than the original crystal representation.
BO instead operates over $z$, and the selected points are decoded or mapped back into material candidates. \revised{The strategy was established for molecules by Gómez-Bombarelli et al., who ran Gaussian-process BO in the continuous latent space of a variational autoencoder and decoded the selected latent points back into candidate compounds [\ref{cit:ref2018}]; the crystalline analogue is BO or gradient ascent over the latent space of a crystal generator, which is how latent-variable models enter inverse design at all (Section~\ref{sec:latent_opt}) [\ref{cit:Xie2021}, \ref{cit:Ren2022}, \ref{cit:Xiao2023}]. Two further design choices matter as much as the embedding. Categorical and mixed variables (element identity, space group) are handled by kernels or samplers built for them, as in Phoenics and Gryffin [\ref{cit:Hase2018}-\ref{cit:Hase2021}]. Where measurements are noisy or unreliable, as they are in most experimental campaigns, robust and noise-aware BO variants are needed to keep the surrogate from chasing measurement error [\ref{cit:Diwale2022}].}

Another issue is that materials data are often noisy, sparse, and method-dependent. Experimental measurements can vary across conditions, computed properties depend on the level of theory, and many targets are observed only for a small subset of materials. A simple observation model,
\begin{equation}
y = f(x) + \varepsilon, \qquad \varepsilon \sim \mathcal{N}(0,\sigma_n^2)
\label{eq:observation_model}
\end{equation}
where $y$ is the observed (noisy) measurement, $f(x)$ is the true objective value, and $\varepsilon$ is Gaussian noise with variance $\sigma_n^2$,
is useful as a starting point, but real uncertainty in materials discovery is often heteroscedastic, biased, or source-dependent. In this setting, uncertainty is not merely a statistical convenience; it is part of the physical problem. Reliable BO therefore depends on well-calibrated surrogates and realistic treatment of noise, because poor uncertainty estimates lead directly to poor acquisition decisions [\ref{cit:Shahriari2016}-\ref{cit:Snoek2012}, \ref{cit:Diwale2022}].

\subsubsection*{\revised{Target-oriented BO: hitting a property window, not a maximum}}
A closely related extension is target-oriented BO, including approaches such as t-EGO. Here the objective is to identify candidates most likely to fall within a desired property window, which is a different target from the maximum of a scalar score. If the design goal is a target region $\mathcal{T}$, the acquisition can be written as
\begin{equation}
x_{t+1}=\arg\max_{x \in \mathcal{X}} P\!\big(f(x)\in \mathcal{T}\big)
\label{eq:target_oriented_bo}
\end{equation}
where $\mathcal{T}$ is the desired property window, $P(f(x) \in \mathcal{T})$ is the predicted probability that candidate $x$ falls within the target region, and $\mathcal{X}$ is the design space.
This can optionally be combined with feasibility terms. That framing sits closer to practical inverse design, where the goal is a material meeting a specified set of performance criteria while remaining feasible, which does not usually coincide with the absolute maximum of any one property [\ref{cit:Shahriari2016}, \ref{cit:Chitturi2024}, \ref{cit:Tian2025}]. \revised{Two studies illustrate this approach. Tian et al.\ formulate t-EGO by replacing expected improvement with an acquisition that scores the probability of landing inside a user-specified property window, and report faster identification of on-target candidates than EGO, constrained EGO, and multi-objective acquisition baselines [\ref{cit:Tian2025}]. Chitturi et al.\ generalize the idea from a window to an arbitrary target \emph{subset}: the user writes a short filtering algorithm that would return the desired subset if the property map were known, and this is converted automatically into an acquisition function (InfoBAX, MeanBAX, or SwitchBAX) [\ref{cit:Chitturi2024}]. Their examples are exactly the goals that maximization cannot express, including a library of monodisperse TiO$_2$ nanoparticles at several specified radii with polydispersity below 5\%, and Fe-Co-Ni compositions whose Kerr rotation and coercivity both fall in prescribed bands; the goal-aware acquisitions collected more on-target samples than random sampling, uncertainty sampling, and expected hypervolume improvement [\ref{cit:Chitturi2024}].}

\subsection{Reinforcement Learning: Policy Search with Explicit Constraints}\label{sec:rl}

Reinforcement learning (RL) is best suited to design problems that are genuinely sequential. Instead of directly predicting a final material candidate in one step, RL learns a policy that takes actions in a design environment so as to maximize a cumulative reward. In materials settings, these actions may correspond to adding or replacing atoms, modifying composition, changing structural motifs, or navigating a latent design space. In simplified form, the policy seeks to maximize
\begin{equation}
{\color{revred}
J(\pi)=\mathbb{E}_{\pi}\!\left[\sum_{t=0}^{T}\gamma^{t} r_t\right],
\qquad
s_t=\big(\mathbf{L}_t,\mathbf{Z}_t,\mathbf{S}_t\big)^{\mathrm{partial}} \in \mathcal{C}}
\label{eq:rl_policy}
\end{equation}
where $\pi$ is the policy, $r_t$ is the reward at step $t$, $\gamma \in (0,1]$ is a discount factor, and $T$ is the episode length. \revised{The state $s_t$ is a partially built crystal, written in the variables of Eq.~\ref{eq:crystal_tuple}, the actions are chemical edits, and $\mathcal{C}$ is the feasible set of charge-balanced, symmetry-consistent, non-overlapping structures; requiring $s_t \in \mathcal{C}$ at every step, and not only at termination, keeps the whole trajectory inside it (Eq.~\ref{eq:constrained_rl}).} The conceptual appeal of RL is that it can naturally encode sequential design decisions and can bias exploration toward rare but potentially high-value regions of candidate space [\ref{cit:Karpovich2024}].

It suits inverse design cast as a multi-step construction process. For example, one may wish to build a structure step by step, refine a candidate through successive modifications, or choose among actions that trade off immediate improvement against longer-term feasibility. In such cases, RL provides a natural language for design: the \emph{state} describes the current candidate, the \emph{action} modifies it, and the \emph{reward} measures whether the modification moves the search toward the desired objective.

A simple constrained RL view can be written as
\begin{equation}
\max_{\pi} \; \mathbb{E}_{\pi}\!\left[\sum_{t=0}^{T}\gamma^{t} r_t\right]
\qquad \text{s.t.} \qquad
x_t \in \mathcal{C} \;\; \text{for all } t
\label{eq:constrained_rl}
\end{equation}
where $\mathcal{C}$ denotes feasibility constraints such as composition rules, structural validity, charge balance, symmetry requirements, or chemistry plausibility. In practice, the reward may combine several terms,
\begin{equation}
r(x)=\sum_{k=1}^{K} w_k f_k(x)-\lambda\,\Phi(x)
\label{eq:rl_reward}
\end{equation}
where $f_k(x)$ are desired properties, $w_k$ are their weights, $\lambda > 0$ controls the infeasibility penalty strength, and $\Phi(x)$ penalizes infeasibility or undesired characteristics.

\subsubsection*{\revised{Reward hacking: maximizing the stand-in instead of the objective}}
The main difficulty is that the reward function is seldom the true scientific objective itself. More often, it is only a computable approximation based on fast predictors, surrogate models, or simplified heuristics. The central failure mode is therefore not mysterious: the policy may learn to maximize the \emph{reward signal} without achieving the \emph{true design goal}. The behavior is commonly called \emph{reward hacking}. In the present context, phrases such as \emph{proxy exploitation} or \emph{surrogate gaming} refer to essentially the same issue: optimizing the stand-in objective instead of the real physical target.

The problem bites hardest in inverse design, because many rewards are based on approximate property predictors, with no direct physics-level validation behind them. For example, an RL agent may discover candidates that receive very high predicted reward because they exploit regions where the surrogate is overconfident or inaccurate. Such candidates may look optimal to the policy, yet fail once they are relaxed, screened for stability, or evaluated experimentally. The danger extends past poor generalization to systematic drift toward scientifically meaningless optima. \revised{The published materials-RL work makes the countermeasures visible. Karpovich et al.\ define a multi-objective reward over predicted band gap, formation energy, and bulk and shear moduli \emph{together with} synthesis objectives (low calcination and sintering temperatures), and report that the policy recovers chemical rules such as negative formation energy, charge neutrality, and electronegativity balance while retaining compositional diversity and uniqueness [\ref{cit:Karpovich2024}]. Two features of that design act as the safeguards in practice. The reward is multi-term, so no single surrogate can be maximized in isolation, and diversity and uniqueness are tracked alongside reward so that a collapsed policy is visible in the reported numbers. Because the agent works in composition space, structures for the proposed compositions must still be supplied by template-based structure prediction and checked afterwards [\ref{cit:Karpovich2024}], which is the physics-level validation step that the reward itself does not provide.}

\subsubsection*{\revised{Safeguards, and the conditions under which RL earns its cost}}
For this reason, discovery-oriented RL pipelines must control the action space and the reward definition carefully. One important safeguard is to impose \emph{constrained action spaces}, so that the policy is prevented from proposing obviously invalid moves. A second safeguard is to use \emph{robust reward design}, in which hard constraints are separated from soft objectives and never left to reward shaping alone. A third safeguard is to include \emph{periodic physics-level validation}, so that the policy is repeatedly checked against higher-fidelity calculations or experiments, and therefore no surrogate is optimized indefinitely.

In practical terms, RL becomes most valuable when three conditions hold. First, the design process is genuinely sequential, so that a policy has an advantage over one-shot generation. Second, the action space can be restricted enough that many invalid candidates are excluded by construction. Third, reward signals are recalibrated against higher-fidelity evaluation often enough to prevent drift toward proxy-driven solutions. Under these conditions, RL can be a powerful search strategy for inverse design. Without them, it can easily become an expensive way of optimizing the wrong objective [\ref{cit:Karpovich2024}]. \revised{A second mode of use has emerged that sits closer to generation than to search. Instead of building a structure action by action, the policy is the generative sampler itself, and RL fine-tunes it. Chemeleon-2 applies group relative policy optimization to a latent denoising diffusion model with verifiable multi-objective rewards that trade creativity, stability, and diversity against one another, and reports that this reaches under-represented regions of materials space that likelihood-based sampling does not, together with property-guided design that preserves chemical validity [\ref{cit:Park2026}]. Two features of that report bear on the failure modes above. The reward is \emph{verifiable}, computed from declared criteria and not from an opaque learned score, which narrows the room for reward hacking; and diversity enters the reward explicitly, because policy-gradient fine-tuning of a generator otherwise collapses its output distribution, an effect reported across the RL-on-diffusion crystal work [\ref{cit:Park2026}]. The approach also depends on the latent formulation of Section~\ref{sec:latent_opt}, since policy-gradient methods need access to the sampler's score.}

\subsection{Active Learning with Generative Loops: Closing the Loop Responsibly}\label{sec:active_learning}

Active learning (AL) operationalizes feedback in inverse design by repeatedly cycling through candidate generation, screening, selective validation, and model update. At each iteration, the loop proceeds through four stages in sequence: generating new candidate proposals, screening and diversifying the candidate set, validating selected candidates at high fidelity, and updating the model with the resulting outcomes. Formally, the accumulated dataset is updated as
\begin{equation}
\mathcal{D}_{t+1} = \mathcal{D}_t \cup \{(x, y(x)) : x \in \mathcal{S}_t\}
\label{eq:al_dataset_update}
\end{equation}
where $\mathcal{D}_t$ is the labeled dataset at iteration $t$, $\mathcal{S}_t$ is the selected batch for expensive validation, and $y(x)$ denotes the newly obtained high-fidelity outcome (for example, relaxed energy, stability, or experimental measurement). Active learning is one of the most realistic ways to make inverse design cumulative, because expensive labels are allocated where they are expected to reduce uncertainty, improve the proposal model, or sharpen the decision boundary most effectively [\ref{cit:Lookman2019}, \ref{cit:Settles2009}-\ref{cit:Hase2021}, \ref{cit:Han2025}].

\subsubsection*{\revised{Two failure modes of letting a model choose its own training data}}
The appeal of active learning is therefore straightforward: instead of spending validation budget uniformly across candidate space, the loop uses the current model to decide where additional evidence will be most useful. In principle, this can accelerate discovery substantially when high-fidelity labels are scarce and expensive. In practice, however, active learning in inverse design is not automatically beneficial. Once the model begins selecting its own future training data, the loop can become progressively biased in ways that improve short-term hit rates while degrading long-term discovery value. Two failure modes dominate.

\paragraph{Selection bias:}
As the loop progresses, the labeled dataset no longer reflects the broader candidate space; instead, it becomes increasingly concentrated around the regions preferred by the current model or acquisition rule. In other words, the training distribution gradually shifts toward what the policy already believes to be promising. If the policy is miscalibrated, this can reinforce early mistakes and produce a self-confirming search trajectory. Formally, if the original candidate distribution is $p(x)$ but the active learner selects according to some acquisition rule $\alpha_t(x)$, the labeled data are effectively drawn from a biased distribution
\begin{equation}
q_t(x) \propto p(x)\,\alpha_t(x)
\label{eq:biased_distribution}
\end{equation}
where $p(x)$ is the original candidate distribution over the design space, $\alpha_t(x)$ is the acquisition rule at iteration $t$, and $q_t(x)$ is the resulting biased sampling distribution induced by the active learner.
This means that the model is increasingly trained on its own preferences, and progressively less on the broader design space. Such feedback can be useful when the policy is well calibrated, but dangerous when surrogate errors or blind spots are present, because it may hide valuable regions that were never sampled in the first place. \revised{The active-learning literature has long noted that a learner which selects its own training data produces a labeled set that is not a random sample of the input distribution, so the usual guarantees no longer apply and a miscalibrated learner can reinforce its own early errors [\ref{cit:Settles2009}]. The materials-specific version is set out by Lookman et al., whose central point is that adaptive design succeeds or fails on the quality of the uncertainty estimate used to trade exploration against exploitation, since it is the uncertainty that decides which unexplored regions are ever visited [\ref{cit:Lookman2019}]. Closed-loop platforms show the same dependence in operation: CAMEO's Bayesian active-learning loop reaches the target composition quickly precisely because its acquisition is driven by uncertainty in the phase map as well as by property value [\ref{cit:Kusne2020}].}

\paragraph{Diversity collapse:}
A second risk is that repeated fine-tuning narrows the generator toward a small mode of candidate space. Once a few regions begin to look promising, the loop may keep returning to near-duplicates of those candidates, gradually reducing structural diversity, compositional novelty, and exploratory coverage. This can happen even if the measured hit rate improves, because the model is effectively learning to exploit a narrow local basin while the discovery frontier stays where it was. In generative inverse design, this problem runs deeper, because the proposal model itself is updated by the validation outcomes. \revised{Retraining a generator on data it selected and largely produced is a recognized failure mode outside materials science as well: repeated training on model-generated data narrows the learned distribution and erases its tails, so the low-density regions disappear first [\ref{cit:Shumailov2024}]. In inverse design the tails are the point, since that is where unfamiliar chemistry lives.} Without explicit novelty or coverage criteria, the loop can converge toward a restricted family of candidates that are easy for the model to optimize but scientifically uninteresting. \revised{The reported countermeasure in materials loops is to make coverage part of the acquisition and part of the report: InvDesFlow-AL keeps a generate-screen-label-fine-tune cycle honest by applying novelty thresholds and diversity filters at the screening stage [\ref{cit:Han2025}], and continuous uniqueness and novelty metrics make the resulting spread measurable instead of anecdotal [\ref{cit:Negishi2025}, \ref{cit:Widdowson2025}].}

These two problems are closely related: selection bias concentrates the data, and diversity collapse narrows the proposal distribution. Together, they create a misleading picture of progress. A loop may appear to improve because the fraction of candidates that satisfy the current target increases, yet the search may simultaneously become less representative, less novel, and less likely to discover truly new materials. For this reason, reporting only an improved hit rate is not sufficient.

\subsubsection*{\revised{What a credible active-learning report must contain}}
A stronger evaluation of active-learning inverse design should report how often the loop finds target-matching candidates and, just as importantly, how that performance was achieved. At minimum, reporting should include: (i) pass rates under consistent validation criteria, (ii) diversity or uniqueness of the retained candidates, and (iii) novelty with respect to the training or known reference set\revised{, quantified by the minimum distance to that set (Eq.~\ref{eq:novelty})}. \revised{Diversity should be summarized through pairwise distances or coverage statistics computed over the batches $\mathcal{S}_t$ accumulated across rounds, since a single batch is too small a sample to characterize the coverage of a loop,} and not by reporting only the number of successful candidates.

In practical terms, responsible active learning in inverse design therefore requires more than iterative retraining. It requires acquisition strategies that balance utility with coverage, proposal mechanisms that do not collapse too quickly, and evaluation protocols that distinguish genuine discovery from repeated exploitation of a narrow region. \revised{Two distinct things are needed here and they are easily conflated. Diversity can act as a \emph{constraint on acquisition}, which changes what the loop does, since diversity-aware batch selection, novelty thresholds, and explicit exploration terms in the acquisition rule keep the loop probing underexplored regions while it exploits current successes. Diversity can also act as a \emph{reported quantity}, which changes what a reader can check, and here the instruments are coverage statistics over chemistry and structure together with distance-to-nearest-known distributions over the candidates the loop retained. The two are not substitutes. A loop can apply a diversity-aware selection rule and still report nothing about the diversity it achieved, and a loop can report coverage faithfully while acquiring greedily. The continuous stable-unique-novel metrics of Section~\ref{sec:ledger} largely cover the second role, since they already express uniqueness and novelty as distances instead of binary counts and can be aggregated over a set [\ref{cit:Negishi2025}]; what they do not do is enter the acquisition rule, which remains a separate design decision.}

The broader lesson is that active learning is not simply an efficiency trick for labeling fewer points. In inverse design, it is a decision-making layer that shapes what the model will be allowed to learn next. Used carefully, it can make discovery increasingly data-efficient and targeted. Used carelessly, it can amplify the current model’s blind spots and create a closed loop around a narrow and potentially misleading notion of success [\ref{cit:Lookman2019}, \ref{cit:Settles2009}-\ref{cit:Hase2021}, \ref{cit:Han2025}].

\subsection{Challenges in Sampling Optimization}\label{sec:sampling_challenges}

\revised{Sections~\ref{sec:latent_opt}-\ref{sec:active_learning} described five search strategies with different mechanics. They share a single failure mechanism, and it is worth stating once in general terms instead of five times in particular ones.

Every strategy in this section optimizes against a learned stand-in for physics. The stand-in occupies a different slot in each case, being a differentiable property head in latent-space optimization, a guidance predictor in guided diffusion, a probabilistic surrogate in Bayesian optimization, a reward model in reinforcement learning, and whichever model supplies the acquisition score in an active-learning loop. In every case the optimizer is rewarded for finding inputs on which the stand-in returns a high value, and it has no way to distinguish inputs on which the stand-in is right from inputs on which it is wrong. Optimization pressure therefore searches, as a matter of course, for the regions where the stand-in departs from the physics it approximates.

Three properties of materials surrogates make that search unusually productive. First, they are trained on curated computational databases dominated by near-equilibrium, already-known chemistry, so their error grows on the unusual candidates that aggressive generation produces. Second, their uncertainty is miscalibrated in the same region, and typically in the optimistic direction, so the natural defense of declining low-confidence candidates degrades exactly where it is needed [\ref{cit:Ovadia2019}-\ref{cit:Guo2017}]. Third, and least often acknowledged, the reference against which a candidate is judged degrades alongside the model. A convex hull is only as good as the set of competing phases entered into it, and that set is sparsest in under-explored chemistries, so a novel candidate is scored against the weakest available baseline while being credited with the largest apparent margin.

These three failures are correlated and not independent, which makes the compound effect dangerous. Proposal model, surrogate, and screening oracle are typically trained on overlapping snapshots of the same databases, so their blind spots coincide. A candidate can therefore clear surrogate scoring, physical prescreening, and a hull comparison in sequence while being wrong, and no single in-distribution metric will reveal it, because each stage agrees with the others for the same reason it is mistaken.

The strategy-specific consequences follow from this one mechanism, and the countermeasures are correspondingly similar. Latent-space optimization drifts off the decodable manifold, so it needs validity priors, hard gates after decoding, and periodic re-anchoring on candidates that survived higher-fidelity checks (Section~\ref{sec:latent_opt}). Guided diffusion loses diversity and amplifies predictor bias as the guidance weight rises, so the weight should be swept and reported instead of tuned once (Section~\ref{sec:guided_diffusion}). Bayesian optimization inherits the calibration problem directly into its acquisition rule, since a miscalibrated posterior misallocates the very budget the method exists to allocate (Section~\ref{sec:bayesian_opt}). Reinforcement learning makes the substitution explicit as a reward, which is why verifiable multi-objective rewards and constrained action spaces matter more there than elsewhere (Section~\ref{sec:rl}). Active learning compounds the problem across rounds, because the loop chooses the data that will train its own next surrogate (Section~\ref{sec:active_learning}). The common requirement is that the loop must at intervals consult something that was not fit to the same data, which in practice means periodic high-fidelity validation distributed across the accepted set and not concentrated on its top scorers. The machine-learned-interatomic-potential tier deserves separate treatment on exactly these grounds, and Section~\ref{sec:mlip_oracle} gives it, because a screening oracle drawn from the same databases as the generator it judges is the sharpest instance of the correlated-blind-spot problem described here.}

\section{From Failure Modes to Reliable Deployment}\label{sec:failure_modes}

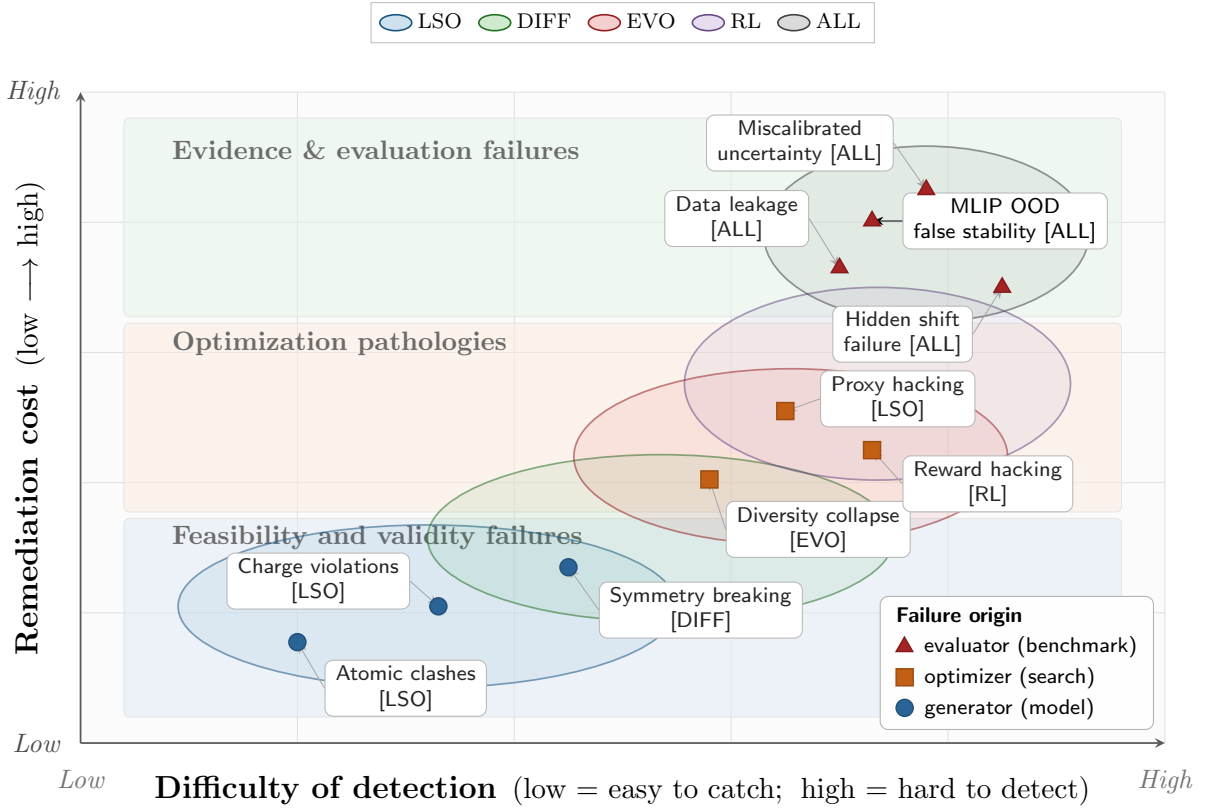
\begin{figure}[t]
\centering
\definecolor{markNeutral}{RGB}{70,70,70}
\definecolor{markGen} {RGB}{42,100,150}   %% generator: steel blue
\definecolor{markOpt} {RGB}{195,95,20}    %% optimizer: burnt orange
\definecolor{markEval}{RGB}{165,35,35}    %% evaluation: crimson
\renewcommand{\AlgoCloud}[7]{%
  \path[draw=#5!70!black, draw opacity=0.55, fill=#5, fill opacity=#7, line width=0.65pt]
    (axis cs:#1,#2) ellipse [x radius=#3, y radius=#4];}
\begin{tikzpicture}[font=\sffamily]
\pgfplotsset{
  genpt/.style ={only marks, mark=*,          mark size=3.2pt, forget plot,
                 draw=markGen!80!black,  fill=markGen,  line width=0.55pt},
  optpt/.style ={only marks, mark=square*,    mark size=3.2pt, forget plot,
                 draw=markOpt!80!black,  fill=markOpt,  line width=0.55pt},
  evalpt/.style={only marks, mark=triangle*,  mark size=3.6pt, forget plot,
                 draw=markEval!80!black, fill=markEval, line width=0.55pt},
}
\tikzset{
  callout/.style={
    fill=white, draw=black!28, rounded corners=3pt,
    inner xsep=4pt, inner ysep=2pt,
    font=\scriptsize\sffamily, text=black!86, align=center},
  leader/.style={draw=black!35, line width=0.35pt, -{Stealth[length=1.5mm,width=1.1mm]}}
}
\begin{axis}[
    width=\linewidth,
    height=0.64\linewidth,
    xmin=0, xmax=10,
    ymin=0, ymax=10,
    axis lines=left,
    axis line style={line width=0.65pt, color=black!65},
    tick style={draw=none},
    grid=both,
    grid style={line width=0.18pt, draw=black!8},
    major grid style={line width=0.28pt, draw=gray!25},
    xtick={0,2,4,6,8,10},
    ytick={0,2,4,6,8,10},
    xticklabels={,,,,,},
    yticklabels={,,,,,},
    xlabel={\large\textbf{Difficulty of detection}\enspace%
            \normalsize(low = easy to catch;\; high = hard to detect)},
    ylabel={\large\textbf{Remediation cost}\enspace%
            \normalsize(low $\longrightarrow$ high)},
    xlabel style={font=\normalsize, align=center, yshift=1pt},
    ylabel style={font=\normalsize, align=center},
    axis background/.style={fill=gray!3},
    clip=false,
    legend style={
      font=\scriptsize,
      draw=black!25,
      fill=white, fill opacity=0.97, text opacity=1,
      at={(0.5,1.075)}, anchor=south,
      /tikz/every even column/.append style={column sep=4pt}
    },
    legend columns=6,
    legend cell align=left,
]

\path[draw=black!10, fill=bandFeas, fill opacity=0.56, rounded corners=2pt]
  (axis cs:0.4,0.4) rectangle (axis cs:9.6,3.45);
\path[draw=black!10, fill=bandOpt,  fill opacity=0.56, rounded corners=2pt]
  (axis cs:0.4,3.55) rectangle (axis cs:9.6,6.45);
\path[draw=black!10, fill=bandEval, fill opacity=0.56, rounded corners=2pt]
  (axis cs:0.4,6.55) rectangle (axis cs:9.6,9.6);

\node[anchor=west, font=\small\bfseries, text=black!58]
  at (axis cs:0.75,3.16) {Feasibility and validity failures};
\node[anchor=west, font=\small\bfseries, text=black!58]
  at (axis cs:0.75,6.14) {Optimization pathologies};
\node[anchor=west, font=\small\bfseries, text=black!58]
  at (axis cs:0.75,9.12) {Evidence \& evaluation failures};

\node[font=\small\itshape, text=black!70, anchor=east]
  at (axis description cs:-0.01,1.0) {High};
\node[font=\small\itshape, text=black!70, anchor=east]
  at (axis description cs:-0.01,0.0) {Low};
\node[font=\small\itshape, text=black!55, anchor=north]
  at (axis description cs:0.0,-0.025) {Low};
\node[font=\small\itshape, text=black!55, anchor=north]
  at (axis description cs:1.0,-0.025) {High};

\AlgoCloud{3.2}{2.10}{2.30}{1.25}{colGD}  {LSO} {0.08}
\AlgoCloud{5.35}{3.15}{2.15}{1.28}{colDIFF}{DIFF}{0.08}
\AlgoCloud{6.55}{4.40}{2.00}{1.35}{colEVO} {EVO} {0.08}
\AlgoCloud{7.35}{5.52}{1.78}{1.48}{colRL}  {RL}  {0.08}
\AlgoCloud{7.80}{7.82}{1.50}{1.35}{colALL} {ALL} {0.08}

\addplot[genpt] coordinates {(2.0,1.55)};
\node[callout, anchor=north west] (atomic) at (axis cs:2.25,1.28) {Atomic clashes\\{\scriptsize [LSO]}};
\draw[leader] (atomic.west) -- (axis cs:2.0,1.55);

\addplot[genpt] coordinates {(3.3,2.10)};
\node[callout, anchor=south east] (charge) at (axis cs:3.04,2.05) {Charge violations\\{\scriptsize [LSO]}};
\draw[leader] (charge.east) -- (axis cs:3.3,2.10);

\addplot[genpt] coordinates {(4.5,2.70)};
\node[callout, anchor=north west] (symm) at (axis cs:4.78,2.48) {Symmetry breaking\\{\scriptsize [DIFF]}};
\draw[leader] (symm.west) -- (axis cs:4.5,2.70);

\addplot[optpt] coordinates {(5.8,4.05)};
\node[callout, anchor=north west] (diversity) at (axis cs:5.95,3.72) {Diversity collapse\\{\scriptsize [EVO]}};
\draw[leader] (diversity.west) -- (axis cs:5.8,4.05);

\addplot[optpt] coordinates {(6.5,5.10)};
\node[callout, anchor=south west] (proxy) at (axis cs:6.82,4.85) {Proxy hacking\\{\scriptsize [LSO]}};
\draw[leader] (proxy.west) -- (axis cs:6.5,5.10);

\addplot[optpt] coordinates {(7.3,4.50)};
\node[callout, anchor=north west] (reward) at (axis cs:7.58,4.42) {Reward hacking\\{\scriptsize [RL]}};
\draw[leader] (reward.west) -- (axis cs:7.3,4.50);

\addplot[evalpt] coordinates {(7.0,7.30)};
\node[callout, anchor=south east] (leakage) at (axis cs:6.72,7.62) {Data leakage\\{\scriptsize [ALL]}};
\draw[leader] (leakage.east) -- (axis cs:7.0,7.30);

\addplot[evalpt] coordinates {(7.30,8.02)};
\node[callout, text=revred, anchor=west] (mlipood) at (axis cs:7.58,8.02) {MLIP OOD\\false stability [ALL]};
\draw[leader, revred] (mlipood.west) -- (axis cs:7.30,8.02);

\addplot[evalpt] coordinates {(7.8,8.50)};
\node[callout, anchor=south east] (uq) at (axis cs:7.50,8.78) {Miscalibrated\\uncertainty [ALL]};
\draw[leader] (uq.east) -- (axis cs:7.8,8.50);

\addplot[evalpt] coordinates {(8.5,7.00)};
\node[callout, anchor=north east] (shift) at (axis cs:8.23,6.72) {Hidden shift\\failure [ALL]};
\draw[leader] (shift.east) -- (axis cs:8.5,7.00);

\node[draw=black!30, fill=white, fill opacity=0.97, text opacity=1,
      rounded corners=3pt, inner xsep=6pt, inner ysep=4pt,
      font=\scriptsize\sffamily, anchor=south east, align=left]
  at (axis description cs:0.99,0.02)
  {\textbf{Failure origin}\\[2pt]
 \tikz[baseline=-0.6ex]{\draw[draw=markEval!80!black, fill=markEval, line width=0.55pt]
   plot[mark=triangle*, mark size=3.6pt] coordinates {(0,0)};}\enspace evaluator (benchmark)\\[2pt]
 \tikz[baseline=-0.6ex]{\draw[draw=markOpt!80!black, fill=markOpt, line width=0.55pt]
   plot[mark=square*, mark size=3.2pt] coordinates {(0,0)};}\enspace optimizer (search)\\[2pt]
 \tikz[baseline=-0.6ex]{\draw[draw=markGen!80!black, fill=markGen, line width=0.55pt]
   plot[mark=*, mark size=3.2pt] coordinates {(0,0)};}\enspace generator (model)};

\addlegendimage{legend image code/.code={\draw[draw=colGD!70!black,  fill=colGD,  fill opacity=0.22, line width=0.6pt](0,0) ellipse [x radius=6pt, y radius=3pt];}}\addlegendentry{LSO}
\addlegendimage{legend image code/.code={\draw[draw=colDIFF!70!black,fill=colDIFF,fill opacity=0.22, line width=0.6pt](0,0) ellipse [x radius=6pt, y radius=3pt];}}\addlegendentry{DIFF}
\addlegendimage{legend image code/.code={\draw[draw=colEVO!70!black, fill=colEVO, fill opacity=0.22, line width=0.6pt](0,0) ellipse [x radius=6pt, y radius=3pt];}}\addlegendentry{EVO}
\addlegendimage{legend image code/.code={\draw[draw=colRL!70!black,  fill=colRL,  fill opacity=0.22, line width=0.6pt](0,0) ellipse [x radius=6pt, y radius=3pt];}}\addlegendentry{RL}
\addlegendimage{legend image code/.code={\draw[draw=colALL!70!black, fill=colALL, fill opacity=0.22, line width=0.6pt](0,0) ellipse [x radius=6pt, y radius=3pt];}}\addlegendentry{ALL}

\end{axis}
\end{tikzpicture}
\caption{\textbf{Failure-mode map for crystal inverse-design systems with algorithm-exposure regions.}
Each labeled callout box marks a representative failure mode positioned according to two qualitative axes.
The \textit{x}-axis encodes \textit{difficulty of detection}, increasing from low difficulty/easy-to-catch failures on the left, such as gross atomic overlaps, to high difficulty/hard-to-detect failures on the right, such as hidden distribution shift.
The \textit{y}-axis encodes remediation cost, increasing from low cost at the bottom, such as applying a simple post-filter or local structural repair, to high cost at the top, such as retraining, redesigning the objective function, or modifying the full discovery pipeline.
Marker shapes and colors jointly encode the primary source of each failure:
blue filled circles ($\bullet$) indicate generator-origin failures (model design or representation);
orange filled squares ($\blacksquare$) indicate optimizer-origin failures (search or guidance pathologies);
and red filled triangles ($\blacktriangle$) indicate evaluation-pipeline-origin failures (benchmark or data issues).
The translucent colored ellipses in the legend and figure denote approximate algorithm-family exposure regions, indicating which classes of inverse-design method are most exposed to related failure modes:
\revised{LSO, latent-space optimization, an umbrella covering gradient descent in a latent space and its stochastic variants such as preconditioned Langevin dynamics, which differ in how they explore but share the same exposure to decoding and surrogate failures;}
DIFF, guided diffusion generation;
EVO, evolutionary optimization such as CMA-ES;
RL, reinforcement-learning policy optimization;
and ALL, failure modes shared across algorithm families.
The upper-right quadrant represents a critical failure regime; here, the convergence of high detection difficulty and extreme remediation cost poses the greatest threat to the credibility of automated discovery claims.
\label{fig:benchmark_stack_rev}}
\end{figure}

The ability to generate crystal candidates at scale is no longer the main bottleneck in inverse design. The more persistent challenge is evaluation: benchmarks must remain reliable under strong optimization pressure and comparable across studies [\ref{cit:DeBreuck2025}-\ref{cit:Dunn2020}]. Evaluation should be treated as a systematic exercise in identifying, exposing, and controlling failure modes. Figure~\ref{fig:benchmark_stack_rev} summarizes this perspective by organizing common weaknesses according to how easily they can be detected and how costly they are to correct.

Failures differ in their scientific consequence. Some can be removed early with inexpensive filters, whereas others remain hidden until prospective validation or deployment and can invalidate the discovery claim itself. The labels  \revised{latent-space optimization (LSO),} guided diffusion generation (DIFF), evolutionary optimization (EVO), reinforcement-learning policy optimization (RL), and failure modes shared across all algorithm families (ALL) should therefore be taken as indicative labels only. Certain pathologies are more common under particular optimization regimes, but the most damaging failures are often benchmark-level and method-agnostic.

\subsubsection*{\revised{Band 1: feasibility and validity failures, cheap to catch and cheap to fix}}
The lowest band corresponds to \emph{feasibility and validity failures}. \revised{They are tagged [LSO] in Figure~\ref{fig:benchmark_stack_rev}, for latent-space optimization, because gradient-based and Langevin-type search in a continuous latent space are the methods most exposed to them: both move a latent code toward a high surrogate score with no guarantee that the decoder maps the endpoint onto a physical structure, so the failure surfaces only after decoding (Section~\ref{sec:latent_opt}). Guided diffusion is less exposed, since each denoising step stays near the learned manifold, and space-group-constrained generators are less exposed still, having removed part of the failure at the representation level (Section~\ref{sec:symmetry_axis}).} These are usually the easiest failures to detect and, when handled early, the least expensive to correct. In crystal-generation pipelines, they include atomic overlap, unrealistic interatomic distances, chemically implausible compositions, and symmetry inconsistency [\ref{cit:Xie2021}, \ref{cit:DeBreuck2025}]. Such checks are standard, but the literature is equally clear that passing them does not establish thermodynamical stability, synthesizability, or practical relevance [\ref{cit:DeBreuck2025}]. They should therefore be treated as necessary entry conditions.

Atomic clashes are a typical example. They arise when optimization in latent or surrogate space reaches a formally high-scoring region that decodes into unphysical structures. Because they can usually be identified with simple geometric rules, they should be removed immediately through minimum-distance filters, cell-sanity checks, and, where feasible, low-cost pre-relaxation [\ref{cit:Xie2021}, \ref{cit:DeBreuck2025}]. Closely related are charge violations and broader forms of chemical implausibility. In inorganic screening, Semiconducting Materials by Analogy and Chemical Theory (SMACT)-based oxidation-state and electronegativity rules remain useful first-pass filters because they cheaply remove many impossible candidates before expensive validation [\ref{cit:Davies2019}]. Symmetry breaking is another important failure, and diffusion-based generators are the usual site of it. Recent work has shown that, when symmetry is central to the task, it is more reliable to encode it directly in the model than to repair it afterwards [\ref{cit:Jiao2023}].

\subsubsection*{\revised{Band 2: optimization pathologies, where the optimizer looks successful}}
The middle band contains \emph{optimization pathologies}. These are harder to see, because the optimizer appears to be working. The score improves and the loop converges, while the scientific objective stands still. The cause is overoptimization of imperfect proxies. \revised{Matbench Discovery makes the point concrete for stability screening. It scores models prospectively, training on the Materials Project and testing on the later WBM set of substitution-derived structures, and it grades them by classification of stability against the convex hull and by discovery acceleration factor, and not by regression error alone. On that protocol the ranking by regression accuracy and the ranking by discovery utility come apart, so a model with strong retrospective energy predictions need not be the model that selects the best candidates to compute [\ref{cit:Riebesell2025}]. This is why the ledger of Section~\ref{sec:ledger} asks for decisions and survival rates alongside accuracies.} This is Goodhart's law in practical form. Once a proxy becomes the optimization target, the search exploits its blind spots and returns nothing physically meaningful.

The point labeled \emph{proxy hacking} reflects this risk. In gradient-driven workflows, the search may exploit surrogate error while leaving the underlying physical objective untouched. The most effective safeguard is staged evaluation, with hard validity checks first, surrogate-based triage second, and periodic high-fidelity validation throughout the loop [\ref{cit:DeBreuck2025}, \ref{cit:Riebesell2025}]. \revised{Reward hacking in reinforcement-learning settings is a more explicit version of the same problem, and the AI literature has given it a precise shape. Amodei et al.\ list reward hacking among a small set of recurring accident modes that follow from specifying an objective by proxy, alongside negative side effects and unsafe exploration [\ref{cit:Amodei2016}]. Skalse et al.\ sharpen this into a formal criterion: a proxy reward is \emph{unhackable} with respect to a true reward only if no policy improvement under the proxy can correspond to a decline under the true objective, and they show that for most pairs of interest this condition fails, so a proxy that is merely well correlated with the objective is not enough [\ref{cit:Skalse2022}]. Read against an inverse-design loop, the consequence is direct. Correlation between a surrogate and DFT on a held-out set does not license optimizing the surrogate hard, because the optimizer searches precisely for the configurations where the correlation breaks.} For inverse-design workflows, this means that the operational reward should be treated as an approximation, not as ground truth. Independent physical endpoints, auditing against simpler baselines, and explicit control of optimization pressure are therefore essential.

Diversity collapse is another common optimization pathology, most visible in evolutionary or population-based search. CMA-ES and related methods are powerful because they can navigate rugged, non-differentiable landscapes, but they can also collapse into narrow basins if diversity is not actively maintained [\ref{cit:Hansen2016}]. In inverse design, this wastes validation budget by repeatedly proposing near-duplicates. Diversity should therefore be built into the objective or the selection stage through novelty terms, archives, restart strategies, clustering, or explicit coverage metrics over chemistry and structure.

\subsubsection*{\revised{Band 3: evidence and evaluation failures, which invalidate the benchmark}}
The highest band represents \emph{evidence and evaluation failures}. These are the most serious because they often invalidate the benchmark itself, with consequences well beyond any single optimizer. Hidden shift failure arises when a model appears reliable on retrospective data but degrades once the candidate distribution changes. This problem is central to uncertainty-aware learning under dataset shift [\ref{cit:Ovadia2019}] and is equally visible in materials benchmarks that distinguish retrospective performance from prospective utility [\ref{cit:Riebesell2025}]. For that reason, inverse-design studies should move beyond random shuffled splits and include more deployment-relevant protocols, such as temporally separated tests, chemically grouped splits, prospective hold-outs, or external validation sets.

Data leakage belongs in the same high-risk category because it can inflate performance silently while remaining invisible in the final metric table. Leakage is a recognized and persistent problem across machine-learning-based science [\ref{cit:Kapoor2023}-\ref{cit:Kapoor2024}]. \revised{The mechanisms are concrete and each has a documented example. The major inorganic databases are not independent samples of chemistry: MP, OQMD, JARVIS, and AFLOW descend substantially from ICSD-derived structures, re-relaxed at different levels of theory, assigned different identifiers, and deduplicated by different conventions [\ref{cit:Jain2013}, \ref{cit:Choudhary2020}-\ref{cit:Saal2013}, \ref{cit:Zagorac2019}]. Novelty assessed against a database other than the training one is therefore contaminated by \emph{cross-database twins}, the same phase carried under a different identifier or cell setting. \emph{Composition-level leakage} is subtler: the same reduced formula straddles training and test as different polymorphs, so models credited with generalization partly perform recall. Matching verdicts are themselves tolerance-dependent: the widely used \texttt{StructureMatcher} defaults (length tolerance $0.2$, site tolerance $0.3$, angle tolerance $5^\circ$) are conventions, and a novelty or uniqueness rate reported without its tolerances is unfalsifiable. The most common artifact is \emph{novelty by perturbation}: strained lattices or single-element substitutions of known prototypes that defeat binary matching, together with \emph{order-disorder aliasing}, where an ordered approximation of a known disordered phase is counted as new. These are not hypothetical. A chemical audit of the GNoME release concluded that most of its claimed new compounds are ordered approximations of known disordered phases or compositional variants of known prototypes [\ref{cit:Cheetham2024}], and a reanalysis of the A-Lab campaign argued that most of its 41 claimed syntheses correspond to known or misidentified phases once compositional disorder and powder-XRD ambiguity are accounted for [\ref{cit:Leeman2024}]. Both are framed by their authors as failure modes of novelty \emph{assessment}, and not as indictments of the underlying methods. One reporting rule follows from this. \emph{Uniqueness, memorization, and rediscovery are three distinct quantities}, corresponding to within-batch duplicates, training-set matches, and external-database matches, and no single \emph{novelty rate} can carry all three. A candidate's ladder position (Section~\ref{sec:ladder}, Table~\ref{tab:ladder}) is meaningful only once they have been separated.} Once detected, these problems often require protocol redesign and rerunning the study. Split audits, duplicate checks, strict preprocessing isolation, \revised{chemically and temporally grouped splits,} and transparent reporting should therefore be treated as part of the benchmark itself.

\subsubsection*{\revised{Out-of-distribution generalization and the calibration of uncertainty}}
\revised{Closely related, and increasingly central to generative crystal design, is the question of \emph{out-of-distribution} (OOD) generalization, the regime in which inverse design is actually valuable, since a model that only interpolates its training distribution rediscovers known chemistry. The difficulty is that every component of the loop degrades in the same OOD region, and often silently. Property surrogates and universal MLIPs are trained on near-equilibrium structures from a handful of overlapping databases, so their errors grow on the unusual candidates that aggressive generation produces, and their uncertainty is typically miscalibrated there too [\ref{cit:Riebesell2025}, \ref{cit:Ovadia2019}-\ref{cit:Guo2017}, \ref{cit:Deng2025}]; the convex-hull reference against which stability is judged is itself sparsest in under-explored chemistries, so an OOD candidate is scored against the weakest available baseline. Generators inherit the same limitation from the input side: rare elements receive undertrained embeddings, and lattice-composition correlations learned on common chemistries extrapolate unreliably to the rest. Because these failures are correlated across the loop, an OOD candidate can pass surrogate scoring, MLIP relaxation, and even a hull check while being physically wrong, giving an over-optimistic pass rate that no single in-distribution metric reveals. \revised{Each link in that chain is separately documented: universal potentials systematically soften the potential-energy surface away from their training manifold, which lowers computed energy differences and makes marginal structures look stable [\ref{cit:Deng2025}]; neural predictors are miscalibrated under dataset shift, and the miscalibration grows with the size of the shift [\ref{cit:Ovadia2019}-\ref{cit:Guo2017}]; and prospective benchmarking shows that in-distribution accuracy does not transfer to selection quality on a later, differently distributed candidate set [\ref{cit:Riebesell2025}]. The audit of the GNoME release is the empirical form of the same chain reaching the reporting stage [\ref{cit:Cheetham2024}].} Diagnosing OOD behavior therefore requires evaluation protocols that deliberately break the assumption that test candidates are drawn from the same distribution as the training data. Chemically grouped splits hold out whole element combinations or structural prototypes; temporally grouped splits train on structures known before a cutoff date and test on those discovered after, mirroring the prospective Matbench Discovery protocol [\ref{cit:Riebesell2025}]; and leave-one-chemistry-out audits report a system's extrapolation ability in place of its interpolation skill. On the modeling side, the mitigations are equally concrete: quantify and act on extrapolative uncertainty [\ref{cit:Ovadia2019}-\ref{cit:Guo2017}]; fine-tune foundation MLIPs on target-region DFT labels with before/after pass rates reported [\ref{cit:Zeni2025}, \ref{cit:PyzerKnapp2025}, \ref{cit:Batatia2022}, \ref{cit:Batatia2025}]; and, for conditional generators, report the guidance-strength trade-off (Section~\ref{sec:guided_diffusion}), since the same conditioning that pushes a model off its data manifold toward a target also drives it OOD. We therefore recommend the following. Any discovery claim made in a chemistry that is poorly represented in the training data should state, alongside its headline metrics, which of these protocols was used to evaluate it. A metric reported without that information implicitly assumes that the candidates being tested resemble the training data. For genuine discovery, this is the assumption most likely to be false.}

Miscalibrated uncertainty is among the hardest failures to diagnose directly. Neural predictors are often poorly calibrated even in standard settings, and calibration can degrade further under shift [\ref{cit:Ovadia2019}-\ref{cit:Guo2017}]. Calibration matters because uncertainty is used operationally in inverse design: it determines which candidates advance to DFT, which regions are explored next, and how validation budget is allocated. Poor calibration therefore degrades interpretation and decision quality alike. A credible evaluation should test calibration under realistic distribution shift, compare several uncertainty methods before settling on a single heuristic, and ask whether uncertainty improves downstream selection; in-distribution calibration scores alone do not answer that question [\ref{cit:Riebesell2025}, \ref{cit:Ovadia2019}-\ref{cit:Guo2017}].

\subsection{Benchmarking Through a Staged Pass-Rate Ledger \revised{and a Minimum Reporting Standard}}\label{sec:ledger}

Inverse-design systems are inherently vulnerable to weak evaluation. Because they optimize aggressively, they will exploit any gap in the benchmark, often producing apparent gains that disappear once candidates are subjected to structural relaxation, phase-stability screening, or synthesis-relevant constraints [\ref{cit:DeBreuck2025}]. For this reason, benchmarking should not stop at a single final score. It should show, stage by stage, where candidates survive, where they fail, and how much effort is required to obtain credible hits. A practical way to do this is to report a \emph{pass-rate ledger} across a minimal evaluation stack:

\begin{itemize}[leftmargin=*]
\item \textit{Validity.} The first question is whether a generated candidate is physically and chemically reasonable in the most basic sense. Screening at this level covers atomic overlap, interatomic distances, density or volume, and no obvious violations of chemical plausibility.

\item \textit{Uniqueness.} The next step is to establish whether the surviving candidates are genuinely distinct under a clearly stated matching criterion, such as declared \texttt{StructureMatcher} tolerances or reduced-formula rules. Uniqueness should not be conflated with novelty: without it, a pipeline can appear effective simply by reproducing the same structural motif many times [\ref{cit:Wagner2026}].

\item \textit{Novelty and diversity.} Once duplicates have been removed, the important question becomes whether the remaining candidates extend beyond what is already known. Novelty should therefore be assessed relative to both the training set and relevant external databases, with transparent definitions of compositional and structural novelty. Continuous similarity-based metrics, in place of binary match thresholds, provide a more reliable basis for this evaluation, as they quantify the degree of uniqueness and novelty within a generated set [\ref{cit:Negishi2025}]. Crystallographic family relationships, including structures connected by group-subgroup transitions or order-disorder variants of a shared parent phase, should also be considered when defining structural novelty, because symmetry-lowered rediscoveries of known phases can otherwise pass standard matching criteria [\ref{cit:Yamazaki2026}]. Local novelty distance measures further support rapid nearest-neighbor lookup against large structural databases, enabling geographic-style navigation of materials space and real-time novelty assessment [\ref{cit:Widdowson2025}]. Diversity should be reported through distributions or coverage statistics, such as chemical spread or symmetry and structure variation, and not through a few selected examples [\ref{cit:Wagner2026}, \ref{cit:Negishi2025}, \ref{cit:Widdowson2025}, \ref{cit:Yamazaki2026}].

\item \textit{Stability.} Stability should be reported at the level actually being measured, distinguishing between \emph{proxy} stability inferred from surrogate scores, \emph{relaxed} stability obtained after structural optimization, \emph{phase} stability assessed from quantities such as convex-hull proximity or energy above hull, and \emph{dynamical} stability assessed through vibrational analysis using machine-learned interatomic potentials. Dynamical stability screening can detect imaginary phonon modes in candidates that survive static energy-based screening, and automated workflows can further guide structural remediation of dynamically unstable candidates toward stable polymorphs [\ref{cit:Gouvea2026}].

\item \textit{Efficiency.} A strong inverse-design workflow finds promising candidates at a reasonable cost. Benchmarks should therefore report wall-clock time, compute budget, and the number of high-fidelity validations required per credible hit, since methods often differ as much in validation burden as in proposal quality [\ref{cit:Wagner2026}].
\end{itemize}

\subsubsection*{\revised{From ledger to a minimum reporting standard}}
\revised{We state the implication of this ledger prescriptively. Much of the generative-crystal literature reports pairwise-distance and density \emph{structural validity} rates that approach saturation and presents them as evidence of success; such rates are near-trivial entry statistics that certify almost nothing about whether a generated structure is a material. \emph{A passed validity filter is an entry condition, not a result.} A structure can satisfy every distance and density filter and still be thermodynamically uncompetitive, dynamically unstable, or chemically impossible. Credible evaluation must therefore rest on the criteria practicing materials scientists already apply, and not on proxies chosen for machine-learning convenience. Those criteria are convex-hull energetics and formation energies at DFT fidelity, dynamical (phonon) stability, and confirmation of the target property at DFT or experimental fidelity. Evaluation protocols should be co-designed with domain experts, and not inherited from generic generative-modeling benchmarks, whose notions of sample quality were developed for images and text and do not transfer to periodic solids [\ref{cit:DeBreuck2025}-\ref{cit:Dunn2020}, \ref{cit:Riebesell2025}]. By the same logic, a discovery claim requires reporting the full distribution of outcomes across these levels for all generated candidates, and not just the properties of the best surviving ones. Table~\ref{tab:report} converts this position into an explicit multi-level minimum reporting standard, grouped into evidence levels of increasing evidentiary strength (annotated with the credibility-ladder rung of Section~\ref{sec:ladder} that each certifies); a passed level is an entry condition for the next. None of these items is exotic, and several of them (declared tolerances, hull versions, database snapshots) cost nothing beyond disclosure [\ref{cit:Gouvea2026}-\ref{cit:Negishi2025}, \ref{cit:Widdowson2025}, \ref{cit:Kapoor2024}, \ref{cit:Yamazaki2026}]. We propose that new-method papers, benchmarks, and reviews treat this list as the minimum for a defensible discovery claim.}

\revised{\begin{table}[htbp]\color{revred}
\centering
\footnotesize
\caption{\revised{Multi-level minimum reporting standard for generative crystal-discovery studies. A passed filter at any level is an entry condition for the next, not a result; discovery claims require the full distribution of outcomes across all levels. Ladder rungs (Section~\ref{sec:ladder}) are noted in the level-group headers.}}
\label{tab:report}
\setlength{\tabcolsep}{4pt}
\renewcommand{\arraystretch}{1.25}
\begin{tabularx}{\textwidth}{
  >{\raggedright\arraybackslash}p{0.17\textwidth}
  >{\raggedright\arraybackslash}p{0.31\textwidth}
  >{\raggedright\arraybackslash}p{0.20\textwidth}
  >{\raggedright\arraybackslash}Y}
\toprule
\textbf{Reporting item} & \textbf{What must be reported} & \textbf{Tools} & \textbf{Failure mode it guards against} \\
\midrule
\multicolumn{4}{@{}l}{\textit{Level 1: Chemical and structural plausibility (entry conditions; ladder rung L1)}}\\
1. Composition validity & Fraction passing charge neutrality and oxidation-state/electronegativity plausibility & SMACT [\ref{cit:Davies2019}] & Chemically impossible candidates inflating headline validity \\
2. Structural validity beyond pairwise distance & Coordination and bonding sanity; symmetry consistency with the declared space group & pymatgen [\ref{cit:Ong2013}] & Near-trivial distance/density checks presented as success \\
\multicolumn{4}{@{}l}{\textit{Level 2: Set-level integrity}}\\
3. Uniqueness & Duplicate rate under a \emph{declared} \texttt{StructureMatcher} tolerance & \texttt{StructureMatcher} [\ref{cit:Ong2013}] & One motif reproduced many times, counted as many hits \\
4. Novelty & Similarity to training \emph{and} external references (ICSD, MP, OQMD), continuous metrics not binary thresholds & SUN-style [\ref{cit:Negishi2025}]; local novelty [\ref{cit:Widdowson2025}]; group-subgroup [\ref{cit:Yamazaki2026}] & \emph{Novel} structures that are perturbed or substituted known prototypes \\
\multicolumn{4}{@{}l}{\textit{Level 3: Stability under relaxation and thermodynamics (rungs L2-L4)}}\\
5. Relaxation success rate & Fraction surviving MLIP/DFT relaxation without collapse, with structural drift & M3GNet [\ref{cit:Chen2022}], CHGNet [\ref{cit:Deng2023}], MACE [\ref{cit:Batatia2022}] & Fragile structures counted as candidates \\
6. Energy above hull & The \emph{full} $E_\text{hull}$ distribution, with functional, corrections, and hull version & MP/OQMD hulls [\ref{cit:Jain2013}, \ref{cit:Kirklin2015}-\ref{cit:Saal2013}]; Matbench Discovery [\ref{cit:Riebesell2025}] & Cherry-picking the best surviving candidate \\
\multicolumn{4}{@{}l}{\textit{Level 4: Dynamical stability (rung L5)}}\\
7. Dynamical stability & Fraction of top candidates free of imaginary phonon modes & MLIP phonons [\ref{cit:Loew2025}]; automated remediation workflows, e.g.\ VibroML [\ref{cit:Gouvea2026}] & Statically converged but dynamically unstable claims \\
\multicolumn{4}{@{}l}{\textit{Level 5: Functional confirmation at high fidelity (rung L6)}}\\
8. Target-property validation & Conditioned property confirmed at DFT/experimental fidelity; surrogate-vs-high-fidelity agreement & DFT or experiment & Surrogate exploitation, proxy hacking (Sec.~5.8) \\
\multicolumn{4}{@{}l}{\textit{Transparency and cost accounting}}\\
9. Compute cost & High-fidelity validations, DFT calls, wall-clock per credible hit & Efficiency stage (Sec.~6.1) & Validation burden hidden behind hit counts \\
10. Code and data availability & Code, generated structures, evaluation scripts, database versions & REFORMS-style [\ref{cit:Kapoor2024}] & Irreproducible discovery claims \\
\bottomrule
\end{tabularx}
\end{table}}

\subsubsection*{\revised{Auditing representative generators against that standard}}
\revised{Table~\ref{tab:quant_A} and Table~\ref{tab:quant_B} apply this standard retrospectively to seven of the generators surveyed here, reporting the claims of each study under its own protocol. Two warnings govern their use. First, the numbers are not mutually comparable: match rates depend on the number of samples $k$ and on tolerances; stability rates depend on the hull reference, its snapshot, and the functional; validity depends on filter definitions. These are claims with provenance, not a leaderboard, and reading them as one would reproduce the very benchmark pathologies discussed above. Second, the most informative feature is the pattern of absent entries: phonon validation is reported by two models [\ref{cit:Park2025}, \ref{cit:Babu2026}], experimental validation once (a single candidate [\ref{cit:Zeni2025}]), declared-tolerance uniqueness almost never, and post-relaxation stability by the three most recent models only [\ref{cit:Zeni2025}, \ref{cit:Park2025}, \ref{cit:Babu2026}]. Every empty cell corresponds to an item Table~\ref{tab:report} asks authors to supply; together the tables show that the field can generate at scale but cannot yet account for what it generates.}

\revised{\begin{table}[htbp]\color{revred}
\centering
\footnotesize
\caption{\revised{Provenance and design of the crystal generators surveyed in this review, as reported in the original studies.}}
\label{tab:quant_A}
\setlength{\tabcolsep}{4pt}
\renewcommand{\arraystretch}{1.2}
\begin{tabularx}{\textwidth}{
  >{\raggedright\arraybackslash}p{0.11\textwidth}
  >{\raggedright\arraybackslash}p{0.24\textwidth}
  >{\raggedright\arraybackslash}p{0.20\textwidth}
  >{\raggedright\arraybackslash}Y
  >{\raggedright\arraybackslash}p{0.10\textwidth}}
\toprule
\textbf{Model} & \textbf{Training data (size)} & \textbf{Representation} & \textbf{Conditioning} & \textbf{Code\slash data} \\
\midrule
\multicolumn{5}{@{}l}{\textit{Crystal-structure-prediction models (evaluated by match-rate to known structure)}}\\
CDVAE [\ref{cit:Xie2021}] & MP-20 (45{,}231); Perov-5, Carbon-24 & Periodic graph; VAE with score decoder & Latent-space property optimization & Yes \\
DiffCSP [\ref{cit:Jiao2023}] & MP-20 (45{,}231); MPTS-52 (40{,}476) & Equivariant joint diffusion (lattice and coordinates) & Composition (CSP) & Yes \\
CrystalFlow [\ref{cit:Luo2025}] & MP-20; MPTS-52; MP-CALYPSO-60 (657{,}377) & Flow matching (lattice and coordinates) & Composition; pressure & Yes \\
\multicolumn{5}{@{}l}{\textit{De novo generators (evaluated by validity / stability / SUN)}}\\
MatterGen [\ref{cit:Zeni2025}] & Alex-MP-20 (about 600k, up to 20 atoms) & Joint diffusion (types, coords, lattice) & Chemistry, space group, property via adapters & Yes \\
Chemeleon [\ref{cit:Park2025}] & MP (32{,}525, below 0.25 eV/atom, chronological) & Text-aligned embedding with diffusion & Natural-language prompts & Yes \\
CrystaLLM [\ref{cit:Antunes2024}] & About 2.3M CIFs (MP, OQMD and NOMAD) & CIF tokens; autoregressive Transformer & Composition; optional space group & Yes \\
MEIDNet [\ref{cit:Babu2026}] & Perov-5 (about 19k); MP-20 (about 45k); Carbon-24 (about 10k); Perov-5 is the inverse-generation benchmark, the other two assess the equivariant encoder & $E(3)$-equivariant crystal autoencoder; shared 128-dimensional structure-property latent embedding & Target bandgap and formation enthalpy via multimodal latent-space optimization & Yes \\
\bottomrule
\end{tabularx}
\end{table}}

\revised{\begin{table}[htbp]\color{revred}
\centering
\footnotesize
\caption{\revised{Evaluation outcomes for the crystal generators surveyed in this review, as reported in the original studies under their own protocols. \emph{n.r.} denotes not reported; numbers are not mutually comparable across rows. The final three columns state whether the study reports \emph{phonon} (dynamical-stability) validation, \emph{DFT}-level validation, and \emph{experimental} synthesis of generated candidates: Y denotes reported and N denotes not reported. Where these validations are reported, the scale is small and is stated in the text and not in the cells: MatterGen synthesized a single generated candidate [\ref{cit:Zeni2025}], and Chemeleon reports phonon checks for four structures [\ref{cit:Park2025}]. Rows are grouped as in Table~\ref{tab:quant_A}: CSP models are scored by match rate to a known target structure, whereas de novo generators are scored by validity/uniqueness/novelty-type rates; the two blocks answer different questions and must not be read as one ranking (Section~\ref{sec:ledger}).}}
\label{tab:quant_B}
\setlength{\tabcolsep}{4pt}
\renewcommand{\arraystretch}{1.2}
\begin{tabularx}{\textwidth}{
  >{\raggedright\arraybackslash}p{0.115\textwidth}
  >{\raggedright\arraybackslash}Y
  >{\raggedright\arraybackslash}p{0.22\textwidth}
  >{\raggedright\arraybackslash}p{0.13\textwidth}
  >{\centering\arraybackslash}p{0.08\textwidth}
  >{\centering\arraybackslash}p{0.05\textwidth}
  >{\centering\arraybackslash}p{0.05\textwidth}}
\toprule
\textbf{Model} & \textbf{Validity / uniqueness / novelty (as reported)} & \textbf{Post-relaxation stability} & \textbf{$E_\text{hull}$ threshold} & \textbf{Phonon} & \textbf{DFT} & \textbf{Exp.} \\
\midrule
\multicolumn{7}{l}{\itshape Crystal-structure-prediction models (match rate to a known target structure)}\\
\midrule
CDVAE [\ref{cit:Xie2021}] & Struct.\ 100\%, comp.\ 86.7\%; CSP match 33.9\% (one attempt; re-benchmarked in [\ref{cit:Jiao2023}]); no declared-tolerance uniqueness & n.r. & n.r. & N & N & N \\
DiffCSP [\ref{cit:Jiao2023}] & Struct.\ 100\%, comp.\ 83.25\%; CSP match 51.49\% (one attempt) and 77.93\% (20 attempts) & n.r. & n.r. & N & N & N \\
CrystalFlow [\ref{cit:Luo2025}] & Struct.\ 96.8\%, comp.\ 92.4\%; CSP match 51.89\% (one attempt) & CHGNet and DFT on subsets; rate not headline & DFT at target pressure & N & Y & N \\
\midrule
\multicolumn{7}{l}{\itshape De novo generators (validity, stability and SUN rates)}\\
\midrule
MatterGen [\ref{cit:Zeni2025}] & SUN more than twice prior; uniqueness 52\%, novelty 61\% at 10M samples & 78\% below threshold after DFT; 95\% within RMSD 0.076~\AA & 0.1 eV/atom vs Alex-MP-ICSD & N & Y & Y \\
Chemeleon [\ref{cit:Park2025}] & Validity 98-99\%; uniqueness above 90\%; 20\% of unseen GT generated & DFT campaigns (e.g.\ TiO$_2$: 549 reduced to 122 unique metastable) & Hull-ref (MP); train filter below 0.25 & Y & Y & N \\
CrystaLLM [\ref{cit:Antunes2024}] & 88.1\% of held-out compositions matched within 3 attempts & 102 novel DFT-relaxed: 20 within threshold, 3 on hull & 0.1 eV/atom vs MP & N & Y & N \\
MEIDNet [\ref{cit:Babu2026}] & SUN 13.6\% (19/140 generated structures; individual rates n.r.) & eSEN-30M-OAM relaxation; stable candidates validated by DFT and phonon analysis\slash VibroML & 0.1 eV/atom vs MP & Y & Y & N \\
\bottomrule
\end{tabularx}
\end{table}}

\revised{A shared, model-agnostic protocol for these quantities also exists on the generative side. LeMat-GenBench [\ref{cit:Betala2025}] evaluates crystal generative models under a single pipeline that combines validity filtering, an ensemble of machine-learned potentials for energetics, and standardized uniqueness, novelty, and diversity metrics. It scores them against a reference set of roughly five million structures, which tightens both the convex hull and the novelty comparison relative to MP-20-based reporting. Its central finding measures the tension this review has argued for throughout. Across the models benchmarked, gains in stability came on average at the cost of novelty and diversity, and no model led on every axis. This is the empirical case for a ledger in place of a headline number, since a single reported rate can always be raised by moving along the trade-off. The axes must therefore be reported together.}

\subsubsection*{\revised{Why high scores do not always mean good materials}}
\revised{A generative model is easy to score and hard to judge. Most reported numbers say whether a
generated structure passed a filter, and not whether it is a material that anyone could make. That gap is
where a high score misleads. The underlying effect has a name: a measure adopted as a target stops working
as a measure [\ref{cit:Strathern1997}], and Manheim and Garrabrant set out how this happens, including the case where the
proxy tracks the goal only loosely, and the case where hard optimization drives the system into a regime
in which the two stop tracking each other at all [\ref{cit:Manheim2018}]. An inverse-design loop applies exactly that
pressure, because the generator is trained toward validity, steered by property scores, and filtered by
surrogate rankings. Four specific effects follow, and each has a check that exposes it.}

\revised{\emph{A validity rate close to 100\% carries almost no information.} Validity filters test minimum
interatomic distances and cell density. Ordinary materials pass these easily, so the test separates
structures that are obviously broken from everything else, and says nothing about the frontier where new
candidates sit. A saturated validity rate should therefore be read together with the fraction of those
structures that survive relaxation.}

\revised{\emph{A model that keeps producing the same motif can look productive.} If a generator collapses
onto a small family of structures, every copy is counted again, and the total number of hits grows while
the number of distinct materials does not. Uniqueness measured under a declared matching tolerance
separates the two, which is why hits and unique hits should be plotted together.}

\revised{\emph{Small distortions of known structures can be counted as new.} Matching a candidate against a
database is a yes-or-no test at a chosen tolerance, so a strained lattice or a single substituted element
can fall on the far side of the threshold and register as novel. This is not hypothetical: an audit of the
GNoME release concluded that most of its claimed new compounds are ordered approximations of known
disordered phases or compositional variants of known prototypes [\ref{cit:Cheetham2024}]. Reporting the distance to the
nearest known material as a distribution, in place of a count above a cutoff, removes the dependence on
the threshold and shows how far the new structures really sit from the old ones [\ref{cit:Negishi2025}, \ref{cit:Widdowson2025}, \ref{cit:Yamazaki2026}].}

\revised{\emph{Reporting only the best candidate reports mostly noise.} Screening scores carry error, so the
top-ranked candidate in a large batch is the one whose error happened to point the right way. Its value is
an extreme of a noisy distribution and does not describe what the model typically produces. Reporting the
full distribution of outcomes over all generated candidates avoids this (Table~\ref{tab:report}, item~6).}

\revised{A related difficulty arises when these metrics move between tasks. Validity, uniqueness, novelty
and the composite S.U.N.\ (stable, unique, novel) rate were defined for a generator that proposes both
composition and structure from scratch [\ref{cit:Xie2021}, \ref{cit:Zeni2025}, \ref{cit:Negishi2025}]. Their meaning depends on that setting. When part of
the answer is fixed in advance, part of the score is fixed with it. If the composition is given, as in
crystal structure prediction, a novelty rate records whether that formula appeared in the training set, and
uniqueness is often saturated as well: a recent symmetry-driven CSP framework reports uniqueness at 100\% on
all three standard benchmarks, both for its own model and for the baseline it is compared against [\ref{cit:Mu2026}]. If
candidates are scored over a fixed pool, as in retrieval, a uniqueness rate describes the pool. If a model
is fine-tuned from a released base model, a S.U.N.\ gain starts from whatever the base model already
reached, so the contribution of the fine-tuning is visible only when the base-model value is given beside
it. This last point asks for a baseline and does not question the value of targeted fine-tuning studies;
the reader simply cannot separate the improvement from the starting point unless both are shown.}

\revised{Three questions make any headline number checkable. \emph{What was held fixed?} A score computed
with the composition, space group or prototype supplied carries no evidence about the part that was
supplied. \emph{Against which reference, and at which tolerance?} A novelty or uniqueness rate quoted
without a database snapshot and a stated \texttt{StructureMatcher} tolerance cannot be checked or
reproduced, and the continuous distance measures of Section~\ref{sec:ledger} avoid the choice of threshold
altogether [\ref{cit:Negishi2025}, \ref{cit:Widdowson2025}, \ref{cit:Yamazaki2026}]. \emph{Does it survive physics?} A validity or novelty rate given without a
relaxation-survival rate and an energy-above-hull distribution describes what the model produced and not
what it found. Reporting each metric together with its task type, its trivial or base-model baseline, and
the tolerances and databases used is the reporting-level counterpart of the credibility ladder of
Section~\ref{sec:ladder}: a number becomes interpretable once the conditions that give it meaning are stated.}

\revised{\subsection{The MLIP Oracle Layer}\label{sec:mlip_oracle}
A third validation tier sits between inexpensive validity filters and scarce DFT calls. Universal machine-learned interatomic potentials (Table~\ref{tab:mlip_models}) supply relaxation, energies, forces and phonon estimates three to five orders of magnitude below DFT cost (geometric filter $<1$~ms; surrogate $\sim$ms; universal MLIP $\sim$0.1-10~s; DFT $\sim 10^2$-$10^4$ core-hours; experiment days to months), and the tier has improved quickly: stability classification on the Matbench Discovery test set has risen from $F_1 \approx 0.6$ for the first universal potentials to $F_1 \approx 0.92$ [\ref{cit:Riebesell2025b}-\ref{cit:Bigi2026}]. Its verdicts are nonetheless conditional, for a structural reason. Universal MLIPs are trained overwhelmingly on near-equilibrium configurations drawn from the same databases that train the generators they judge, and their documented out-of-distribution failure is systematic softening, in which energy differences and restoring forces off the training manifold are underestimated [\ref{cit:Deng2025}]. Unusual structures therefore appear more stable, and their relaxations more successful, than DFT finds. Because generator and oracle share training distributions, the oracle is least reliable where the generator is most adventurous. \revised{Because this bias belongs to the potential and its near-equilibrium training data, and not to any one optimizer, it is placed with the evaluation failures in Figure~\ref{fig:benchmark_stack_rev}.} \revised{The tier also involves a choice the literature often leaves implicit: a universal potential trained across the periodic table, or a domain-specific one fitted to the chemistry at hand. A bespoke potential is more accurate inside its training domain and silently unreliable outside it, which makes it a poor match for generators whose value lies precisely in leaving that domain; a universal potential is less accurate but fails more diagnosably. Foundation-scale models now make fine-tuning on the loop's own DFT labels cheap enough [\ref{cit:Batatia2025}] that the practical question is how much loop-generated data has accumulated in a given chemistry, rather than which of the two to adopt once and for all.}

Fidelity within the tier is task-specific. Harmonic phonons of near-equilibrium structures are within reach [\ref{cit:Loew2025}], whereas anharmonic observables such as lattice thermal conductivity show errors only weakly correlated with stability-ranking skill [\ref{cit:Pota2024}]. The tier should therefore serve for triage and not for confirmation: candidates are routed by extrapolative uncertainty, and DFT spot checks are distributed across the whole accepted set instead of concentrated on its top scorers. Table~\ref{tab:mlip} sets out, task by task, what the tier can be trusted to decide, how each verdict fails, and the guardrail it requires.

Matbench Discovery [\ref{cit:Riebesell2025b}] supplies a common prospective protocol for this tier, covering relaxation, hull-distance ranking and lattice thermal conductivity, and Table~\ref{tab:mlip_models} summarizes the potentials in current use against it. New families continue to appear, among them models trained to span several atomistic domains within a single architecture [\ref{cit:Wood2025}]. Two properties of that evaluation matter more than the ranking itself. First, the potentials differ only modestly in energy ranking but by more than an order of magnitude in curvature, so the choice of potential should follow the property being claimed. Second, models that predict forces directly, without differentiating an energy, buy speed at the cost of a rough potential-energy surface near minima [\ref{cit:Fu2025}, \ref{cit:Rhodes2025}], and that is the regime on which phonon and transport verdicts depend. A prescreening tier should therefore be specified as a pair, naming both the potential and the claim it supports, and reported as such under the standard of Table~\ref{tab:report}.}

\revised{\revised{\begin{table}[htbp]\color{revred}
\centering
\footnotesize
\caption{\revised{Universal machine-learned interatomic potentials in current use as first-line physical prescreeners for generated crystals, ordered by combined performance score. Metrics are taken from the public Matbench Discovery leaderboard [\ref{cit:Riebesell2025b}]. CPS is the combined performance score; F1 scores stability classification on the WBM set; $\kappa_\text{SRME}$ is the symmetric relative mean error on thermal conductivity (lower is better, ceiling 2.0) and probes second- and third-order derivatives of the potential energy surface, not of energies. The final two rows show that a model can rank energies well and still be unusable for curvature.}}
\label{tab:mlip_models}
\setlength{\tabcolsep}{4pt}
\renewcommand{\arraystretch}{1.25}
\begin{tabularx}{\textwidth}{
  >{\raggedright\arraybackslash}p{0.17\textwidth}
  >{\raggedright\arraybackslash}p{0.26\textwidth}
  >{\raggedright\arraybackslash}p{0.15\textwidth}
  >{\raggedright\arraybackslash}Y}
\toprule
\textbf{Potential} & \textbf{Architecture and training data} & \textbf{CPS / F1 / $\kappa_\text{SRME}$} & \textbf{Prescreening role and limitation} \\
\midrule
PET-OAM-XL [\ref{cit:Bigi2026}] & Unconstrained point-edge transformer; OMat24, sAlex and MPtrj; 730M parameters & 0.898 / 0.924 / 0.119 & Current leaderboard leader and the best available surrogate for curvature-sensitive triage; cost per relaxation is the highest in the table \\
eSEN-30M-OAM [\ref{cit:BarrosoLuque2026}-\ref{cit:Fu2025}] & Smooth, energy-conserving equivariant network; OMat24 pre-training, MPtrj and sAlex fine-tuning; 30M parameters & 0.888 / 0.925 / 0.170 & Highest F1 in the table at a twenty-fifth of the parameters; a reasonable default for $E_\text{hull}$ triage at generative scale \\
Orb-v3 [\ref{cit:Rhodes2025}] & Graph-network simulator, conservative variant; MPtrj, Alexandria and OMat24 & 0.860 / 0.905 / 0.210 & Fastest of the top tier; the non-conservative variants are faster still but degrade the very derivatives phonons depend on \\
SevenNet-Omni [\ref{cit:Park2024}] & Equivariant message passing, multi-fidelity training & 0.873 / 0.906 / 0.192 & Parallel molecular-dynamics implementation; useful when the loop needs dynamics beyond single-point relaxation \\
GRACE-2L-OAM [\ref{cit:Lysogorskiy2026}] & Graph atomic cluster expansion; OMat24, sAlex and MPtrj & 0.865 / 0.883 / 0.169 & Systematically improvable basis; good accuracy-per-parameter and competitive curvature error \\
MACE-MPA-0 [\ref{cit:Batatia2022}, \ref{cit:Batatia2025}] & Higher-order equivariant message passing; MPtrj and sAlex & 0.795 / 0.852 / 0.412 & Mature tooling and the best-documented fine-tuning path; mid-table on accuracy \\
MatterSim v1 5M [\ref{cit:Yang2024}] & Equivariant network trained on a proprietary set spanning temperature and pressure & 0.767 / 0.862 / 0.575 & Broad thermodynamic coverage; the training set is not public, so distribution shift cannot be audited by the user \\
\midrule
\multicolumn{4}{@{}l}{\textit{Cautionary rows: good energies, unusable curvature}}\\
eqV2 M [\ref{cit:BarrosoLuque2026}] & EquiformerV2 with non-conservative (direct) force prediction; OMat24 and MPtrj & 0.558 / 0.917 / 1.771 & F1 within 0.01 of the leader yet a thermal-conductivity error near the ceiling: acceptable for hull ranking, unsafe for phonons or dynamics \\
M3GNet [\ref{cit:Chen2022}]; CHGNet [\ref{cit:Deng2023}] & First-generation universal potentials; MPF and MPtrj & 0.310 / 0.569 / 2.000; 0.343 / 0.613 / 2.000 & Still widely deployed and still useful as cheap filters, but both sit at the $\kappa_\text{SRME}$ ceiling; a dynamical-stability claim resting on either would be hard to defend \\
\bottomrule
\end{tabularx}
\end{table}}}

\revised{\begin{table}[htbp]\color{revred}
\centering
\footnotesize
\caption{\revised{The MLIP validation tier: what it delivers, how it fails, and the guardrail for each task. Every guardrail is reportable under the standard of Table~\ref{tab:report}.}}
\label{tab:mlip}
\setlength{\tabcolsep}{4pt}
\renewcommand{\arraystretch}{1.25}
\begin{tabularx}{\textwidth}{
  >{\raggedright\arraybackslash}p{0.17\textwidth}
  >{\raggedright\arraybackslash}p{0.24\textwidth}
  >{\raggedright\arraybackslash}Y
  >{\raggedright\arraybackslash}p{0.26\textwidth}}
\toprule
\textbf{Task} & \textbf{What universal MLIPs deliver} & \textbf{Failure mode (mechanism)} & \textbf{Operational guardrail} \\
\midrule
Relaxation / pre-screening & High-throughput relaxation; geometries within tens of meV/atom of DFT for near-database chemistries [\ref{cit:Riebesell2025}, \ref{cit:Batatia2022}, \ref{cit:Chen2022}-\ref{cit:Deng2023}]\revised{; current foundation models reach hull-distance MAEs near 20~meV/atom on the standard benchmark [\ref{cit:Riebesell2025b}-\ref{cit:Bigi2026}, \ref{cit:Mazitov2025}]} & Unphysical outputs relax into spurious minima; softening flattens the PES off-distribution [\ref{cit:Deng2025}] & DFT single-point spot checks across the accepted set, not only top scorers; report MLIP-DFT disagreement (Table~\ref{tab:report}, item 5) \\
$E_\text{hull}$ / stability ranking & Hull prescreening at generative scale; strongest prescreeners in prospective benchmarking [\ref{cit:Riebesell2025}] & Energy underestimation biases hull distances optimistic off-distribution, correlated with generator novelty & Chemistry-stratified error bars; every sub-threshold candidate DFT-verified before any stability claim (item 6) \\
Phonons / dynamical stability & Imaginary-mode screening at scale; harmonic phonons of near-equilibrium structures [\ref{cit:Gouvea2026}, \ref{cit:Loew2025}] & Curvature noisier than energy off-distribution; anharmonic observables much harder [\ref{cit:Pota2024}]. \revised{Energy ranking and curvature are measurably decoupled: on the leaderboard's thermal-conductivity task, models separated by less than 0.01 in F1 differ by more than an order of magnitude in $\kappa_\text{SRME}$ [\ref{cit:Riebesell2025b}]} & DFT phonon confirmation for discovery-claim candidates (item 7); state which property each verdict refers to. \revised{Never inherit a phonon verdict from a potential selected on F1 alone} \\
\revised{Forces and dynamics} & \revised{Conservative (gradient) forces give a well-defined PES; direct-force models are faster per step [\ref{cit:Rhodes2025}]} & \revised{Non-conservative forces are not exact energy gradients; the PES is rough near minima, so vibrational and transport observables degrade even when test-set errors look competitive [\ref{cit:Fu2025}]} & \revised{Use a conservative model for anything past single-point ranking; report which force convention produced each number} \\
Uncertainty estimation & Committee/ensemble variance; feature-space extrapolation grades & Single-model confidence miscalibrated under shift [\ref{cit:Ovadia2019}-\ref{cit:Guo2017}] & Route: accept / escalate to DFT single point / treat as unlabeled \\
Active-learning retraining & Fine-tuning on the loop's own DFT labels; foundation models make this cheap [\ref{cit:Batatia2025}] & Loop-selected labels biased by the loop (Section~\ref{sec:active_learning}) & Independent, chemically grouped audit sets; before/after pass rates at each ledger level \\
\bottomrule
\end{tabularx}
\end{table}}

\subsection{Metrics for Selection Under Uncertainty}\label{sec:metrics_uncertainty}

Inverse design is ultimately a question of which candidates are worth expensive follow-up. Because only a limited number of generated materials can be advanced to DFT validation or experiment, uncertainty is useful only if it improves those choices by reducing wasted evaluations and increasing the hit rate within a fixed budget [\ref{cit:Lookman2019}]. Benchmarking should therefore go beyond property-prediction accuracy and ask whether a method supports good decisions under realistic constraints. In particular, uncertainty estimates should remain meaningful under distribution shift and under optimization or guidance pressure, and performance should be reported as a function of validation budget, for example, as successful candidates per relaxation, DFT call, or experiment, alongside the best candidate found. The stability-synthesizability gap must also be stated explicitly. Convex-hull stability does not guarantee practical synthesis: kinetic barriers, competing phases, and narrow processing windows can still prevent realization [\ref{cit:Kononova2019}, \ref{cit:Olivetti2020}-\ref{cit:Venugopal2024}, \ref{cit:DeBreuck2025}]\revised{; Table~\ref{tab:synth} and Section~\ref{sec:synth} give concrete tools for closing this gap}.

\begin{figure}[t]
\centering
\begin{tikzpicture}[
  font=\sffamily, x=1cm, y=1cm,
  %% ---- module box styles --------------------------------------------------
  mod/.style    ={rounded corners=5pt, draw=black!35, line width=1.0pt,
                  fill=white, minimum width=5.0cm, minimum height=1.40cm,
                  align=center},
  modData/.style={mod, draw=cData!80!black,  fill=fillData},
  modGen/.style ={mod, draw=cModel!80!black, fill=fillModel},
  modLab/.style ={mod, draw=cLab!85!black,   fill=fillLab},
  modEval/.style={mod, draw=cEval!80!black,  fill=fillEval},
  %% ---- banner style -------------------------------------------------------
  banner/.style ={rounded corners=4pt, draw=black!28, line width=0.85pt,
                  fill=cMid, minimum width=13.5cm, minimum height=0.78cm,
                  align=center},
  %% ---- scheduler circle ---------------------------------------------------
  sched/.style  ={circle, draw=cEval!85!black, line width=1.80pt,
                  fill=fillEval, minimum size=2.80cm, align=center},
  %% ---- arrow styles (three visually distinct types) -----------------------
  %% dfArrow: data/results flowing TO the scheduler (blue, solid, thick)
  dfArrow/.style={-{Stealth[length=3.2mm,width=2.4mm]},
                  line width=1.60pt, color=cModel!100!black},
  %% ctArrow: control signals FROM the scheduler (orange, long-dash)
  ctArrow/.style={-{Stealth[length=3.0mm,width=2.2mm]},
                  line width=1.10pt, color=cLab!95!black,
                  dashed, dash pattern=on 9pt off 4pt},
  %% fbArrow: module-to-module feedback loop (dark grey, solid, thin)
  fbArrow/.style={-{Stealth[length=2.5mm,width=1.8mm]},
                  line width=0.95pt, color=black!52},
]

%% =========================================================================
%% BANNERS
%% =========================================================================
\node[banner] (topbanner) at (6.95, 6.10)
  {\small\textbf{Interface contracts:}~inputs $\cdot$ outputs $\cdot$
   uncertainty $\cdot$ provenance};
\node[banner] (botbanner) at (6.95, -0.82)
  {\small\textbf{Discovery reporting:}~pass rates $\cdot$ compute $\cdot$
   novelty/uniqueness $\cdot$ failures};

%% =========================================================================
%% MODULE NODES
%% =========================================================================
\node[modData] (enc)  at (2.50, 4.50)
  {\textbf{Multimodal}\\[-1pt]\textbf{encoder}\\[-2pt]{\small(retrieval)}};
\node[modGen]  (gen)  at (2.50, 2.20)
  {\textbf{Generator}\\[-2pt]{\small(proposal)}};
\node[modLab]  (char) at (11.40, 4.50)
  {\textbf{Characterization}\\[-2pt]{\small(measure)}};
\node[modLab]  (lab)  at (11.40, 2.20)
  {\textbf{Robotic lab}\\[-2pt]{\small(synthesis)}};
\node[modEval] (ver)  at (6.95, 0.40)
  {\textbf{Verifier}\\[-2pt]{\small(relaxation and stability)}};

%% =========================================================================
%% SCHEDULER (center)
%% =========================================================================
\node[sched] (sch) at (6.95, 3.40)
  {\textbf{Scheduler}\\[-2pt]{\small budget and}\\[-2pt]{\small stop rules}};

%% =========================================================================
%% BIDIRECTIONAL ARROWS
%% Arrows target the (sch) NODE so TikZ clips them at the circle boundary;
%% arrowheads sit ON the boundary, never inside the circle.
%% Using the same bend direction on both arrows in a pair separates them
%% automatically: the two arrows travel in opposite directions, so "bend L"
%% on a rightward arrow arcs UP, while "bend L" on the return leftward
%% arrow arcs DOWN — naturally opposite sides of the direct path.
%% Left-side modules use bend left; right-side and bottom use bend right.
%% Blue solid (thick)  = data/results flowing TO the scheduler.
%% Orange dash (thin)  = control signals issued FROM the scheduler.
%% =========================================================================

%% --- Multimodal encoder (upper-left)  <-->  Scheduler --------------------
\draw[dfArrow]
  ($(enc.east)+(0pt,+8pt)$)    to[bend left=22]   (sch);
\draw[ctArrow]
  (sch)                         to[bend left=22]   ($(enc.east)+(0pt,-8pt)$);

%% --- Generator (lower-left)  <-->  Scheduler -----------------------------
\draw[dfArrow]
  ($(gen.east)+(0pt,+8pt)$)    to[bend left=22]   (sch);
\draw[ctArrow]
  (sch)                         to[bend left=22]   ($(gen.east)+(0pt,-8pt)$);

%% --- Characterization (upper-right)  <-->  Scheduler ---------------------
\draw[dfArrow]
  ($(char.west)+(0pt,+8pt)$)   to[bend right=22]  (sch);
\draw[ctArrow]
  (sch)                         to[bend right=22]  ($(char.west)+(0pt,-8pt)$);

%% --- Robotic lab (lower-right)  <-->  Scheduler --------------------------
\draw[dfArrow]
  ($(lab.west)+(0pt,+8pt)$)    to[bend right=22]  (sch);
\draw[ctArrow]
  (sch)                         to[bend right=22]  ($(lab.west)+(0pt,-8pt)$);

%% --- Verifier (bottom)  <-->  Scheduler ----------------------------------
\draw[dfArrow]
  ($(ver.north)+(+8pt,0pt)$)   to[bend right=22]  (sch);
\draw[ctArrow]
  (sch)                         to[bend right=22]  ($(ver.north)+(-8pt,0pt)$);

%% =========================================================================
%% MODULE-TO-MODULE FEEDBACK LOOP  (lab -> char -> enc -> gen)
%% =========================================================================
\draw[fbArrow] (lab.north)  -- (char.south);
\draw[fbArrow] (char.west)  to[out=150, in=30] (enc.east);
\draw[fbArrow] (enc.south)  -- (gen.north);

%% =========================================================================
%% BANNER CONNECTORS
%% =========================================================================
\draw[fbArrow] (enc.north)  -- ($(topbanner.south west)+(1.0,0)$);
\draw[fbArrow] (char.north) -- ($(topbanner.south east)+(-1.0,0)$);
\draw[fbArrow] (ver.south)  -- (botbanner.north);

%% =========================================================================
%% LEGEND
%% =========================================================================
\begin{scope}[yshift=-1.45cm]
  %% entry 1 -- data flow
  \draw[dfArrow] (0.50, -0.32) -- (1.35, -0.32);
  \node[anchor=west, font=\footnotesize\sffamily, text=black!70]
    at (1.42,-0.32) {Data to scheduler};
  %% entry 2 -- control
  \draw[ctArrow] (5.00, -0.32) -- (5.85, -0.32);
  \node[anchor=west, font=\footnotesize\sffamily, text=black!70]
    at (5.92,-0.32) {Control from scheduler};
  %% entry 3 -- feedback
  \draw[fbArrow] (10.60,-0.32) -- (11.45,-0.32);
  \node[anchor=west, font=\footnotesize\sffamily, text=black!70]
    at (11.52,-0.32) {Feedback};
\end{scope}

\end{tikzpicture}
\caption{\textbf{The self-driving laboratory as an orchestrated closed-loop system for inverse materials design.}
The architecture comprises five functional modules coordinated by a central \textit{Scheduler}.
\textit{(i) Multimodal encoder}: maps evidence from multiple modalities into a shared representation space to enable cross-modal retrieval of relevant precedents.
\textit{(ii) Generator}: proposes new candidates based on design intent and retrieved context.
\textit{(iii) Verifier}: performs rapid structural relaxation and thermodynamical stability screening to filter candidates prior to costly experimental validation.
\textit{(iv) Robotic laboratory}: executes automated synthesis protocols.
\textit{(v) Characterization}: measures properties and returns experimental observations.
The central \textit{Scheduler} manages the loop by allocating the validation budget, applying stop rules, and determining the next optimal action across all modules.
Solid blue arrows indicate data and results flowing to the scheduler; dashed orange arrows indicate control signals issued from the scheduler; dark grey arrows indicate the internal module-to-module feedback loop.}

\label{fig:sdl_orchestrated_system}
\end{figure}
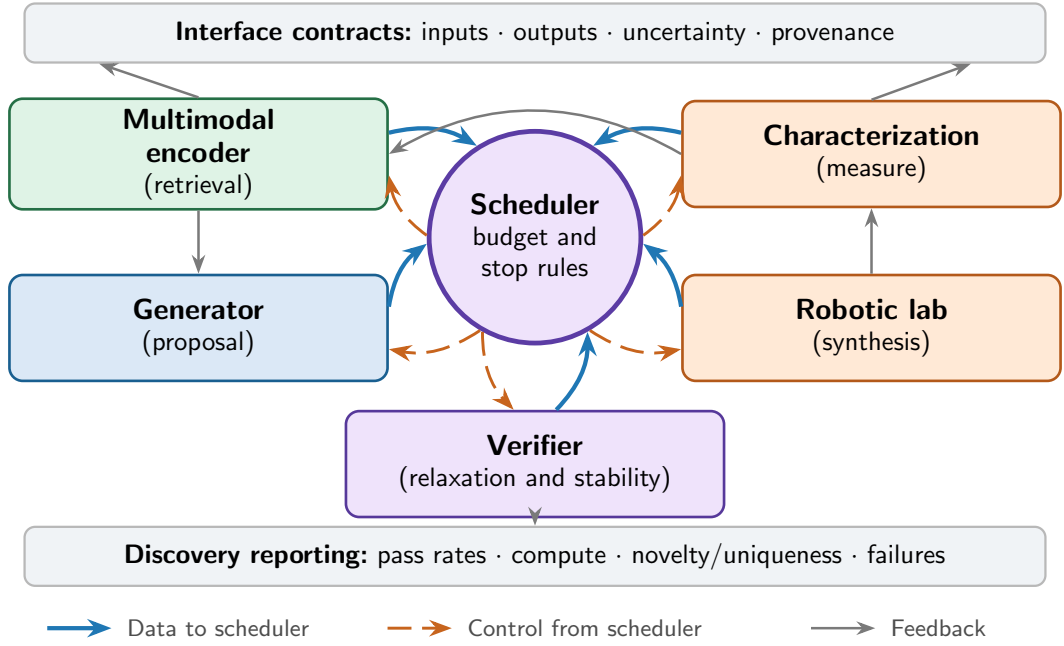

Figure~\ref{fig:sdl_orchestrated_system} illustrates a concrete orchestration architecture that realises
these requirements: each module exposes explicit interface contracts, and a central scheduler coordinates
the full proposal-to-validation loop under explicit budget constraints.

\subsection{Bridging to Experiment: Synthesizability and Multimodal Design Interfaces}
\label{sec:synth}

As materials proposal frameworks become increasingly effective at generating large numbers of DFT-plausible candidates, the central bottleneck shifts from \emph{proposal} to \emph{execution}. The harder task is to convert promising in-silico suggestions into validated outcomes, to identify candidates that can be synthesized, processed, and evaluated within realistic experimental constraints. For this reason, synthesis and processing considerations should not remain informal afterthoughts; they need to be incorporated explicitly into model inputs, feasibility criteria, and decision-making pipelines [\ref{cit:Kononova2019}, \ref{cit:Olivetti2020}-\ref{cit:Venugopal2024}, \ref{cit:DeBreuck2025}, \ref{cit:Davies2019}].

Three near-term priorities follow from this shift. First, synthesis and process constraints should be incorporated directly into inverse-design workflows, both as conditioning signals during generation and as feasibility filters during candidate screening. Second, text-mined synthesis knowledge, literature-derived procedural information, and experimental metadata should be integrated into multimodal representations so that models can reason jointly over structure, properties, and process context. Third, benchmarking protocols should move beyond energetic plausibility alone and, where data permit, assess whether generated candidates are compatible with empirically observed synthesis outcomes [\ref{cit:Kononova2019}, \ref{cit:DeBreuck2025}].

\subsubsection*{\revised{Practical routes for assessing synthesizability}}
\revised{Several concrete tools already exist for turning these priorities into practice, and Table~\ref{tab:synth} organizes them by the question each answers, because they are complementary and not interchangeable. Two limitations should be stated explicitly. First, text-mined corpora record procedures (what was heated, for how long, and in which atmosphere) and not reaction pathways, so they cannot by themselves supply a thermodynamic driving force or a kinetic barrier. Their proper role is precedent retrieval and prior construction, and not proof of realizability [\ref{cit:Kononova2019}, \ref{cit:Olivetti2020}-\ref{cit:Venugopal2024}, \ref{cit:He2023}]. Second, the toolboxes answer different questions. Computational thermodynamics and kinetics address whether a compound can exist and through which reactions it could form, using competing-phase and convex-hull analysis [\ref{cit:Sun2016}-\ref{cit:Aykol2018}], reaction-energy and precursor-pathway calculations [\ref{cit:McDermott2021}], Pourbaix and electrochemical-stability analysis [\ref{cit:Singh2017}], and activation-barrier estimates along candidate reaction coordinates, increasingly affordable through MLIP-accelerated transition-state searches [\ref{cit:Batatia2022}, \ref{cit:Chen2022}-\ref{cit:Deng2023}, \ref{cit:Henkelman2000}] (although saddle points are the configurations furthest from MLIP training data, where softening [\ref{cit:Deng2025}] is most severe). Machine-learned synthesizability scores add a third, data-driven estimate of realizability, trained by positive-unlabeled learning on the corpus of known synthesized crystals at the structure [\ref{cit:Jang2020}] or composition [\ref{cit:Antoniuk2023}] level. These enter the closed loop either as a screening gate in the staged ledger (Table~\ref{tab:report}) or as a conditioning signal that biases generation toward synthesizable regions from the outset, or as an additional factor in a multi-objective acquisition alongside the property target and the feasibility term of Eq.~\ref{eq:constrained_acquisition}.}

\revised{For the thermodynamic question, empirical calibration matters as much as the tool. Among roughly 30{,}000 experimentally observed inorganic phases, half ($50.5 \pm 4\%$) are metastable, but the observed metastability is tightly bounded (median $15$~meV/atom, 90th percentile near $67$~meV/atom) and strongly chemistry-dependent, scaling with cohesive energy (observed metastable nitrides show a median of $63$ and a 90th percentile of $195$~meV/atom) [\ref{cit:Sun2016}-\ref{cit:Aykol2018}]. A universal $0.1$~eV/atom cutoff is therefore simultaneously too strict for strongly bonded chemistries and too lenient for weakly bonded ones; hull thresholds should be calibrated per chemistry against the observed distribution. For a shortlisted candidate, a minimal realizability screen can then be run before any laboratory commitment: (i) judge $E_{\text{hull}}$ against the chemistry-calibrated metastability distribution in place of a universal cutoff [\ref{cit:Sun2016}-\ref{cit:Aykol2018}]; (ii) enumerate competing phases and reaction energies, flagging thermodynamic sinks that would trap the synthesis [\ref{cit:McDermott2021}]; (iii) for aqueous or electrochemical applications, compute Pourbaix stability across the operating window [\ref{cit:Singh2017}]; (iv) where a mechanism can be hypothesized, estimate the activation barrier with MLIP-accelerated transition-state searches, confirming the saddle at DFT level [\ref{cit:Batatia2022}, \ref{cit:Chen2022}-\ref{cit:Deng2023}, \ref{cit:Deng2025}, \ref{cit:Henkelman2000}]; and (v) retrieve nearest synthesized analogues and their recipes to anchor a processing window [\ref{cit:Kononova2019}, \ref{cit:He2023}]. A candidate surviving all five enters the laboratory as a graded hypothesis with a dossier, not a bare structure file.}

\subsubsection*{\revised{How large the computation-to-experiment gap actually is}}
\revised{Most current inverse-design workflows terminate at DFT or MLIP validation and not at experimental realization, and the numbers make the gap concrete. A single large-scale screening campaign reported on the order of $381{,}000$ new computationally stable structures [\ref{cit:Merchant2023}]; the most advanced autonomous-laboratory campaign to date attempted $58$ targets and realized $41$ over $17$ days [\ref{cit:Szymanski2023}]; and flagship generative models have so far reported experimental validation of individual candidates: MatterGen synthesized one generated oxide and measured its target property to within $20\%$ of the design goal [\ref{cit:Zeni2025}]. The ratio of computational stability claims to experimental attempts thus stands at four to five orders of magnitude. The corpus problem compounds this gap. Text-mined synthesis datasets record almost exclusively successful procedures [\ref{cit:Kononova2019}, \ref{cit:He2023}], so every synthesizability model trained on them inherits a survivorship bias. \revised{That the missing half carries information is not a conjecture: Raccuglia et al.\ recovered failed and partially successful hydrothermal syntheses from laboratory notebooks and trained on them a model that predicted reaction outcomes for new organically templated inorganic products more accurately than human intuition [\ref{cit:Raccuglia2016}].} Reporting and sharing failed syntheses (Section~\ref{sec:missing_data}) is therefore both good practice and a source of training data that these models currently lack.}

\revised{\begin{table}[htbp]\color{revred}
\centering
\footnotesize
\caption{\revised{Practical routes for incorporating synthesizability into inverse-design workflows, grouped by the kind of evidence each tool provides. Each row lists one line of work, the signal it provides, where it enters the closed loop, and its key limitation.}}
\label{tab:synth}
\setlength{\tabcolsep}{4pt}
\renewcommand{\arraystretch}{1.25}
\begin{tabularx}{\textwidth}{
  >{\raggedright\arraybackslash}p{0.16\textwidth}
  >{\raggedright\arraybackslash}p{0.13\textwidth}
  >{\raggedright\arraybackslash}Y
  >{\raggedright\arraybackslash}p{0.22\textwidth}
  >{\raggedright\arraybackslash}p{0.19\textwidth}}
\toprule
\textbf{Signal} & \textbf{Example work} & \textbf{What it provides} & \textbf{How to adopt in the loop} & \textbf{Key limitation} \\
\midrule
\multicolumn{5}{@{}l}{\textit{Literature-derived}}\\
Text-mined recipes & Kononova [\ref{cit:Kononova2019}] & Structured solid-state procedures from the literature & Training corpus; feasibility prior at screening & Success-only; no thermo/kinetic grounding \\
Precursor recommendation & He [\ref{cit:He2023}] & Data-driven precursor selection via materials similarity & Ranking synthesis routes before lab attempts & Restricted to chemistries with precedent \\
\multicolumn{5}{@{}l}{\textit{Data-driven realizability estimate}}\\
Synthesizability score (structure) & Jang [\ref{cit:Jang2020}] & PU-learned probability a structure is synthesizable & Gate in the ledger (Table~\ref{tab:report}); rank before DFT & Trained on successes; degrades OOD \\
Synthesizability score (composition) & Antoniuk [\ref{cit:Antoniuk2023}] & Composition-level likelihood, no structure needed & Cheap pre-filter; conditioning signal at generation & No structural resolution; database bias \\
\multicolumn{5}{@{}l}{\textit{Computational thermodynamics and kinetics}}\\
Metastability bounds & Sun [\ref{cit:Sun2016}]; Aykol [\ref{cit:Aykol2018}] & Chemistry-dependent realizable $E_\text{hull}$; amorphous limit & Calibrate hull thresholds per chemistry & Thermodynamic proxy; no explicit kinetics \\
Reaction-network analysis & McDermott [\ref{cit:McDermott2021}] & Precursor pathways and competing intermediate phases & Kinetic-plausibility check for top candidates & Needs reliable free energies for all phases \\
Activation barriers & Henkelman [\ref{cit:Henkelman2000}]; MLIP-accelerated [\ref{cit:Batatia2022}, \ref{cit:Chen2022}-\ref{cit:Deng2023}] & Reaction-coordinate barriers for formation\slash decomposition & Kinetic go/no-go before synthesis & MLIP accuracy weakest at saddle points [\ref{cit:Deng2025}] \\
Pourbaix\slash electrochemical & Singh [\ref{cit:Singh2017}] & Aqueous\slash electrochemical stability windows & Operating-condition filter (e.g.\ (photo)electrocatalysis) & Medium-specific; sensitive to reference states \\
\multicolumn{5}{@{}l}{\textit{Experimental feedback}}\\
Autonomous synthesis & CAMEO\slash Kusne [\ref{cit:Kusne2020}]; A-Lab\slash Szymanski [\ref{cit:Szymanski2023}] & Success/failure labels at scale from closed-loop labs & Feedback stage: failures retrain synthesizability models & Expensive; failure attribution often ambiguous \\
\bottomrule
\end{tabularx}
\end{table}}

In this context, multimodal learning is emerging as a design interface: in practice, intent is often expressed through partial characterization data, spectra, microscopy, processing context, prior literature, and textual constraints. Multimodal models are therefore well positioned to align these heterogeneous inputs, retrieve relevant precedents, and propose candidates or operating regions that are more likely to be experimentally feasible. A realistic near-term strategy is therefore hybrid control, in which multimodal models support constraint specification, candidate triage, and next-step selection under calibrated uncertainty, well short of unrestricted end-to-end text-to-material generation [\ref{cit:Park2025}, \ref{cit:Wu2025}, \ref{cit:Olivetti2020}-\ref{cit:Venugopal2024}, \ref{cit:PyzerKnapp2025}].

\subsection{Connecting to Self-Driving Laboratories: Deployment Constraints Become Design Constraints}\label{sec:sdl}

Inverse design changes in a fundamental way once it is embedded in a self-driving laboratory. In a purely computational workflow, large numbers of candidates can be explored quickly and realism can be deferred to a final validation stage. In an autonomous experimental loop, by contrast, the pace and reliability of discovery are governed by the physical system itself. Instrument availability, queueing delays, calibration drift, batch-to-batch variability, failed syntheses, and heterogeneous measurement noise all shape what it actually means to optimize a candidate. Under these conditions, feasibility is not only a question of structure or thermodynamics; it also depends on whether a proposal can be executed robustly, at reasonable cost, and with reproducible outcomes.

For this reason, the self-driving-laboratory literature increasingly treats automation, robustness, and reproducibility as core algorithmic constraints in closed-loop materials discovery, and no longer as peripheral engineering concerns [\ref{cit:Stach2021}-\ref{cit:FloresLeonar2020}, \ref{cit:Fudge2024}-\ref{cit:Rank2024}]. Benchmarking studies likewise show that credible progress requires reporting what the laboratory actually experiences: throughput, validation budget, failure and retry rates, calibration and quality-control procedures, and evaluation protocols that reflect real operating conditions in place of idealized assumptions [\ref{cit:Wagner2026}]. These considerations carry directly into generative inverse design, because the practical value of an optimizer is ultimately measured by how efficiently it converts proposals into \emph{validated} materials under real-world constraints, and never by how many candidates it can produce.\cut{ Figure~\ref{fig:sdl_orchestrated_system} illustrates a concrete orchestration architecture that captures these requirements.}

\revised{Figure~\ref{fig:sdl_orchestrated_system} is a schematic, and its modules do not operate as cleanly as the diagram suggests. A working self-driving laboratory is a noisy, drifting, resource-constrained measurement instrument with its own error model, far from an oracle returning ground truth, and a loop that does not model the instrument will attribute the instrument's failures to the candidates. Six idealizations that computational treatments make silently, and the design consequence each imposes, are summarized in Table~\ref{tab:sdl}. The highest-stakes challenge is characterization ambiguity, as it gates rung~L7 of the credibility ladder. Automated phase identification from powder diffraction is inherently underdetermined: compositional disorder, preferred orientation, amorphous fractions, and multiphase mixtures can easily defeat automated refinement. Indeed, the published reanalysis of the A-Lab campaign [\ref{cit:Leeman2024}] demonstrates how much claimed synthesis success hinges on this single step. Phase assignments should therefore carry confidence grades, raw patterns should be archived for independent reanalysis, and an L7 claim is on firmer ground when the characterization has been reviewed at a confidence the automated pipeline alone cannot yet provide.}

\revised{\begin{table}[htbp]\color{revred}
\centering
\footnotesize
\caption{\revised{Six idealizations in closed-loop laboratory treatments, the corresponding reality, and the design consequence each imposes on the loop.}}
\label{tab:sdl}
\setlength{\tabcolsep}{4pt}
\renewcommand{\arraystretch}{1.25}
\begin{tabularx}{\textwidth}{
  >{\raggedright\arraybackslash}p{0.24\textwidth}
  >{\raggedright\arraybackslash}Y
  >{\raggedright\arraybackslash}Y}
\toprule
\textbf{Idealized assumption} & \textbf{Laboratory reality} & \textbf{Design consequence for the loop} \\
\midrule
Every attempted synthesis returns a usable label & Failures go unlogged or unattributed; success-only corpora inherit survivorship bias (Section~\ref{sec:synth}) & Structured failure logging with an attribution taxonomy (target infeasible / recipe wrong / execution fault / characterization inconclusive) [\ref{cit:Kononova2019}, \ref{cit:Wagner2026}, \ref{cit:Szymanski2023}] \\
Synthesizability is a binary property of the target & Outcomes depend on narrow process windows; metastable targets need kinetically routed recipes & Report process windows, not success bits; acquire over the joint space of candidates and conditions [\ref{cit:Szymanski2023}, \ref{cit:Kusne2020}] \\
Any candidate can be attempted & Precursor inventory, purity, cost, and air sensitivity bound the reachable set & Inventory-aware acquisition: the feasibility term (Eq.~\ref{eq:constrained_acquisition}) includes precursor availability and cost \\
Samples are clean and independent & Crucible reactions, cross-contamination in shared robotics, air/moisture exposure & Blanks and controls in batch design; contamination flags in metadata [\ref{cit:Fudge2024}-\ref{cit:Rank2024}] \\
The instrument is stationary & Calibration drift, precursor aging, ambient variation, giving a nonstationary oracle that breaks the constant-noise assumption of Eq.~\ref{eq:observation_model} & Periodic reference samples and control charts; drift-triggered recalibration; time-stamped provenance [\ref{cit:Wagner2026}] \\
Characterization returns ground truth & Automated powder-XRD phase ID is underdetermined: disorder, preferred orientation, amorphous fractions, multiphase mixtures [\ref{cit:Leeman2024}] & Confidence-graded phase assignment; raw patterns archived; human review before any L7 claim [\ref{cit:Leeman2024}] \\
\bottomrule
\end{tabularx}
\end{table}}

\subsection{Toward Recursive Self-Improvement in Materials Discovery}\label{sec:recursive}

The longer-term opportunity is to make the loop between model and experiment progressively more informative over time. In this view, an inverse-design system does not only generate candidates; it also updates its own search strategy as new experimental outcomes, failure cases, and process knowledge accumulate. Each cycle can then improve multiple components at once: the proposal mechanism becomes better calibrated, surrogate models become more reliable in relevant regions of chemical space, uncertainty estimates become more informative for decision-making, and the system develops a sharper understanding of which synthesis routes and operating windows are realistically accessible.

This recursive self-improvement perspective is increasingly central to the state of the art in autonomous materials discovery. The goal has widened from one-shot inverse prediction to an adaptive discovery system that learns from success, failure, and ambiguity in a continuous manner. In such a framework, generated candidates become hypotheses, validation outcomes become training signals, and the laboratory itself becomes a source of structured feedback for improving future decisions. The key challenge is therefore to design loops that genuinely improve as they run: they must preserve diversity, remain robust to model bias and distribution shift, and allocate experimental effort in ways that maximize both immediate progress and long-term learning.

From this perspective, the most compelling inverse-design systems will be those that couple generative modeling, multimodal reasoning, uncertainty-aware selection, and experimental feedback into a unified improvement cycle. Here inverse design, multimodal modeling, and self-driving laboratories begin to converge into a single discovery paradigm.
% ==========================================================
\section{Outlook}\label{sec:outlook}
Inverse materials design is moving beyond proof-of-concept model demonstrations toward deployable discovery systems. At this stage, progress is increasingly determined by how proposal models, multimodal constraints, uncertainty-aware selection, and verification pipelines function together under finite computational and experimental budgets, and only marginally by any single architectural advance. This shift also makes several long-standing scientific challenges impossible to treat as secondary, including faithful representation of constraints, reliable uncertainty estimation under distribution shift, realistic treatment of validation cost, and actionable connections between generated candidates and feasible synthesis routes [\ref{cit:FloresLeonar2020}-\ref{cit:Wagner2026}].
A generated structure should therefore be regarded as a hypothesis awaiting test. Converting such hypotheses into reliable advances requires a pipeline-level perspective in which each component is designed for closed-loop operation (Sections~\ref{sec:closedloop}-\ref{sec:failure_modes}). In effective workflows, candidates progress through a transparent hierarchy of validation, beginning with rapid feasibility checks, followed by structural relaxation and stability screening, and, where possible, experimental verification. Reporting pass rates across these stages is essential, because it makes systems more comparable and reveals where they fail under realistic operating conditions [\ref{cit:Wagner2026}]. In the same spirit, strong inverse-design systems should be judged by how efficiently they translate limited validation resources into validated hits, and not by the sheer number of candidates generated. Cumulative improvement requires explicit selection strategies that allocate computational and experimental effort on the basis of calibrated uncertainty and expected information gain [\ref{cit:Wagner2026}]. Equally important, failure should not be discarded as noise. Invalid structures, unstable relaxations, and near-miss candidates can all provide useful supervision for updating proposal policies, recalibrating surrogate models, and refining feasibility filters, thereby improving subsequent rounds of search [\ref{cit:Wagner2026}].
Because inverse design systems actively exploit weaknesses in metrics and surrogates, \emph{new materials discovery} claims
should be supported by (i) staged verification ledgers, (ii) robustness checks under distribution shift and label noise,
(iii) explicit accounting of validation budgets, and (iv) reporting practices that preserve negative results and failure
modes as part of the scientific record [\ref{cit:Wagner2026}-\ref{cit:Dunn2020}].

\subsection{Missing Data, Missing Modalities, and Negative Results}\label{sec:missing_data}

Progress in multimodal inverse design is currently limited as much by data availability as by model architectures. Many modalities that would be most useful for modeling: synthesis conditions, process-property relationships, experimental failure outcomes, and microstructure under realistic conditions, are sparsely available or systematically underreported. Negative results, such as failed synthesis attempts and unstable predictions, carry high information value yet are seldom shared. \revised{This is not merely a matter of good practice. Raccuglia~\textit{et al.}~[\ref{cit:Raccuglia2016}] recovered failed and partially successful hydrothermal syntheses from laboratory notebooks and trained on them a model that predicted reaction outcomes for new organically templated inorganic products more accurately than human intuition, so the unreported half of the record demonstrably carries signal. Because text-mined corpora record almost exclusively successful procedures [\ref{cit:Kononova2019}, \ref{cit:He2023}], every synthesizability model trained on them inherits a survivorship bias that no quantity of additional positive data will correct (Section~\ref{sec:synth}). Reporting failures under a structured attribution — target infeasible, recipe wrong, execution fault, characterization inconclusive (Table~\ref{tab:sdl}) — would convert them into the training data these models currently lack, and belongs in the reporting standard of Table~\ref{tab:report} alongside pass rates.} The field would benefit from curated datasets that pair computational predictions with experimental outcomes, benchmarks that explicitly reward the discovery of genuinely novel materials over rediscovery of known stable phases, and reporting standards that require disclosure of validation budgets and failure rates alongside positive results [\ref{cit:DeBreuck2025}-\ref{cit:Dunn2020}].

\subsection{Toward Experimentally Grounded Multimodal Inverse Design}\label{sec:exp_grounded}

The deepest open challenge is closing the gap between computational plausibility and experimental realization. Current pipelines optimize for properties predictable from structure, but the true bottleneck for materials discovery lies in whether a proposed structure can be synthesized, characterized, and deployed under practical conditions. Multimodal learning offers a path toward experimentally grounded inverse design: by incorporating synthesis metadata, processing conditions, and characterization data as explicit conditioning signals, future systems can propose structures that are synthesizable as well as stable. Realizing this vision will require tighter integration of laboratory automation, structured experimental metadata, and uncertainty-aware selection strategies that account for the full cost of physical validation.

\subsection{Reproducibility and Reporting Standards}\label{sec:repro}

As inverse design systems grow increasingly complex, reproducibility has emerged as a first-order scientific concern. Reported metrics (such as validity, novelty, stability) are sensitive to matching criteria, database versions, relaxation protocols, and evaluation budgets; subtle variations in these parameters can induce significant discrepancies in performance. Consequently, the community should adopt standardized reporting protocols that encompass: the full staged pass-rate ledger, explicit matching criteria and tolerances, validation budgets measured in DFT calls or wall-clock time per hit, and the survival fraction of candidates through each pipeline stage [\ref{cit:Wagner2026}-\ref{cit:Dunn2020}]. \revised{These items are precisely the discovery-credibility ladder of Section~\ref{sec:ladder} (Table~\ref{tab:ladder}) and the minimum reporting standard of Section~\ref{sec:ledger} (Table~\ref{tab:report}), read as a reproducibility and audit requirement: a claim is reproducible only when the rung it reaches, and the protocol under which it reached it, are both stated.}

\section*{Data and code availability}
No new datasets or code were generated for this review.

\section*{Acknowledgments}
This study was financed by Université catholique de Louvain (Ref. No.: ARH/MKK/01155003).

\section*{Author contributions}
A.B. conceived the review, performed the literature analysis, developed the manuscript structure, and wrote the initial draft. R.A.G. contributed to the literature analysis, scientific framing, and manuscript revision. G.M.R. supervised the work, contributed to the conceptual framing, and revised the manuscript. All authors discussed the content and approved the final manuscript.

\section*{Declaration of interests}
The authors declare no competing interests.

\section*{References}

\begingroup
\small
\sloppy
\begin{enumerate}[leftmargin=2.7em, labelsep=0.6em, align=left]
\item \label{cit:Curtarolo2013} Curtarolo, S. et al. The high-throughput highway to computational materials design. \textit{Nature Materials} \textbf{12}, 191-201 (2013). doi:10.1038/nmat3568. \url{https://doi.org/10.1038/nmat3568}
\item \label{cit:Jain2013} Jain, A. et al. Commentary: The Materials Project: A materials genome approach to accelerating materials innovation. \textit{APL Materials} \textbf{1}, 011002 (2013). doi:10.1063/1.4812323. \url{https://doi.org/10.1063/1.4812323}
\item \label{cit:Ward2018} Ward, L. et al. Matminer: An open source toolkit for materials data mining. \textit{Computational Materials Science} \textbf{152}, 60-69 (2018). doi:10.1016/j.commatsci.2018.05.018. \url{https://doi.org/10.1016/j.commatsci.2018.05.018}
\item \label{cit:Choudhary2020} Choudhary, K. et al. The joint automated repository for various integrated simulations (JARVIS) for data-driven materials design. \textit{npj Computational Materials} \textbf{6}, 173 (2020). doi:10.1038/s41524-020-00440-1. \url{https://doi.org/10.1038/s41524-020-00440-1}
\item \label{cit:Kirklin2015} Kirklin, S. et al. The Open Quantum Materials Database (OQMD): assessing the accuracy of DFT formation energies. \textit{npj Computational Materials} \textbf{1}, 15010 (2015). doi:10.1038/npjcompumats.2015.10. \url{https://doi.org/10.1038/npjcompumats.2015.10}
\item \label{cit:Saal2013} Saal, J. E. et al. Materials design and discovery with high-throughput density functional theory: The Open Quantum Materials Database (OQMD). \textit{JOM} \textbf{65}, 1501-1509 (2013). doi:10.1007/s11837-013-0755-4. \url{https://doi.org/10.1007/s11837-013-0755-4}
\item \label{cit:Ong2013} Ong, S. P. et al. Python Materials Genomics (pymatgen): A robust, open-source Python library for materials analysis. \textit{Computational Materials Science} \textbf{68}, 314-319 (2013). doi:10.1016/j.commatsci.2012.10.028. \url{https://doi.org/10.1016/j.commatsci.2012.10.028}
\item \label{cit:Lookman2019} Lookman, T., Balachandran, P. V., Xue, D. and Yuan, R. Active learning in materials science with emphasis on adaptive sampling using uncertainties for targeted design. \textit{npj Computational Materials} \textbf{5}, 21 (2019). doi:10.1038/s41524-019-0153-8. \url{https://doi.org/10.1038/s41524-019-0153-8}
\item \label{cit:Shahriari2016} Shahriari, B., Swersky, K., Wang, Z., Adams, R. P. and de Freitas, N. Taking the human out of the loop: A review of Bayesian optimization. \textit{Proceedings of the IEEE} \textbf{104}, 148-175 (2016). doi:10.1109/JPROC.2015.2494218. \url{https://doi.org/10.1109/JPROC.2015.2494218}
\item \label{cit:Snoek2012} Snoek, J., Larochelle, H. and Adams, R. P. Practical Bayesian optimization of machine learning algorithms. \textit{Advances in Neural Information Processing Systems} \textbf{25} (2012). \url{https://arxiv.org/abs/1206.2944}
\item \label{cit:Frazier2018} Frazier, P. I. A tutorial on Bayesian optimization. \textit{arXiv} (2018). arXiv:1807.02811. \url{https://arxiv.org/abs/1807.02811}
\item \label{cit:Settles2009} Settles, B. Active Learning Literature Survey. University of Wisconsin-Madison, Computer Sciences Technical Report 1648 (2009). \url{https://minds.wisconsin.edu/handle/1793/60660}
\item \label{cit:Hase2018} H\"ase, F., Roch, L. M. and Aspuru-Guzik, A. Phoenics: A Bayesian optimizer for chemistry. \textit{ACS Central Science} \textbf{4}, 1134-1145 (2018). doi:10.1021/acscentsci.8b00307. \url{https://doi.org/10.1021/acscentsci.8b00307}
\item \label{cit:Hase2021} H\"ase, F., Roch, L. M. and Aspuru-Guzik, A. Gryffin: An algorithm for Bayesian optimization of categorical variables informed by physical intuition with applications to chemistry. \textit{ACS Central Science} \textbf{7}, 1230-1237 (2021). doi:10.1021/acscentsci.0c01520. \url{https://doi.org/10.1021/acscentsci.0c01520}
\item \label{cit:Luo2024} Luo, X. et al. Deep learning generative model for crystal structure prediction. \textit{npj Computational Materials} \textbf{10}, 254 (2024). doi:10.1038/s41524-024-01443-y. \url{https://doi.org/10.1038/s41524-024-01443-y}
\item \label{cit:Xie2021} Xie, T. et al. Crystal Diffusion Variational Autoencoder for periodic material generation. arXiv:2110.06197 (2021). \url{https://arxiv.org/abs/2110.06197}
\item \label{cit:Jiao2023} Jiao, R. et al. Crystal Structure Prediction by Joint Equivariant Diffusion. arXiv:\allowbreak2309.04475 (2023). \url{https://arxiv.org/abs/2309.04475}
\item \label{cit:Zhao2023} Zhao, Y. et al. Probabilistic constrained graph variational autoencoders for crystal generation. \textit{npj Computational Materials} \textbf{9}, 30 (2023). doi:10.1038/s41524-023-00987-9. \url{https://doi.org/10.1038/s41524-023-00987-9}
\item \label{cit:Qiu2025} Qiu, J. et al. VQCrystal: A vector-quantized diffusion framework for crystal generation. \textit{npj Computational Materials} \textbf{11}, 63 (2025). doi:10.1038/s41524-025-01613-6. \url{https://doi.org/10.1038/s41524-025-01613-6}
\item \label{cit:Zeni2025} Zeni, C. et al. A generative model for inorganic materials design (MatterGen). \textit{Nature} (2025). \url{https://www.nature.com/articles/s41586-025-08628-5}
\item \label{cit:Luo2025} Luo, X. et al. CrystalFlow: a flow-based generative model for crystal structure prediction and materials discovery. \textit{Nature Communications} (2025). \url{https://www.nature.com/articles/s41467-025-64364-4}
\item \label{cit:Park2025} Park, H., Onwuli, A. and Walsh, A. Exploration of crystal chemical space using text-guided generative artificial intelligence (Chemeleon). \textit{Nature Communications} \textbf{16} (2025). doi:10.1038/s41467-025-59636-y. \url{https://www.nature.com/articles/s41467-025-59636-y}
\item \label{cit:Ren2022} Ren, Z. et al. An invertible crystallographic representation for general inverse design of inorganic crystals with targeted properties. \textit{Matter} \textbf{5}, 314-335 (2022). doi:10.1016/j.matt.2021.11.032. \url{https://doi.org/10.1016/j.matt.2021.11.032}
\item \label{cit:Ho2020} Ho, J., Jain, A. and Abbeel, P. Denoising diffusion probabilistic models. \textit{arXiv} (2020). arXiv:2006.11239. \url{https://arxiv.org/abs/2006.11239}
\item \label{cit:Song2020} Song, Y., Sohl-Dickstein, J., Kingma, D. P., Kumar, A., Ermon, S. and Poole, B. Score-based generative modeling through stochastic differential equations. \textit{arXiv} (2020). arXiv:2011.13456. \url{https://arxiv.org/abs/2011.13456}
\item \label{cit:Wu2025} Wu, Y. et al. A versatile multimodal learning framework bridging multiscale knowledge for material design. \textit{npj Computational Materials} \textbf{11}, 276 (2025). doi:10.1038/s41524-025-01767-3. \url{https://doi.org/10.1038/s41524-025-01767-3}
\item \label{cit:Moro2025} Moro, V. et al. Multimodal foundation models for material property prediction and discovery. \textit{Newton} \textbf{1}, 100016 (2025). doi:10.1016/j.newton.2025.100016. \url{https://doi.org/10.1016/j.newton.2025.100016}
\item \label{cit:Babu2026} Babu, A., Gouv\^ea, R. A., Vandergheynst, P. and Rignanese, G.-M. MEIDNet: multimodal generative AI framework for inverse materials design. \textit{npj Computational Materials} (2026). doi:10.1038/s41524-026-02153-3. \url{https://doi.org/10.1038/s41524-026-02153-3}
\item \label{cit:Kononova2019} Kononova, O. et al. Text-mined dataset of inorganic materials synthesis recipes. \textit{Scientific Data} \textbf{6}, 203 (2019). doi:10.1038/s41597-019-0224-1. \url{https://doi.org/10.1038/s41597-019-0224-1}
\item \label{cit:Tshitoyan2019} Tshitoyan, V. et al. Unsupervised word embeddings capture latent knowledge from materials science literature. \textit{Nature} \textbf{571}, 95-98 (2019). doi:10.1038/s41586-019-1335-8. \url{https://doi.org/10.1038/s41586-019-1335-8}
\item \label{cit:Olivetti2020} Olivetti, E. A. et al. Data-driven materials research enabled by natural language processing and information extraction. \textit{Applied Physics Reviews} \textbf{7}, 041317 (2020). doi:10.1063/5.0021106. \url{https://doi.org/10.1063/5.0021106}
\item \label{cit:Venugopal2024} Venugopal, V. and Olivetti, E. A. MatKG: A knowledge graph for materials science. \textit{Scientific Data} \textbf{11}, 141 (2024). doi:10.1038/s41597-024-03039-z. \url{https://doi.org/10.1038/s41597-024-03039-z}
\item \label{cit:Stach2021} Stach, E. et al. Autonomous experimentation systems for materials development: A community perspective. \textit{Matter} \textbf{4}, 2702-2726 (2021). doi:10.1016/j.matt.2021.06.036. \url{https://doi.org/10.1016/j.matt.2021.06.036}
\item \label{cit:Burger2020} Burger, B. et al. A mobile robotic chemist. \textit{Nature} \textbf{583}, 237-241 (2020). doi:10.1038/s41586-020-2442-2. \url{https://doi.org/10.1038/s41586-020-2442-2}
\item \label{cit:MacLeod2020} MacLeod, B. P. et al. Self-driving laboratory for accelerated discovery of thin-film materials. \textit{Science Advances} \textbf{6}, eaaz8867 (2020). doi:10.1126/sciadv.aaz8867. \url{https://doi.org/10.1126/sciadv.aaz8867}
\item \label{cit:FloresLeonar2020} Flores-Leonar, M. M. et al. Materials acceleration platforms: On the way to autonomous experimentation. \textit{Nature Reviews Materials} \textbf{5}, 575-590 (2020). doi:10.1038/s41578-020-0226-2. \url{https://doi.org/10.1038/s41578-020-0226-2}
\item \label{cit:DeBreuck2025} De Breuck, P.-P. et al. Generative AI for crystal structures: a review. \textit{npj Computational Materials} \textbf{11}, 370 (2025). doi:10.1038/s41524-025-01881-2. \url{https://doi.org/10.1038/s41524-025-01881-2}
\item \label{cit:Wagner2026} Wagner, J. et al. Benchmarking machine learning in self-driving laboratories. \textit{Digital Discovery} (2026). doi:10.1039/D5DD00337G. \url{https://doi.org/10.1039/D5DD00337G}
\item \label{cit:Dunn2020} Dunn, A. et al. Benchmarking materials property prediction methods: the Matbench test set and Automatminer reference algorithm. \textit{npj Computational Materials} \textbf{6}, 138 (2020). doi:10.1038/s41524-020-00406-3. \url{https://doi.org/10.1038/s41524-020-00406-3}
\item \label{cit:PyzerKnapp2025} Pyzer-Knapp, E. O. et al. Foundation models for materials discovery -- current state and future directions. \textit{npj Computational Materials} \textbf{11}, 61 (2025). doi:10.1038/s41524-025-01538-0. \url{https://doi.org/10.1038/s41524-025-01538-0}
\item \label{cit:Butler2018} Butler, K. T., Davies, D. W., Cartwright, H., Isayev, O. and Walsh, A. Machine learning for molecular and materials science. \textit{Nature} \textbf{559}, 547-555 (2018). doi:10.1038/s41586-018-0337-2. \url{https://doi.org/10.1038/s41586-018-0337-2}
\item \label{cit:Schmidt2019} Schmidt, J., Marques, M. R. G., Botti, S. and Marques, M. A. L. Recent advances and applications of machine learning in solid-state materials science. \textit{npj Computational Materials} \textbf{5}, 83 (2019). doi:10.1038/s41524-019-0221-0. \url{https://doi.org/10.1038/s41524-019-0221-0}
\item \label{cit:Batra2021} Batra, R., Song, L. and Ramprasad, R. Emerging materials intelligence ecosystems propelled by machine learning. \textit{Nature Reviews Materials} \textbf{6}, 655-678 (2021). doi:10.1038/s41578-020-00255-y. \url{https://doi.org/10.1038/s41578-020-00255-y}
\item \label{cit:Chen2021} Chen, C. et al. Generative models for inverse design of inorganic solid materials. \textit{Journal of Materials Informatics} (2021). \url{https://www.oaepublish.com/articles/jmi.2021.07}
\item \label{cit:Jha2018} Jha, D. et al. ElemNet: Deep learning the chemistry of materials from only elemental composition. \textit{Scientific Reports} \textbf{8}, 17593 (2018). doi:10.1038/s41598-018-35934-y. \url{https://doi.org/10.1038/s41598-018-35934-y}
\item \label{cit:Fudge2024} Fudge, B. et al. Design and implementation of self-driving laboratories. \textit{Digital Discovery} (2024). \url{https://doi.org/10.1039/D4DD00059E}
\item \label{cit:Rank2024} Rank, K. C. et al. Self-driving labs perspective. \textit{Digital Discovery} (2024). doi:10.1039/D4DD00040D. \url{https://doi.org/10.1039/D4DD00040D}
\item \label{cit:Han2025} Han, X.-Q. et al. InvDesFlow-AL: active learning-based workflow for inverse design of functional materials. \textit{npj Computational Materials} (2025). doi:10.1038/s41524-025-01830-z. \url{https://doi.org/10.1038/s41524-025-01830-z}
\item \label{cit:Riebesell2025} Riebesell, J., Goodall, R. E. A., Benner, P., Chiang, Y., Deng, B., Ceder, G., Asta, M., Lee, A. A., Jain, A. and Persson, K. A. A framework to evaluate machine learning crystal stability predictions. \textit{Nature Machine Intelligence} \textbf{7}, 836-847 (2025). doi:10.1038/s42256-025-01055-1. \url{https://doi.org/10.1038/s42256-025-01055-1}
\item \label{cit:Sun2016} {\color{revred}Sun, W. et al. The thermodynamic scale of inorganic crystalline metastability. \textit{Sci. Adv.} \textbf{2}, e1600225 (2016). doi:10.1126/sciadv.1600225.}
\item \label{cit:Aykol2018} {\color{revred}Aykol, M., Dwaraknath, S. S., Sun, W. and Persson, K. A. Thermodynamic limit for synthesis of metastable inorganic materials. \textit{Sci. Adv.} \textbf{4}, eaaq0148 (2018). doi:10.1126/sciadv.aaq0148.}
\item \label{cit:Gouvea2026} Gouv\^{e}a, R. A. and Rignanese, G.-M. VibroML: an automated toolkit for high-throughput vibrational analysis and dynamical instability remediation of crystalline materials using machine-learned potentials. \textit{arXiv} (2026). arXiv:2604.27685. \url{https://arxiv.org/abs/2604.27685}
\item \label{cit:Negishi2025} Negishi, M., Park, H., Mastej, K. O. and Walsh, A. Continuous Uniqueness and Novelty Metrics for Generative Modeling of Inorganic Crystals. \textit{arXiv} (2025). arXiv:2510.12405. \url{https://arxiv.org/abs/2510.12405}
\item \label{cit:Antunes2024} Antunes, L. M. et al. CrystaLLM: data-efficient autoregressive generation of inorganic crystal structures. \textit{Nature Communications} \textbf{15}, 10570 (2024). doi:10.1038/s41467-024-54639-7. \url{https://www.nature.com/articles/s41467-024-54639-7}
\item \label{cit:Szymanski2023} Szymanski, N. J. et al. An autonomous laboratory for the accelerated synthesis of novel materials. \textit{Nature} \textbf{624}, 86-91 (2023). doi:10.1038/s41586-023-06734-w. \url{https://doi.org/10.1038/s41586-023-06734-w}
\item \label{cit:Chenebuah2024} Chenebuah, C., Qi, Y. and Ghosh, P. Learning crystal morphology with graph autoencoders for inverse design. \textit{npj Computational Materials} \textbf{10}, 129 (2024). doi:10.1038/s41524-024-01381-9. \url{https://doi.org/10.1038/s41524-024-01381-9}
\item \label{cit:Davies2019} Davies, D. W. et al. SMACT: Semiconducting materials by analogy and chemical theory. \textit{Journal of Open Source Software} \textbf{4}, 1361 (2019). doi:10.21105/joss.01361. \url{https://doi.org/10.21105/joss.01361}
\item \label{cit:Wang2021} Wang, A. Y.-T. et al. Compositionally restricted attention-based network for materials property prediction. \textit{npj Computational Materials} \textbf{7}, 77 (2021). doi:10.1038/s41524-021-00545-1. \url{https://doi.org/10.1038/s41524-021-00545-1}
\item \label{cit:Goodall2020} Goodall, R. E. A. and Lee, A. A. Predicting materials properties without crystal structure: Deep representation learning from stoichiometry. \textit{npj Computational Materials} \textbf{6}, 148 (2020). doi:10.1038/s41524-020-00381-w. \url{https://doi.org/10.1038/s41524-020-00381-w}
\item \label{cit:Court2020} Court, C. J. and Cole, J. M. 3-D inorganic crystal structure generation and property prediction via representation learning. \textit{J. Chem. Inf. Model.} (2020). \url{https://pubs.acs.org/doi/10.1021/acs.jcim.0c00048}
\item \label{cit:Kim2020} Kim, B. et al. Generative adversarial network for crystal structure prediction. \textit{ACS Central Science} (2020). \url{https://pubs.acs.org/doi/10.1021/acscentsci.0c00426}
\item \label{cit:Gao2025} Gao, Y. et al. ConditionCDVAE+: physically informed conditional crystal generation. \textit{Scientific Reports} (2025). \url{https://www.nature.com/articles/s41598-025-06432-9}
\item \label{cit:Karras2022} Karras, T., Aittala, M., Aila, T. and Laine, S. Elucidating the design space of diffusion-based generative models. \textit{arXiv} (2022). arXiv:2206.00364. \url{https://arxiv.org/abs/2206.00364}
\item \label{cit:Park2026} {\color{revred}Park, H. and Walsh, A. Guiding generative models to uncover diverse and novel crystals via reinforcement learning (Chemeleon-2). \textit{Nat. Mach. Intell.} (2026). doi:10.1038/s42256-026-01262-4. arXiv:2511.07158.}
\item \label{cit:Raccuglia2016} {\color{revred}Raccuglia, P., Elbert, K. C., Adler, P. D. F., Falk, C., Wenny, M. B., Mollo, A., Zeller, M., Friedler, S. A., Schrier, J. and Norquist, A. J. Machine-learning-assisted materials discovery using failed experiments. \textit{Nature} \textbf{533}, 73-76 (2016). doi:10.1038/nature17439.}
\item \label{cit:Kohler2020} {\color{revred}K\"ohler, J., Klein, L. and No\'e, F. Equivariant flows: exact likelihood generative learning for symmetric densities. \textit{PMLR} \textbf{119}, 5361-5370 (2020). \url{https://proceedings.mlr.press/v119/kohler20a.html}}
\item \label{cit:Xu2022} {\color{revred}Xu, M., Yu, L., Song, Y., Shi, C., Ermon, S. and Tang, J. GeoDiff: a geometric diffusion model for molecular conformation generation. \textit{ICLR} (2022). arXiv:2203.02923.}
\item \label{cit:Satorras2021} {\color{revred}Satorras, V. G., Hoogeboom, E. and Welling, M. E($n$) equivariant graph neural networks. \textit{PMLR} \textbf{139}, 9323-9332 (2021). \url{https://proceedings.mlr.press/v139/satorras21a.html}}
\item \label{cit:Batatia2022} {\color{revred}Batatia, I., Kov\'acs, D. P., Simm, G. N. C., Ortner, C. and Cs\'anyi, G. MACE: higher order equivariant message passing neural networks for fast and accurate force fields. \textit{Adv. Neural Inf. Process. Syst.} \textbf{35}, 11423-11436 (2022).}
\item \label{cit:Xiao2023} Xiao, H., Li, R., Shi, Q., Chen, Y., Zheng, L., Chen, F. and Xu, L. An invertible, invariant crystal representation for inverse design of solid-state materials using generative deep learning. \textit{Nature Communications} \textbf{14}, 7027 (2023). doi:10.1038/s41467-023-42870-7. \url{https://doi.org/10.1038/s41467-023-42870-7}
\item \label{cit:Jiao2024} {\color{revred}Jiao, R., Huang, W., Liu, Y., Zhao, D. and Liu, Y. Space group constrained crystal generation. \textit{ICLR} (2024). arXiv:2402.03992.}
\item \label{cit:Cao2024} {\color{revred}Cao, Z., Luo, X., Lv, J. and Wang, L. Space group informed transformer for crystalline materials generation (CrystalFormer). arXiv:2403.15734 (2024).}
\item \label{cit:Karpovich2024} Karpovich, C. et al. Deep reinforcement learning for inverse inorganic materials design. \textit{npj Computational Materials} (2024). doi:10.1038/s41524-024-01474-5. \url{https://doi.org/10.1038/s41524-024-01474-5}
\item \label{cit:Isayev2017} Isayev, O. et al. Universal fragment descriptors for predicting properties of inorganic crystals. \textit{Nature Communications} \textbf{8}, 15679 (2017). doi:10.1038/ncomms15679. \url{https://doi.org/10.1038/ncomms15679}
\item \label{cit:Huh2024} {\color{revred}Huh, M., Cheung, B., Wang, T. and Isola, P. Position: the platonic representation hypothesis. \textit{PMLR} \textbf{235} (2024). arXiv:2405.07987.}
\item \label{cit:Suzuki2025} Suzuki, Y. et al. Bridging text and crystal structures: literature-driven contrastive learning for materials science. \textit{Machine Learning: Science and Technology} \textbf{6}, 035006 (2025). doi:10.1088/2632-2153/ade58c. \url{https://doi.org/10.1088/2632-2153/ade58c}
\item \label{cit:Ozawa2024} Ozawa, N. et al. CLICS. \textit{STAM Methods} (2024). doi:10.1080/27660400.2024.2406219. \url{https://doi.org/10.1080/27660400.2024.2406219}
\item \label{cit:Radford2021} {\color{revred}Radford, A. et al. Learning transferable visual models from natural language supervision (CLIP). \textit{PMLR} \textbf{139}, 8748-8763 (2021).}
\item \label{cit:Das2023} Das, K. et al. CrysMMNet. \textit{Proceedings of Machine Learning Research (UAI)} \textbf{216} (2023). \url{https://proceedings.mlr.press/v216/das23a.html}
\item \label{cit:Weston2021} Weston, L. et al. \textit{Chemistry of Materials} (2021). doi:10.1021/acs.chemmater.1c01368. \url{https://pubs.acs.org/doi/10.1021/acs.chemmater.1c01368}
\item \label{cit:Chitturi2024} Chitturi, K. et al. Targeted materials discovery using Bayesian algorithm execution. \textit{npj Computational Materials} \textbf{10}, 126 (2024). doi:10.1038/s41524-024-01326-2. \url{https://doi.org/10.1038/s41524-024-01326-2}
\item \label{cit:Diwale2022} Diwale, S. et al. Bayesian optimization for material discovery processes with noisy and unreliable measurements. \textit{Molecular Systems Design \& Engineering} (2022). \url{https://doi.org/10.1039/D1ME00154J}
\item \label{cit:Tian2025} Tian, K. et al. Materials design with target-oriented Bayesian optimization (t-EGO). \textit{npj Computational Materials} (2025). doi:10.1038/s41524-025-01704-4. \url{https://doi.org/10.1038/s41524-025-01704-4}
\item \label{cit:ref2018} Gómez-Bombarelli, R., Wei, J. N., Duvenaud, D., Hernández-Lobato, J. M., Sánchez-Lengeling, B., Sheberla, D., Aguilera-Iparraguirre, J., Hirzel, T. D., Adams, R. P. and Aspuru-Guzik, A. Automatic Chemical Design Using a Data-Driven Continuous Representation of Molecules. \textit{ACS Central Science} \textbf{4}(2), 268-276 (2018). doi:10.1021/acscentsci.7b00572. \url{https://doi.org/10.1021/acscentsci.7b00572}
\item \label{cit:Hansen2016} Hansen, N. The CMA Evolution Strategy: A Tutorial. \textit{arXiv} (2016). arXiv:1604.00772. \url{https://arxiv.org/abs/1604.00772}
\item \label{cit:Welling2011} Welling, M. and Teh, Y. W. Bayesian Learning via Stochastic Gradient Langevin Dynamics. In \textit{Proceedings of the 28th International Conference on Machine Learning}, 681-688 (2011). \url{https://icml.cc/2011/papers/398_icmlpaper.pdf}
\item \label{cit:Li2016} Li, C., Chen, C., Carlson, D. and Carin, L. Preconditioned Stochastic Gradient Langevin Dynamics for Deep Neural Networks. In \textit{Proceedings of the 30th AAAI Conference on Artificial Intelligence} \textbf{30}(1), 1788-1794 (2016). \url{https://ojs.aaai.org/index.php/AAAI/article/view/10200}
\item \label{cit:Ho2022} Ho, J. and Salimans, T. Classifier-Free Diffusion Guidance. \textit{arXiv preprint} arXiv:2207.12598 (2022). \url{https://arxiv.org/abs/2207.12598}
\item \label{cit:Schulman2017} Schulman, J., Wolski, F., Dhariwal, P., Radford, A. and Klimov, O. Proximal Policy Optimization Algorithms. \textit{arXiv preprint} arXiv:1707.06347 (2017). \url{https://arxiv.org/abs/1707.06347}
\item \label{cit:Banik2023} Banik, S., Dhabal, D., Chan, H., Manna, S., Cherukara, M., Molinero, V. and Sankaranarayanan, S. K. R. S. CATING: Crystal structure generation from composition using attention-based neural networks. \textit{npj Computational Materials} \textbf{9}, 117 (2023). doi:10.1038/s41524-023-01094-5. \url{https://doi.org/10.1038/s41524-023-01094-5}
\item \label{cit:Kusne2020} Kusne, A. G., Yu, H., Wu, C., Zhang, H., Hattrick-Simpers, J., DeCost, B., Sarker, S., Oses, C., Toher, C., Curtarolo, S., Davydov, A. V., Agarwal, R., Bendersky, L. A., Li, M., Mehta, A. and Takeuchi, I. On-the-fly closed-loop materials discovery via Bayesian active learning. \textit{Nature Communications} \textbf{11}, 5966 (2020). doi:10.1038/s41467-020-19597-w. \url{https://doi.org/10.1038/s41467-020-19597-w}
\item \label{cit:Merchant2023} Merchant, A., Batzner, S., Schoenholz, S. S., Aykol, M., Cheon, G. and Cubuk, E. D. Scaling deep learning for materials discovery. \textit{Nature} \textbf{624}, 80-85 (2023). doi:10.1038/s41586-023-06735-9. \url{https://doi.org/10.1038/s41586-023-06735-9}
\item \label{cit:Volk2024} Volk, A. A. and Abolhasani, M. Autonomous flow chemistry platforms: benchmarking and assessment criteria. \textit{Nature Communications} \textbf{15}, 5433 (2024). doi:10.1038/s41467-024-49716-4. \url{https://doi.org/10.1038/s41467-024-49716-4}
\item \label{cit:Chen2022} {\color{revred}Chen, C. and Ong, S. P. A universal graph deep learning interatomic potential for the periodic table. \textit{Nat. Comput. Sci.} \textbf{2}, 718-728 (2022). doi:10.1038/s43588-022-00349-3.}
\item \label{cit:Deng2023} {\color{revred}Deng, B. et al. CHGNet as a pretrained universal neural network potential for charge-informed atomistic modeling. \textit{Nat. Mach. Intell.} \textbf{5}, 1031-1041 (2023). doi:10.1038/s42256-023-00716-3.}
\item \label{cit:Batatia2025} {\color{revred}Batatia, I. et al. A foundation model for atomistic materials chemistry (MACE-MP-0). \textit{J. Chem. Phys.} \textbf{163}, 184110 (2025). arXiv:2401.00096.}
\item \label{cit:Widdowson2025} Widdowson, D. and Kurlin, V. Geographic-style maps with a local novelty distance help navigate in the materials space. \textit{Scientific Reports} \textbf{15}, 27588 (2025). doi:10.1038/s41598-025-10672-0. \url{https://doi.org/10.1038/s41598-025-10672-0}
\item \label{cit:Rombach2022} Rombach, R., Blattmann, A., Lorenz, D., Esser, P. and Ommer, B. High-resolution image synthesis with latent diffusion models. \textit{CVPR} (2022). doi:10.1109/CVPR52688.2022.01042. \url{https://doi.org/10.1109/CVPR52688.2022.01042}
\item \label{cit:Shumailov2024} {\color{revred}Shumailov, I., Shumaylov, Z., Zhao, Y., Papernot, N., Anderson, R. and Gal, Y. AI models collapse when trained on recursively generated data. \textit{Nature} \textbf{631}, 755-759 (2024). doi:10.1038/s41586-024-07566-y.}
\item \label{cit:Ovadia2019} Ovadia, Y., Fertig, E., Ren, J., Nado, Z., Sculley, D., Nowozin, S., Dillon, J. V., Lakshminarayanan, B. and Snoek, J. Can You Trust Your Model's Uncertainty? Evaluating Predictive Uncertainty Under Dataset Shift. \textit{Advances in Neural Information Processing Systems} \textbf{32} (2019). \url{https://arxiv.org/abs/1906.02629}
\item \label{cit:Guo2017} Guo, C., Pleiss, G., Sun, Y. and Weinberger, K. Q. On Calibration of Modern Neural Networks. In \textit{Proceedings of the 34th International Conference on Machine Learning} \textbf{70}, 1321-1330 (2017). \url{https://arxiv.org/abs/1706.04599}
\item \label{cit:Amodei2016} Amodei, D., Olah, C., Steinhardt, J., Christiano, P., Schulman, J. and Man\'e, D. Concrete Problems in AI Safety. \textit{arXiv} (2016). arXiv:1606.06565. \url{https://arxiv.org/abs/1606.06565}
\item \label{cit:Skalse2022} Skalse, J., Howe, N. H. R., Krasheninnikov, D. and Krueger, D. Defining and Characterizing Reward Gaming. \textit{Advances in Neural Information Processing Systems} \textbf{35} (2022). \url{https://arxiv.org/abs/2209.13085}
\item \label{cit:Kapoor2023} Kapoor, S. and Narayanan, A. Leakage and the reproducibility crisis in machine-learning-based science. \textit{Patterns} \textbf{4}(9), 100804 (2023). doi:10.1016/j.patter.2023.100804. \url{https://doi.org/10.1016/j.patter.2023.100804}
\item \label{cit:Kapoor2024} Kapoor, S. et al. REFORMS: Consensus-based Recommendations for Machine-learning-based Science. \textit{Science Advances} \textbf{10}(18), eadk3452 (2024). doi:10.1126/sciadv.adk3452. \url{https://doi.org/10.1126/sciadv.adk3452}
\item \label{cit:Zagorac2019} {\color{revred}Zagorac, D., M\"uller, H., Ruehl, S., Zagorac, J. and Rehme, S. Recent developments in the Inorganic Crystal Structure Database. \textit{J. Appl. Crystallogr.} \textbf{52}, 918-925 (2019). doi:10.1107/S160057671900997X.}
\item \label{cit:Cheetham2024} {\color{revred}Cheetham, A. K. and Seshadri, R. Artificial intelligence driving materials discovery? Perspective on the article: Scaling deep learning for materials discovery. \textit{Chem. Mater.} \textbf{36}, 3490-3495 (2024). doi:10.1021/acs.chemmater.4c00643.}
\item \label{cit:Leeman2024} {\color{revred}Leeman, J., Liu, Y., Stiles, J., Lee, S. B., Bhatt, P., Schoop, L. M. and Palgrave, R. G. Challenges in high-throughput inorganic materials prediction and autonomous synthesis. \textit{PRX Energy} \textbf{3}, 011002 (2024). doi:10.1103/PRXEnergy.3.011002.}
\item \label{cit:Deng2025} {\color{revred}Deng, B. et al. Systematic softening in universal machine learning interatomic potentials. \textit{npj Comput. Mater.} \textbf{11}, 9 (2025). doi:10.1038/s41524-024-01500-6.}
\item \label{cit:Yamazaki2026} Yamazaki, S., Huang, Y., Petersen, M. H., Nong, W. and Hippalgaonkar, K. Navigating Order-(Dis)Order Family Trees via Group-Subgroup Transitions. \textit{arXiv} (2026). arXiv:2604.21386. \url{https://arxiv.org/abs/2604.21386}
\item \label{cit:Betala2025} {\color{revred}Betala, S. et al. LeMat-GenBench: a unified evaluation framework for crystal generative models. arXiv:2512.04562 (2025). Leaderboard: \url{https://huggingface.co/spaces/LeMaterial/lemat-genbench}}
\item \label{cit:Strathern1997} {\color{revred}Strathern, M. ``Improving ratings'': audit in the British University system. \textit{European Review} \textbf{5}, 305-321 (1997).}
\item \label{cit:Manheim2018} {\color{revred}Manheim, D. and Garrabrant, S. Categorizing variants of Goodhart's law. arXiv:1803.04585 (2018).}
\item \label{cit:Mu2026} {\color{revred}Mu, J., He, L., Zhu, X. and Yin, S. NextCrystal: a symmetry-driven generative framework for crystal structure prediction. arXiv:2602.17176 (2026).}
\item \label{cit:Riebesell2025b} {\color{revred}Riebesell, J., Goodall, R. E. A., Benner, P., Chiang, Y., Deng, B., Ceder, G., Asta, M., Lee, A. A., Jain, A. and Persson, K. A. A framework to evaluate machine learning crystal stability predictions (Matbench Discovery). \textit{Nat. Mach. Intell.} \textbf{7}, 836-847 (2025). doi:10.1038/s42256-025-01055-1. Leaderboard: \url{https://matbench-discovery.materialsproject.org}}
\item \label{cit:BarrosoLuque2026} {\color{revred}Barroso-Luque, L., Shuaibi, M., Fu, X., Wood, B. M., Dzamba, M., Gao, M., Rizvi, A., Zitnick, C. L. and Ulissi, Z. W. The Open Materials 2024 (OMat24) inorganic materials dataset and models. \textit{Nat. Comput. Sci.} \textbf{6}, 642-652 (2026). doi:10.1038/s43588-026-00996-w.}
\item \label{cit:Fu2025} {\color{revred}Fu, X., Wood, B. M., Barroso-Luque, L., Levine, D. S., Gao, M., Dzamba, M. and Zitnick, C. L. Learning smooth and expressive interatomic potentials for physical property prediction (eSEN). \textit{PMLR} \textbf{267}, 17875-17893 (2025). arXiv:2502.12147.}
\item \label{cit:Bigi2026} {\color{revred}Bigi, F., Pegolo, P., Mazitov, A. and Ceriotti, M. Pushing the limits of unconstrained machine-learned interatomic potentials (PET-OAM). arXiv:2601.16195 (2026).}
\item \label{cit:Loew2025} {\color{revred}Loew, A. et al. Universal machine learning interatomic potentials are ready for phonons. \textit{npj Comput. Mater.} \textbf{11}, 178 (2025). doi:10.1038/s41524-025-01650-1.}
\item \label{cit:Pota2024} {\color{revred}P\'ota, B., Ahlawat, P., Cs\'anyi, G. and Simoncelli, M. Thermal conductivity predictions with foundation atomistic models. arXiv:2408.00755 (2024).}
\item \label{cit:Wood2025} {\color{revred}Wood, B. M. et al. UMA: a family of universal models for atoms. arXiv:2506.23971 (2025).}
\item \label{cit:Rhodes2025} {\color{revred}Rhodes, B., Vandenhaute, S., \v{S}imkus, V., Gin, J., Godwin, J., Duignan, T. and Neumann, M. Orb-v3: atomistic simulation at scale. arXiv:2504.06231 (2025).}
\item \label{cit:Park2024} {\color{revred}Park, Y., Kim, J., Hwang, S. and Han, S. Scalable parallel algorithm for graph neural network interatomic potentials in molecular dynamics simulations (SevenNet). \textit{J. Chem. Theory Comput.} \textbf{20}, 4857-4868 (2024).}
\item \label{cit:Lysogorskiy2026} {\color{revred}Lysogorskiy, Y., Bochkarev, A. and Drautz, R. Graph atomic cluster expansion for foundational machine learning interatomic potentials (GRACE). \textit{npj Comput. Mater.} (2026). doi:10.1038/s41524-026-01979-1.}
\item \label{cit:Yang2024} {\color{revred}Yang, H. et al. MatterSim: a deep learning atomistic model across elements, temperatures and pressures. arXiv:2405.04967 (2024).}
\item \label{cit:Mazitov2025} {\color{revred}Mazitov, A., Bigi, F., Kellner, M., Pegolo, P., Tisi, D., Fraux, G., Pozdnyakov, S., Loche, P. and Ceriotti, M. PET-MAD as a lightweight universal interatomic potential for advanced materials modeling. \textit{Nat. Commun.} \textbf{16}, 10653 (2025). doi:10.1038/s41467-025-65662-7.}
\item \label{cit:He2023} {\color{revred}He, T. et al. Precursor recommendation for inorganic synthesis by machine learning materials similarity from scientific literature. \textit{Sci. Adv.} \textbf{9}, eadg8180 (2023). doi:10.1126/sciadv.adg8180.}
\item \label{cit:McDermott2021} {\color{revred}McDermott, M. J., Dwaraknath, S. S. and Persson, K. A. A graph-based network for predicting chemical reaction pathways in solid-state materials synthesis. \textit{Nat. Commun.} \textbf{12}, 3097 (2021). doi:10.1038/s41467-021-23339-x.}
\item \label{cit:Singh2017} {\color{revred}Singh, A. K. et al. Electrochemical stability of metastable materials. \textit{Chem. Mater.} \textbf{29}, 10159-10167 (2017). doi:10.1021/acs.chemmater.7b03980.}
\item \label{cit:Henkelman2000} {\color{revred}Henkelman, G., Uberuaga, B. P. and J\'onsson, H. A climbing image nudged elastic band method for finding saddle points and minimum energy paths. \textit{J. Chem. Phys.} \textbf{113}, 9901-9904 (2000). doi:10.1063/1.1329672.}
\item \label{cit:Jang2020} {\color{revred}Jang, J., Gu, G. H., Noh, J., Kim, J. and Jung, Y. Structure-based synthesizability prediction of crystals using partially supervised learning. \textit{J. Am. Chem. Soc.} \textbf{142}, 18836-18843 (2020). doi:10.1021/jacs.0c07384.}
\item \label{cit:Antoniuk2023} {\color{revred}Antoniuk, E. R. et al. Predicting the synthesizability of crystalline inorganic materials from the data of known material compositions. \textit{npj Comput. Mater.} \textbf{9}, 155 (2023). doi:10.1038/s41524-023-01114-4.}
\end{enumerate}
\endgroup

\end{document}